\documentclass[conference]{IEEEtran}
\IEEEoverridecommandlockouts
\usepackage{cite}
\usepackage{amsmath,amssymb,amsfonts}
\usepackage[linesnumbered,ruled,vlined]{algorithm2e}
\usepackage[noend]{algpseudocode}
\usepackage{graphicx}
\usepackage{textcomp}
\usepackage{xcolor}

\usepackage[section]{placeins}
\usepackage{float}
\usepackage{subfigure}
\usepackage{mathrsfs}
\usepackage{enumitem}
\usepackage{threeparttable}
\usepackage{soul}
\usepackage{times}
\usepackage{booktabs} 
\usepackage{bbm}
\usepackage{graphicx}
\usepackage{booktabs}
\usepackage{url}
\usepackage{multirow}
\usepackage{balance}

\newtheorem{defn}{Definition}
\newenvironment{proof}{\quad{\it Proof:}}{\hfill $\square$\par}

\newtheorem{Lemma}{Lemma}
\newtheorem{example}{Example}

\newtheorem{obser}{Observation}
\newtheorem{theorem}{Theorem}[section]

\newcommand{\kw}[1]{{\ensuremath {\mathsf{#1}}}\xspace}

\newcommand{\stitle}[1]{\vspace{0.5ex} \noindent{\bf #1}}
\long\def\comment#1{}
\newcommand{\eop}{\hspace*{\fill}\mbox{$\Box$}}

\newcommand{\etal}{\emph{et~al.}\xspace}

\newcommand\figref[1]{Fig.~\ref{#1}}
\newcommand\lamref[1]{Lemma~\ref{#1}}

\newcommand\secref[1]{Section~\ref{#1}}

\newcommand{\uonesidebc}{\kw{Single}-\kw{Side~Fair~BiClique}}
\newcommand{\onesidebc}{\kw{single}-\kw{side~fair~biclique}}
\newcommand{\onesidebcs}{\kw{single}-\kw{side~fair~bicliques}}
\newcommand{\uuonesidebc}{\kw{Single}-\kw{side~fair~biclique}}
\newcommand{\uuonesidebcpro}{\kw{Proportion}~\kw{single}-\kw{side~fair~biclique}}
\newcommand{\uonesidebcpro}{\kw{Proportion}~\kw{Single}-\kw{Side~Fair~BiClique}}
\newcommand{\osbc}{\kw{SSFBC}}
\newcommand{\osbcp}{\kw{PSSFBC}}
\newcommand{\nonesidebc}{single-side~fair~biclique}
\newcommand{\nonesidebcpro}{proportion~single-side~fair~biclique}
\newcommand{\utwosidebc}{\kw{Bi}-\kw{Side~Fair~BiClique}}
\newcommand{\twosidebc}{\kw{bi}-\kw{side~fair~biclique}}
\newcommand{\twosidebcs}{\kw{bi}-\kw{side~fair~bicliques}}
\newcommand{\uutwosidebc}{\kw{Bi}-\kw{side~fair~biclique}}
\newcommand{\uutwosidebcpro}{\kw{Proportion}~\kw{bi}-\kw{side~fair~biclique}}
\newcommand{\utwosidebcpro}{\kw{Proportion}~\kw{Bi}-\kw{Side~Fair~BiClique}}

\newcommand{\tsbc}{\kw{BSFBC}}
\newcommand{\tsbcp}{\kw{PBSFBC}}
\newcommand{\ntwosidebc}{bi-side~fair~biclique}
\newcommand{\ntwosidebcpro}{proportion~bi-side~fair~biclique}

\newcommand{\onesideFBCEM}{\kw{FairBCEM}}
\newcommand{\inonesideFBCEM}{\kw{BackTrackFBCEM}}

\newcommand{\twosideFBCEM}{\kw{BFairBCEM}}
\newcommand{\naivesearchtree}{\kw{NSF}}
\newcommand{\tsnaivesearchtree}{\kw{BNSF}}

\newcommand{\maximalfairset}{\kw{MFSCheck}}

\newcommand{\onesideFBCEMPP}{{\kw{FairBCEM}}\text{++}}
\newcommand{\inonesideFBCEMPP}{{\kw{BackTrackFBCEM}}\text{++}}
\newcommand{\onesideFBCEMPPPRO}{{\kw{FairBCEMPro}}\text{++}}
\newcommand{\inonesideFBCEMPPPRO}{{\kw{BackTrackFBCEMPro}}\text{++}}

\newcommand{\twosideFBCEMPP}{{\kw{BFairBCEM}}\text{++}}
\newcommand{\twosideFBCEMPPPRO}{{\kw{BFairBCEMPro}}\text{++}}
\newcommand{\combination}{{\kw{Combination}}}
\newcommand{\combinationpro}{{\kw{CombinationPro}}}

\newcommand{\fcore}{\kw{FCore}}
\newcommand{\cfcore}{\kw{CFCore}}
\newcommand{\buildhopgraph}{\kw{Construct2HopGraph}}

\newcommand{\bfcore}{\kw{BFCore}}
\newcommand{\bcfcore}{\kw{BCFCore}}
\newcommand{\bbuildhopgraph}{\kw{BiConstruct2HopGraph}}

\newcommand{\degreeorder}{\kw{DegOrd}}
\newcommand{\idorder}{\kw{IDOrd}}

\newcommand{\wiki}{{\kw{Wiki}}-{\kw{cat}}}
\newcommand{\twi}{\kw{Twitter}}
\newcommand{\imdb}{\kw{IMDB}}
\newcommand{\youtube}{\kw{Youtube}}
\newcommand{\dblp}{\kw{DBLP}}
\newcommand{\movies}{\kw{Movies}}
\newcommand{\job}{\kw{Jobs}}
\newcommand{\dbds}{\kw{DBDS}}
\newcommand{\dbda}{\kw{DBDA}}

\usepackage{times}
\begin{document}
\title{Fairness-aware Maximal Biclique Enumeration on Bipartite Graphs\\
}
\author{{Ziqi Yin$\scriptsize^{\dag}$, Qi Zhang$\scriptsize^{\dag}$, Wentao Zhang$\scriptsize^{\ddag}$, Rong-Hua Li$\scriptsize^{\dag}$, Guoren Wang$\scriptsize^{\dag}$}
	\vspace{1.6mm}\\
	\fontsize{9}{9}\selectfont\itshape
	$\scriptsize^{\dag}$Beijing Institute of Technology, Beijing, China; 
	$\scriptsize^{\ddag}$Peking university, Beijing, China \\
	\fontsize{8}{8}\selectfont\ttfamily\upshape
	ZIQI003@e.ntu.edu.sg; \{qizhangcs,rhli,wanggr\}@bit.edu.cn; wentao.zhang@pku.edu.cn
}

\maketitle

\begin{abstract}
Maximal biclique enumeration is a fundamental problem in bipartite graph data analysis. Existing biclique enumeration methods mainly focus on non-attributed bipartite graphs and also ignore the \emph{fairness} of graph attributes. In this paper, we introduce the concept of fairness into the biclique model for the first time and study the problem of fairness-aware biclique enumeration. Specifically, we propose two fairness-aware biclique models, called \nonesidebc~and \ntwosidebc~respectively. To efficiently enumerate all {\nonesidebc}s, we first present two non-trivial pruning techniques, called fair $\alpha$-$\beta$ core pruning and colorful fair $\alpha$-$\beta$ core pruning, to reduce the graph size without losing accuracy. Then, we develop a branch and bound algorithm, called \onesideFBCEM, to enumerate all single-side fair bicliques on the reduced bipartite graph. To further improve the efficiency, we propose an efficient branch and bound algorithm with a carefully-designed combinatorial enumeration technique. Note that all of our techniques can also be extended to enumerate all bi-side fair bicliques. We also extend the two fairness-aware biclique models by constraining the ratio of the number of vertices of each attribute to the total number of vertices and present corresponding enumeration algorithms. Extensive experimental results on five large real-world datasets demonstrate our methods' efficiency, effectiveness, and scalability. 
\end{abstract}
\vspace{-3.5mm}
\section{Introduction} \label{sec:introduction}
A bipartite graph $G(U,V,E)$ contains two disjoint vertex sets $U$ and $V$ and one edge set $E$ in which each edge links a node in $U$ and a node in $V$. Many real-world networks, such as online user-item networks \cite{WangZWDZW18, WangVR06, ZhuTWCKK17, ColaceS0MP15, Wu2020} and gene co-expression networks \cite{ZhangPRBCL14, Corel2018BipartiteNA, ChiYCH21, XingYLZZGHY22} can be modeled as bipartite graphs. Recently, the problems of analysis of bipartite graphs have attracted much attention due to numerous real-world applications, such as maximal biclique enumeration \cite{AbidiZCL20, ZhangPRBCL14, MaLHYLD22, ChenLZXL22}, butterfly counting \cite{WangFC14, WangLQZZ19, SaneiMehriST18, ZhouWC21}, and  maximum biclique search \cite{ChenL00021, wang2018new, manurangsi2018inapproximability, pardalos1999maximum}.

In recent years, the concept of fairness has also been widely investigated in data analysis related areas \cite{verma2018fairness, hardt2016equality, dwork2012fairness, abs220409888}. Many existing studies reveal that a biased machine learning model may result in discrimination upon a discrimination group, such as the gender bias and the racial bias \cite{zehlike2017fa, serbos2017fairness, beutel2019fairness, MaGTW22}. Various methods (e.g., group fairness and individual fairness \cite{verma2018fairness, BeutelCDQWWHZHC19, abs210509522}, etc.) are proposed to tackle this problem. Despite their effectiveness in data analysis applications, the fairness in graph data analysis \cite{abs210710025} is still under-explored. A notable example is that Pan \etal proposed two fairness-aware maximal clique models to find fair communities in attributed graphs \cite{abs210710025}. Their models, however, are mainly tailored for traditional attributed graphs, and they cannot be directly generalized to other types of graphs, such as bipartite graphs studied in this paper. 

In this work, we focus mainly on attributed bipartite graphs, motivated by the fact that many real-life graphs, such as online customer-product networks, can be modeled as attributed bipartite graphs. We introduce the concept of fairness into the classic biclique model and investigate the problem of mining fairness-aware bicliques on attributed bipartite graphs. Here a biclique is a subgraph of the bipartite graph in which every pair of nodes belonging to two different sides has an edge. Note that nodes at the upper side and lower side of the attributed bipartite graph are often with different types of attributes.  The fairness property can be defined on one side of nodes, and also can be defined on two sides of nodes. Therefore, we propose two new models to characterize the fairness of bicliques in bipartite graphs called \nonesidebc~and \ntwosidebc~respectively. A \nonesidebc~is a biclique that requires one side nodes satisfying the fairness property and also it is a maximal subgraph satisfying such a property. That is, the number of vertices for each attribute is no less than a threshold $\beta$ and the maximum difference between the number of vertices of every attribute is no greater than a threshold $\delta$. Similarly, a \ntwosidebc~is a biclique that guarantees fairness on both sides, and also it is the maximal subgraph that meets such a property. In a \ntwosidebc, the number of vertices in the upper side and the lower side for each attribute is no less than the thresholds $\alpha$ and $\beta$, and the maximum difference between the number of vertices of every attribute is no greater than a threshold $\delta$. Notably, both \nonesidebc~and \ntwosidebc~can be extended to the proportion fair biclique models by introducing a fairness ratio $\theta$. In particular, the threshold $\theta$ requires that on the fair side, the ratio of the number of vertices of each attribute to the total number of vertices is no less than $\theta$.

{\comment{Take \nonesidebc~as an example, the parameter $\theta$ indicates that the ratio of the number of different attribute vertex in the upper vertex set to the total number of the upper vertex set is no less than $\theta$. The \ntwosidebc~definition can be extended to ratio form similarly and easily.In the job recommendation system (e.g., \job), there is always nationality bias, foreigners are always recommended for less popular jobs even if they have a better degree and working experience. The same problem lies in the movie recommendation system(e.g., \movies), in which exposure bias exists. The intuition is that already popular always get more recommendation chances than new movies even if their quality is similar. To alleviate these problems, we can mine the \nonesidebc~on the jobs vertex side and movies vertex side according to the recommendation results, to ensure the recommendation results are not nationality or age sensitive. In scientific collaboration networks (e.g., \dblp), we may wish to find a team of experts that includes a similar number of junior and senior experts and also with different research areas. Such teams can be identified by mining the \ntwosidebc~in author-publication networks (e.g., DBLP), as the \ntwosidebc~can ensure the team contains a similar number of junior and senior researchers and also with different research areas.}}

Mining fair bicliques in bipartite graphs has a variety of applications. For instance, in scientific collaboration networks (e.g., \dblp), we may wish to find a team of experts that includes a similar number of junior and senior experts and also with different research areas. Such teams can be identified by mining the \ntwosidebc~in author-publication networks, as the \ntwosidebc~can ensure the team contains a similar number of junior and senior researchers and also with different research areas. In job recommendation systems (e.g., \job), there may exist nationality bias. That is, foreigners may be recommended for less popular jobs even if they have a better degree and working experience. The same problem lies in movie recommendation systems (e.g., \movies), in which exposure bias exists. The intuition is that already popular movies typically get more recommendation chances than relatively new movies even if they are of equal mass. To eliminate the biases, we can mine one-side fair bicliques by defining the fairness on the job side and movie side, to ensure the recommendation results are not nationality or time sensitive.

Although the practical significance of our fair biclique models, there are no existing solutions that can be used to mine all \onesidebcs~or \twosidebcs in bipartite graphs. Moreover, we show that the problem of enumerating all \onesidebcs~or \twosidebcs on bipartite graphs is NP-hard. To solve this problems, we first propose a branch and bound algorithm, called \onesideFBCEM, with two carefully-designed pruning techniques to enumerate all {\nonesidebc}s. To further improve the efficiency, we propose a novel \onesideFBCEMPP~algorithm which first enumerates all maximal bicliques and then uses a carefully-designed combinatorial enumeration technique to enumerate all results in the set of all maximal bicliques, instead of in the original bipartite graph. We show that all our techniques can also be extended to solve the \ntwosidebc~enumeration problem. To summarize, we make the following contributions.

{\vspace{-0.1mm}}
\underline{\kw{New~models}.} We propose a \nonesidebc~and a \ntwosidebc~models to characterize the fairness of cohesive bipartite subgraphs. Additionally, we also propose \nonesidebcpro~and \ntwosidebcpro~models which take account of the ratio of the number of vertices of each attribute to the total number of vertices. To the best of our knowledge, we are the first to introduce the concept of fairness into bipartite graphs for biclique mining tasks.

{\vspace{-0.1mm}}
\underline{\kw{Novel~algorithms}.} To enumerate all {\nonesidebc}s, we first propose a fair $\alpha$-$\beta$ core pruning technique to prune unpromising nodes in the original bipartite graph. 
Then, we develop a pruning technique, called colorful $\alpha$-$\beta$ core pruning, by first constructing a 2-hop graph on the fair-side vertices and then applying the colorful core pruning technique to reduce the fair-side vertices. 
A branch and bound algorithm, namely, \onesideFBCEM, is proposed to enumerate all {\nonesidebc}s. 
To further boost the performance, we develop a new algorithm called \onesideFBCEMPP~which makes use of maximal bicliques as the candidates, and then enumerates all {\nonesidebc}s in such candidates by using a carefully-devised combinatorial enumeration technique. 
Besides, we also extend the proposed pruning techniques and the enumeration algorithms to handle the \ntwosidebc~enumeration problem, which results in a basic enumeration algorithm \twosideFBCEM~and an improved algorithm \twosideFBCEMPP. Additionally, we also present the algorithms, called \onesideFBCEMPPPRO~and \twosideFBCEMPPPRO, to enumerate all {\nonesidebcpro}s and {\ntwosidebcpro}s.

\underline{\kw{Extensive~experiments}.} We conduct extensive experiments to evaluate the efficiency and effectiveness of our algorithms using five real-world networks. The results show that: (1) the pruning techniques for \onesidebc~enumeration and \twosidebc~enumeration can significantly prune unpromising vertices; (2) for \onesidebc~enumeration, \onesideFBCEMPP~is at least two orders of magnitude faster than that \onesideFBCEM; (3) for \twosidebc~enumeration, \twosideFBCEMPP~is around 3-100 times faster than \twosideFBCEM; (4) both our improved algorithms can process a large bipartite graph with 7,577,304 nodes and 12,282,059 edges. In addition, we conduct three case studies on \dblp, \job~and \movies, to evaluate the effectiveness of our solutions. The results show that both \nonesidebc~and~\ntwosidebc~can find meaningful and interesting fair communities in \dblp and fair recommendation results in \job and \movies. For reproducibility purposes, the source code of this paper is released at \url{https://github.com/Heisenberg-Yin/fairnesss-biclique}.



\comment{
\stitle{Organization.} We introduce some important notations and formulate our problem in ~\secref{sec:preliminaries}. \secref{sec:onesidebc} presents the pruning techniques and search algorithms for \nonesidebc~enumeration problem. We expand these techniques into fair two-side biclique enumeration problem in \secref{sec:twosidebc}. \secref{sec:experiments} reports the experimental results. We survey the related work in~\secref{sec:relatedwork} and conclude this work in~\secref{sec:conclusion}.
}

\vspace{-3.0mm}
\section{Preliminaries} \label{sec:preliminaries}
Let $G = (U, V, E, A)$ be an undirected, unweighted, and attributed bipartite graph, where $U(G)$ and $V(G)$ are two disjoint vertex sets, and $E(G) \subseteq U(G) \times V(G)$ denotes the edge set of $G$. Generally, we call the vertex sets $U(G)$ and $V(G)$ the upper side and lower side of $G$, respectively. $A(G)=\{A_U, A_V\}$ is the attribute set of $G$ in which $A_U$ is the attribute of vertices in $U(G)$ and $A_V$ is that of vertices in $V(G)$. For an arbitrary vertex $u$, we use $u.val$ to indicate the value of its attribute. Let $A(U)$ be the set of all attribute values of $A_U$, i.e., $A(U)=\{u.val|u \in U(G)\}$. Analogously, we denote $A(V)=\{u.val|u \in V(G)\}$. The cardinalities of $A(U)$ and $A(V)$ are $A^U_n$ and $A^V_n$, respectively. We mainly focus on the case of two-dimensional attribute for each side of $G$, i.e., $A^U_n=A^V_n=2$. Without loss of generality, we denote $A(U)=\{a^U_i| 0 \le i < A^U_n\}$ and $A(V)=\{a^V_i| 0 \le i < A^V_n\}$. The set of neighbors of vertex $u$ in graph $G$ is denoted as $N(u,G) = \{v|(u,v) \in E(G)\}$, and the degree of $u$ in $G$ is represented as $D(u, G)=|N(u,G)|$. Given a vertex set $S$, we use $N(S)=\{v|v \in N(u), \forall u \in S \}$ to indicate the set of neighbors of $S$. The number of vertices with attribute value $a^{*}_i$ in the set $S$ is $S_{a^{*}_i}=\{ v|v.val=a^{*}_i\}$ where the symbol ``$*$" is either $U$ or $V$. We omit the symbol $G$ in the above notations when the context is clear.

\begin{defn}\label{def:biclique}
	(\kw{Biclique}) Given an bipartite graph $G(U, V, E)$, a subgraph $C$ is a biclique if: (1) $E(C)=U(C) \times V(C)$; (2) $U(C) \subseteq U(G)$; (3) $V(C) \subseteq V(G)$.
\end{defn}

\comment{
\begin{defn}\label{def:biclique}
	(\kw{Biclique}) Given an bipartite graph $G(U, V, E)$ and a subgraph $C(L, R, E)$, $C$ is a biclique if it satisfies (1) $E(C)=L(C) \times R(C)$; (2) $L(C) \subseteq U(G)$; (3) $R(C) \subseteq V(G)$.
\end{defn}
\begin{defn}\label{def:maximalbiclique}
	(\kw{Maximal~biclique}) Given a bipartite graph $G(U,V,E)$ and a subgraph $C$, $C$ is a maximal biclique if it satisfies (1) $B$ is a biclique. (2) there is no other biclique $C$ satisfies $B$ is a subgraph of $C$.
\end{defn}
\begin{example}
	Consider a bipartite graph $G = (U,V,E)$ in \ref{expgraph1}. By Definition~\ref{def:maximalbiclique}, we can see that the subgraph $B(L,R)$ induced by the vertex set $\{v_3, v_4, v_7, v_9, v_{11}, v_{14} \}$ is a maximal biclique. the subgraph $C(L,R)$ induced by the vertex set $\{v_1, v_3, v_7,v_8, v_{11}\}$ is also a maximal biclique. \eop
\end{example}
}
\begin{defn}\label{def:maximalbiclique}
	(\kw{Maximal~biclique}) Given a bipartite graph $G(U,V,E)$ and a subgraph $C$, $C$ is a maximal biclique if: (1) $C$ is a biclique; (2) there is no other biclique $C' \supset C$ satisfies (1).
\end{defn}

\begin{figure*}[t!]
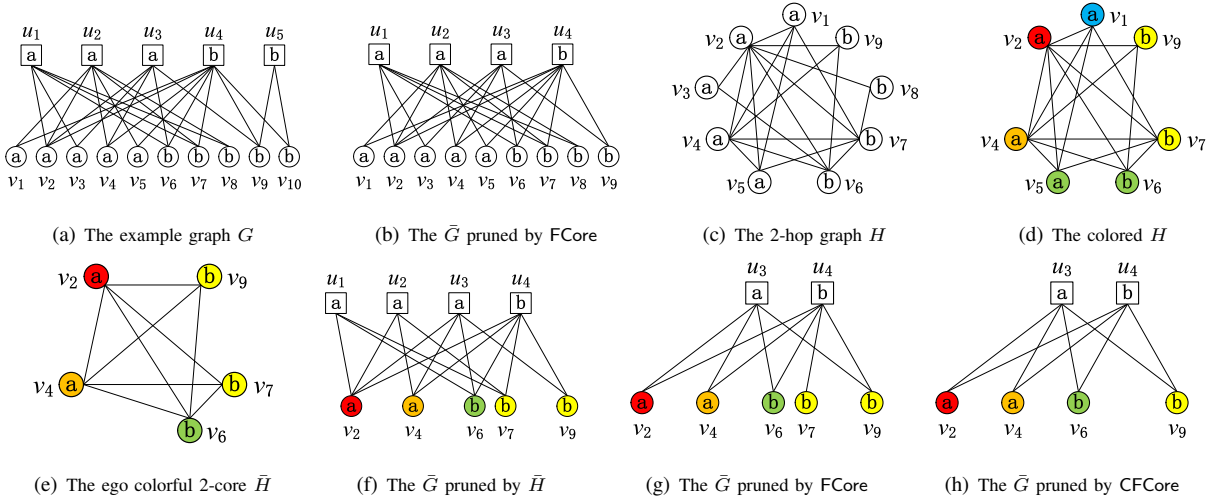
\vspace*{-0.5cm}
	\begin{center}
			\subfigure[\scriptsize{The example graph $G$}]{
				\label{fig:expgraph1}
				\includegraphics[width=0.444\columnwidth]{pic/example_1_1.eps}
			}
			\hspace*{0.2cm}
			\subfigure[\scriptsize{The $\bar G$ pruned by \fcore}]{
				\label{fig:expgraph2}
			\includegraphics[width=0.4 \columnwidth]     {pic/example_1_2.eps}
			}
			\hspace*{0.2cm}
			\subfigure[\scriptsize{The 2-hop graph $H$}]{
				\label{fig:expgraph3}
				\includegraphics[width=0.37\columnwidth]{pic/example_1_3.eps}
			}
			\hspace*{0.2cm}
			\subfigure[\scriptsize{The colored $H$}]{
				\label{fig:expgraph4}
				\includegraphics[width=0.37\columnwidth]{pic/example_1_4.eps}
			}
			\\
			\vspace{-0.1cm}
			\subfigure[\scriptsize{The ego colorful 2-core $\bar H$}]{
				\label{fig:expgraph5}
				\includegraphics[width=0.36\columnwidth]{pic/example_1_5.eps}
			}
			\hspace*{0.2cm}
			\subfigure[\scriptsize{The $\bar G$ pruned by $\bar H$}]{
				\label{fig:expgraph6}
				\includegraphics[width=0.38\columnwidth]{pic/example_1_6.eps}
			}
			\hspace*{0.2cm}
			\subfigure[\scriptsize{The $\bar G$ pruned by \fcore}]{
				\label{fig:expgraph7}
				\includegraphics[width=0.38\columnwidth]{pic/example_1_7.eps}
			}
			\hspace*{0.2cm}
			\subfigure[\scriptsize{The $\bar G$ pruned by \cfcore}]{
				\label{fig:expgraph8}			
				\includegraphics[width=0.38\columnwidth]{pic/example_1_8.eps}
			}
	\end{center}
	\vspace*{-0.5cm}
	\caption{\scriptsize The pruning process of \fcore and \cfcore on the example graph $G$.}
	\vspace*{-0.4cm}
	\label{fig:exp:Core}
\end{figure*}

{\comment{
\begin{example}
	Consider a bipartite graph $G = (U,V,E)$ in \figref{fig:expgraph1}. By Definition~\ref{def:biclique} and Definition~\ref{def:maximalbiclique}, the subgraph $C_1$ induced by the vertex set $\{u_3, u_4, v_2, v_4, v_6, v_9 \}$ is a maximal biclique as there is no larger biclique $C'$ that can contain $C$. We can also check that the biclique $C_2$ induced by the vertex set $\{u_1, u_3, v_2, v_3, v_6\}$ is also a maximal biclique. \eop
\end{example}
}}

Below, we introduce two novel fairness-aware biclique models, namely, \uonesidebc (\osbc) and \utwosidebc (\tsbc). Without losing generality, we consider $V$ as the fair side in the \osbc model and both $U$ and $V$ as the fair sides in \tsbc.



{}

\begin{defn}\label{def:faironesidebiclique}
	(\uuonesidebc) Given an attributed bipartite graph $G(U, V, E, A)$ and three integers $\alpha, \beta, \delta$, a biclique $C(U, V, E, A)$ of $G$ is a \onesidebc if (1) $|C(U)| \geq \alpha$; (2) $\forall a^V_i \in A(V), |C(V)_{a^V_i}| \geq \beta$ and $\forall a^V_i, a^V_j \in A(V)$, $||C(V)_{a^V_i}|-|C(V)_{a^V_j}|| \leq \delta$; (3) there is no biclique $C' \supset C$ satisfying (1) and (2).
\end{defn}

\begin{defn}\label{def:fairtwosidebiclique}
	(\uutwosidebc) Given an attributed bipartite graph $G(U, V, E, A)$ and three integers $\alpha, \beta, \delta$, a biclique $C(U, V, E, A)$ of $G$ is a \twosidebc if (1) $\forall a^U_i \in A(U), |C(U)_{a^U_i}| \geq \alpha$ and $\forall a^U_i, a^U_j \in A(U)$, $||C(U)_{a^U_i}|-|C(U)_{a^U_j}|| \leq \delta$; (2) $\forall a^V_i \in A(V), |C(V)_{a^V_i}| \geq \beta$ and $\forall a^V_i, a^V_j \in A(V)$, $||C(V)_{a^V_i}|-|C(V)_{a^V_j}|| \leq \delta$; (3) there is no biclique $C' \supset C$ satisfying (1) and (2).
\end{defn}

\begin{example}
	Consider an attributed bipartite graph $G = (U, V, E, A)$ in \figref{fig:expgraph1}. For the upper side $U(G)$, the values of attribute $A_U$ are represented as $a$ and $b$ in a square, respectively. And the attribute values of $A_V$ are $a$ and $b$ in a circle for the lower side $V(G)$. Suppose that $\alpha=1, \beta=2$ and $\delta=1$. By Definition~\ref{def:faironesidebiclique}, the subgraph $C_S$ induced by the vertex set $\{u_3, u_4, v_2, v_4, v_6, v_9\}$ is a \osbc of $G$ and the subgraph $C_B$ induced by $\{u_3, u_4, v_2, v_4, v_6, v_9\}$ is a \tsbc. Clearly, $C_B$ is a subgraph of $C_S$, which means that a \tsbc must be contained in {\osbc}s. \eop
\end{example}

{\comment{
\begin{example}
	Consider an attributed bipartite graph $G = (U, V, E, A)$ in \figref{fig:expgraph1}. For the upper side $U(G)$, the values of attribute $A_U$ are represented as $a$ and $b$ in a square, respectively. And the attribute values of $A_V$ are $a$ and $b$ in a circle for the lower side $V(G)$. Suppose that $\alpha=2, \beta=2$ and $\delta=1$. By Definition~\ref{def:faironesidebiclique}, the subgraph $C_S$ induced by the vertex set $\{u_3, u_4, v_2, v_4, v_6, v_9\}$ is a \osbc. This is because there are two vertices with attribute value $a$ in $C_S(V)$ and two vertices with attribute value $b$ in $C_S(V)$, and the difference between the number of vertices of $a, b$ is zero which is no larger than $\delta=1$. Moreover, we cannot find a biclique $C' \supset C_S$ satisfying $|C'(V)_{a}| \ge \alpha=2$, $|C'(V)_{b}| \ge \beta=2$ and $|C'(V)_{a}|-|C'(V)_{b}| \le \delta=1$. In the same parameter setting, we can easily check that there is no \tsbc in $G$ with Definition~\ref{def:fairtwosidebiclique}. Considering the case of $\alpha=1, \beta=2, \delta=1$, the biclique $C_S$ is still a \osbc of $G$ and the subgraph $C_B$ induced by $\{u_3, u_4, v_2, v_4, v_6, v_9\}$ is a \tsbc. Clearly, $C_B$ contains two vertices with $a$ in a circle and two vertices with $b$ in a circle in the lower side $C_B(V)$, and one vertex with $a$ in a square and one vertex with $b$ in a square in the upper side $C_B(U)$. In addition, we can find that $C_B$ is a subgraph of $C_S$, which means that a \tsbc must be contained in {\osbc}s. \eop
\end{example}
}}

In addition, fairness considers not only the number of vertices with each attribute but also the ratio of the number of vertices of each attribute to the total number of vertices on the fair side. Below, we propose two extended models of \osbc~and \tsbc, namely, \uonesidebcpro~(\osbcp) and \utwosidebcpro~(\tsbcp), to further guarantee the fairness by introducing a fairness radio threshold $\theta$.

{\comment{
Sometimes, we need to consider a more general situation, which is that we need to not only consider the number of nodes with a specific attribute but also consider the proportion of nodes with a specific attribute compared to the total vertex set. Below, we introduce two extension models of  to solve the proportion problem, namely \uuonesidebcpro (\osbcp) and \uutwosidebcpro (\tsbcp).
}}

\begin{defn}\label{def:faironesidebicliquepro}
	(\uuonesidebcpro) Given an attributed bipartite graph $G(U, V, E, A)$, three integers $\alpha, \beta, \delta$, and a float $\theta$, a biclique $C(U, V, E, A)$ of $G$ is a \nonesidebcpro~if (1) $|C(U)| \geq \alpha$; (2) $\forall a^V_i \in A(V), |C(V)_{a^V_i}| \geq \beta$ and $\forall a^V_i, a^V_j \in A(V)$, $||C(V)_{a^V_i}|-|C(V)_{a^V_j}|| \leq \delta$; (3) $\forall a^V_i \in A(V), |C(V)_{a^V_i}|/|C(V)| \geq \theta$; (4) there is no biclique $C' \supset C$ satisfying (1), (2) and (3).
\end{defn}

\begin{defn}\label{def:fairtwosidebicliquepro}
	(\uutwosidebcpro) Given an attributed bipartite graph $G(U, V, E, A)$, three integers $\alpha, \beta, \delta$, and a float $\theta$, a biclique $C(U, V, E, A)$ of $G$ is a \ntwosidebcpro~if (1) $\forall a^U_i \in A(U), |C(U)_{a^U_i}| \geq \alpha$ and $\forall a^U_i, a^U_j \in A(U)$, $||C(U)_{a^U_i}|-|C(U)_{a^U_j}|| \leq \delta$; (2) $\forall a^V_i \in A(V), |C(V)_{a^V_i}| \geq \beta$ and $\forall a^V_i, a^V_j \in A(V)$, $||C(V)_{a^V_i}|-|C(V)_{a^V_j}|| \leq \delta$; (3) $\forall a^V_i \in A(V), |C(V)_{a^V_i}|/|C(V)| \geq \theta$, $\forall a^U_i \in A(U), |C(U)_{a^U_i}|/|C(U)| \geq \theta$; (4) there is no biclique $C' \supset C$ satisfying (1), (2) and (3).
\end{defn}

\stitle{Problem statement.} Given an attributed bipartite graph $G(U, V, E, A)$, three integers $\alpha, \beta, \delta$, and a float $\theta$, our goal is to find all {\osbc}s, {\osbcp}s, {\tsbc}s, {\tsbcp}s in $G$.

\stitle{Hardness.} We first discuss the hardness of the \nonesidebc~enumeration problem. Considering a special case: $\alpha = 0, \beta=0, \delta=n$, where $n$ is the graph size. Clearly, with these parameters, the \nonesidebc~enumeration problem degenerates to the traditional maximal biclique enumeration problem, which is NP-hard. Thus, finding all {\nonesidebc}s is also an NP-hard problem. The \ntwosidebc~enumeration problem is more challenging than enumerating all {\nonesidebc}s because the number of {\ntwosidebc}s is often much larger than that of {\nonesidebc}s. By definition, we can see that a \ntwosidebc~is always contained in a~\nonesidebc. On the contrary, a~\nonesidebc~is not necessarily a \ntwosidebc. 

Compared to the traditional biclique enumeration problem, the fairness-aware biclique enumeration problem is harder. First, both \nonesidebc~and \ntwosidebc models do not satisfy the hereditary property. That is, subgraphs of a \osbc or \tsbc are not always fair subgraphs due to the attribute constraint. As a result, it is more difficult to check the maximally for both \nonesidebc~and \ntwosidebc. Second, the number of fairness-aware bicliques is generally larger than that of traditional maximal bicliques, resulting in a higher time cost to enumerate all fairness-aware bicliques. For example, on \imdb, with the parameters $\alpha=8, \beta=10, \delta=2$, the number of maximal bicliques and {\nonesidebc}s are 12,614 and 3,502,746, respectively. In the case of $\alpha=4, \beta=6, \delta=2$, we can find 42,023 maximal bicliques and 11,091,721 {\ntwosidebc}s. 

Below, we analyze the lower bounds of time complexity for finding all {\osbc}s~and {\tsbc}s. We first introduce an important theorem which is proved in \cite{prisner2000bicliques}.

\begin{theorem}\label{theo:bicliquesnum}
Every bipartite graph with $n$ vertices contains at most $2^{n/2}$ bicliques \cite{prisner2000bicliques}.
\end{theorem}

In the worst case, all bicliques can satisfy the $\alpha$ and $\beta$ constraints of Definition \ref{def:faironesidebiclique}, and thus we only consider the parameter $delta$. Given a biclique $C(U, V, E, A)$, without loss of generality, we assume that $|C(V)_{a_{1}^V}|=|C(V)_{a_{2}^V}|+n_1$ and $|C(U)_{a_{1}^U}|=|C(U)_{a_{2}^U}|+n_2$ hold, where $n_1 > \delta$ and $n_2>\delta$. Then, the number of {\osbc}s is $\tbinom{|C(V)_{a_{2}^V}+\delta|}{|C(V)_{a_{1}^V}|}$, whose maximum value is $\tbinom{\lfloor C(V)_{a_{1}^V}/2 \rfloor}{|C(V)_{a_{1}^V}|}$. Similarly, the maximum number of {\tsbc}s is equal to ${\tbinom{\lfloor C(V)_{a_{1}^V}/2 \rfloor}{|C(V)_{a_{1}^V}|}} {\tbinom{\lfloor C(U)_{a_{1}^U}/2 \rfloor}{|C(U)_{a_{1}^U}|}}$. Since there are $2^{n/2}$ bicliques (Theorem~\ref{theo:bicliquesnum}) and $C(V), C(u) \le n$ holds, finding all {\osbc}s~and {\tsbc}s take at least $O(C_{n}^{\lfloor n/2 \rfloor}*2^{n/2})$ and $O(C_{n}^{\lfloor n/2 \rfloor})^2*2^{n/2})$ time respectively as algorithms need to output these fair bicliques.

For enumerating all {\osbcp}s and {\tsbcp}s, the lower bound of time complexity can be easily derived by analogous methods of finding {\osbc}s and {\tsbc}s, we omit the analysis due to the space limit.

\vspace{-1.0mm}
\section{Single-side fair biclique enumeration} \label{sec:onesidebc}
In this section, we first introduce two non-trivial pruning techniques, called fair $\alpha$-$\beta$ core pruning and colorful fair $\alpha$-$\beta$ core pruning, to reduce the scale of a graph. Then, two branch-and-bound enumeration algorithms, called \onesideFBCEM~and \onesideFBCEMPP, are proposed to enumerate all {\nonesidebc}s. Finally, we develop the \onesideFBCEMPPPRO~algorithm to solve the \osbcp~enumeration problem.

\vspace{-1.0mm}
\subsection{Fair $\alpha$-$\beta$ core pruning} \label{sec:Fairalphabetacorepruning}
Below, we first give the definition of \emph{attribute degree} which is important to derive the fair $\alpha$-$\beta$ core pruning technique.

\begin{defn}
	\label{def:attrdeg}
	(\kw{Attribute}~\kw{degree}) Given an attributed bipartite graph $G = (U,V,E,A)$ and an attribute value $a_i \in A(U) \cup A(V)$. The attribute degree of vertex $u$, denoted by $D_{a_i}(u, G)$, is the number of vertices of $u$'s neighbors whose attribute value is $a_i$, i.e., $D_{a_i}(u, G) = | \lbrace v |v \in N(u), v.val = a_i \rbrace |$.
\end{defn}

\begin{defn}
	\label{def:fairalphabetacore}
	(\kw{Fair}~$\alpha$-$\beta$~\kw{core}) Given an attributed bipartite graph $G= (U,V,E,A)$, a subgraph $H=(L,R,E,A)$ is a fair $\alpha$-$\beta$ core if (1) $D_{a_i}(u, H) \ge \beta, u \in L, a_i \in A(V)$; (2) $D(v, H) \ge \alpha, v \in R$; (3) there is no subgraph $H' \supset H$ that satisfies (1) and (2) in $G$.
\end{defn}

With Definition \ref{def:fairalphabetacore}, we have the following lemma. Due to the space limit, all the proofs in this paper are omitted. 

\begin{Lemma}
	\label{lem:coreprune}
	Given an attributed bipartite graph $G = (U, V, E, A)$ and two integers $\alpha, \beta$, any \nonesidebc~must be contained in a fair $\alpha$-$\beta$ core.
\end{Lemma}

\comment{
\begin{proof}
    Given a fair one-side biclique $B(L,R)$ in bipartite graph $G(U,V,E,A)$, we have $|L|\geq \alpha$ and $|R_{a_i}| \geq \beta$. So $B(L,R)$ must be contained in a fair alpha-beta core. 
\end{proof}	
\vspace*{0.05cm}
}

\begin{algorithm}[t]
    \scriptsize
	\caption{\fcore}
	\label{alg:fairalphabetacore}
	\KwIn{$G = (U,V,E,A)$, two integers $\alpha,\beta$}
	\KwOut{The fair $\alpha$-$\beta$ core $\hat {G}$}
	Let ${\mathcal Q} $ be a priority queue; ${\mathcal Q} \leftarrow \emptyset $\;
	\For{$u \in U$}
	{
	    {\bf {for}} $v \in N(u)$ {\bf {do}} $D_{v.val}(u)$++\;
    	$D_{\min}(u) \leftarrow \min \lbrace D_{a^V_i}(u) | a^V_i \in A(V) \rbrace$\;	
	}
	\For{$u \in U$}
	{
	    {\bf {if}} $D_{\min}(u) < \beta$ {\bf {then}} ${\mathcal Q}.push(u)$; Remove $u$ from $G$\;
	}
	\For{$v \in V$}
	{
	    {\bf {for}} $u \in N(v)$ {\bf {do}} $D(v)$++\;
	}
	\For{$v \in V$}
	{
	    {\bf {if}} $D(v) < \alpha$ {\bf {then}} ${\mathcal Q}.push(v)$; Remove $v$ from $G$\;
	}
	\While{${\mathcal Q} \neq \emptyset$}
	{
		$u \leftarrow {\mathcal Q}.pop()$\;
		\For{$v \in N(u)$}
		{
			\If{$v$ is not removed}
			{
			    \If{$v \in U$}
			    {
			        $D_{u.val}(v)${-}{-}\; $D_{\min}(v) \leftarrow \min \lbrace D_{a^V_i}(v) | a^V_i \in A(V) \rbrace$\;	\If{$D_{\min}(v) < \beta$}
					{
						${\mathcal Q}.push(v)$; Remove $v$ from $G$\;
					}
			    }
				\If{$v \in V$}
				{
				    $D(v)${-}{-}\;
				    \If{$D(v) < \alpha$}
					{
						${\mathcal Q}.push(v)$; Remove $v$ from $G$\;
					}
                    ${\mathcal Q}.push(v)$; Remove $v$ from $G$\;		
				}				
			}
		}
	}
	$\hat {G} \leftarrow$ the remaining graph of $G$\;
	{\bf return} $\hat {G}$;
\end{algorithm}

According to \lamref{lem:coreprune}, we propose a fair $\alpha$-$\beta$ core computation algorithm, namely, \fcore, to prune unpromising vertices that definitely do not belong to any \nonesidebc. The pseudo-code of \fcore is outlined in Algorithm \ref{alg:fairalphabetacore}, which is a variant of the classic core decomposition algorithm \cite{csDS0310049, MatulaB83}. Specifically, a priority queue $Q$ is used to maintain the vertices which will be removed during the peeling procedure (line 1). \fcore first calculates the attribute degrees and degrees for vertices in the upper side and lower side, respectively, to initialize $Q$ (lines 2-10). Based on Definition \ref{def:fairalphabetacore}, for a vertex $u \in U$ (i.e., the upper side), \fcore removes $u$ from $G$ once its minimum attribute degree $D_{min}(u)$ is less than $\beta$; and for $v \in V$, (i.e., the lower side), it removes $v$ from $G$ once its degree $D(v)$ is less than $\alpha$. After that, the algorithm computes the fair $\alpha$-$\beta$ core of $G$ by iteratively peeling vertices from the remaining graph based on their degrees and attribute degrees (lines 11-24). Finally, \fcore returns the remaining graph $\hat{G}$ as the fair $\alpha$-$\beta$ core. It is easy to show that \fcore consumes $O(E+V)$ time using $O(U \times A^V_n + V)$ space.


{\comment{
\vspace*{0.05cm}
\begin{example}
		Consider the bipartite graph $G = (U, V, E, A)$ in \figref{fig:expgraph1}. Suppose that $\alpha=2, \beta=2$. We first consider the upper side of $G$. $u_5$ is connected to $v_9$ and $v_{10}$ whose attribute values are $b$ in the lower side, and thus we have $D_a(u_5)=0$ and $D_b(u_5)=2$. Further, $D_{min}(u_5)=D_a(u_5)=0 \le \beta=2$. According to \lamref{lem:coreprune}, $u_5$ must not be contained in a \nonesidebc, thus we can safely remove $u_5$ from $G$. The removal of $u_5$ subsequently updates the degrees of $v_9$ and $v_{10}$ which are equal to $2$ and $1$, respectively. With Definition \ref{def:faironesidebiclique}, $v_{10}$ can be pruned since the \nonesidebc~model requires that each vertex in the lower side has at least $\alpha=2$ neighbors. The \fcore algorithm repeatedly removes vertices until the remaining vertices in the upper side and the lower side satisfy $D_{min}(*) \ge \beta$ and $D(*) \ge \alpha$, respectively. Finally, the remaining graph is shown in \figref{fig:expgraph2}. \eop
\end{example}
\vspace*{0.05cm}
}}



\vspace{-1.0mm}
\subsection{Colorful fair $\alpha$-$\beta$ core pruning} \label{sec:colorfulfairalphabetacorepruning}

The fair $\alpha$-$\beta$ core pruning may not be very effective as it only employs the constraint of attribute degree and ignores the property of cliques. To this end, we present a more powerful pruning technique, called Colorful Fair $\alpha$-$\beta$ core (\cfcore) pruning, by establishing an interesting connection between our problem and the weak fair clique model proposed in \cite{abs210710025}.


Recall that by Definition \ref{def:faironesidebiclique}, in a \nonesidebc~$C$, any two vertices in $C(V)$ share at least $\alpha$ common neighbors. Thus, we can construct a 2-hop graph $H(V, E, A)$ on the fair side of $G$ as follows. We keep the vertices of $H$ as those in the lower side of $G$, i.e., $H(V)=G(V)$ and $A=A_V$. Given two vertices $v_i, v_j \in V(H)$, if the number of common neighbors of $v_i$ and $v_j$ in $G$ is no less than $\alpha$, we connect $v_i$ and $v_j$ in $H$ as $v_i$ and $v_j$ may appear in the same \nonesidebc. With the 2-hop graph $H$, we have the following observation. 

\begin{obser}\label{obs:obstransweak}
Given an attributed bipartite graph $G$ and its 2-hop graph $H$. For an arbitrary \nonesidebc~$C$, the vertices in $C(V)$ form a clique $\hat C$ in $H$ in which the number of vertices whose attribute value equals $a^V_i$ is no less than $\beta$.
\end{obser}

With Observation \ref{obs:obstransweak}, the clique $\hat C$ satisfies the fairness restriction of the weak fair clique model in \cite{abs210710025}. As a weak fair clique is maximal, $\hat C$ must be contained in a weak fair clique. Thus, we can apply the colorful core pruning technique proposed in \cite{abs210710025} to prune unpromising vertices in $H$ that cannot form a weak fair clique. However, the colorful core pruning in \cite{abs210710025} does not consider the attribute value of the vertex itself. Below, we give the variants of colorful degree and colorful core, called \emph{ego colorful degree} and \emph{ego colorful core} by incorporating the vertex attribute.

\begin{defn}
	\label{def:egocolorfuldeg}
	(\kw{Ego}~\kw{colorful}~\kw{degree}) Given an attributed graph $G=(V,E,A)$ and an attribute value $a_i \in A$. The ego colorful degree of vertex $u$, denoted by $ED_{a_i}(u, G)$, is the number of colors of $u$ and $u$'s neighbors whose attribute value is $a_i$, i.e., $ED_{a_i}(u, G) = | \lbrace color(v) |v \in N(u) \cup \{u\}, v.val = a_i \rbrace|$.
\end{defn}

In Definition~\ref{def:egocolorfuldeg}, the color of each node can be obtained by the classic greedy graph coloring algorithm \cite{14spaacolororder}, which ensures that two adjacent nodes have different colors. Let $ED_{\min}(u, G)$ denotes the minimum ego colorful degree of a vertex $u$, i.e., $ED_{\min}(u, G) = \min \lbrace ED_{a_i}(u, G) | a_i \in A \rbrace$. We omit the symbol $G$ in $ED_{a_i}(u, G)$ and $ED_{\min}(u, G)$ when the context is clear.

\begin{defn}
	\label{def:egocolorfulcore}
	(\kw{Ego}~\kw{colorful}~$k$-\kw{core}) Given an attributed graph $G=(V,E,A)$ and an integer $k$, a subgraph $H=(V_H, E_H, A)$ of $G$ is an ego colorful $k$-core if: (1) for each vertex $u \in V_H, ED_{\min}(u, H) \ge k$; (2) there is no subgraph $H'$ that satisfies (1) and $H' \supset H$.
\end{defn}

Based on Definition \ref{def:egocolorfulcore}, we have the following lemma.

\begin{Lemma}
	\label{lem:eccorekwfc}
    Given an attributed bipartite graph $G$, its 2-hop graph $H$, and the parameters $\alpha, \beta, \delta$. For an arbitrary \nonesidebc~$C$, the vertices in $C(V)$ must be contained in the ego colorful $\beta$-core of $H$.
\end{Lemma}

\begin{algorithm}[t]
    \scriptsize
	\caption{\cfcore}
	\label{alg:colorfulprune}
	\KwIn{$G = (U,V,E,A)$, two integers $\alpha,\beta$}
	\KwOut{The pruned graph $\hat{G}$}
	$\bar G(U,V,E,A) \leftarrow \fcore(G, \alpha, \beta)$\;
	Let ${\mathcal Q} $ be a priority queue; ${\mathcal Q} \leftarrow \emptyset $\;
	$H(V,E,A_V) \leftarrow \buildhopgraph(\bar G, \alpha, G(V))$\;
	\For{$u \in H(V)$}
	{
	    {\bf {if}} $D(u, H) < A^V_n \times \beta-1$ {\bf {then}} Remove $u$ from $H$\;
	}
	Color all vertices in $H$ by invoking a degree based greedy coloring algorithm\;
	\For{$u \in H(V)$}
	{
		\For{$v \in N(u) \cup \{u\}$}
		{
			{\bf {if}} $M_u(v.val, color(v)) = 0$ {\bf {then}} $ED_{v.val}(u)$\text{++}\;
			$M_u(v.val, color(v))\text{++}$\;
		}
		$ED_{\min}(u) \leftarrow \min \lbrace ED_{a^V_i}(u) | a^V_i \in A(V) \rbrace$\;
		
	}
	\For{$u \in H(V)$}
	{
	    {\bf {if}} $ED_{\min}(u) < \beta$ {\bf {then}} ${\mathcal Q}.push(u)$; Remove $u$ from $H$\;
	}
	\While{${\mathcal Q} \neq \emptyset$}
	{
		$u \leftarrow {\mathcal Q}.pop()$\;
		\For{$v \in N(u, H)$}
		{
			\If{$v$ is not removed}
			{
				$M_v(u.val, color(u))${-}{-}\;
				\If{$M_v(u.val, color(u)) \le 0$}
				{
					$ED_{u.val}(v) \leftarrow ED_{u.val}(v) - 1$\;
					$ED_{\min}(v) \leftarrow \min \lbrace ED_{a^V_i}(v) | a^V_i \in A(V) \rbrace$\;
					
					\If{$ED_{\min}(v) < \beta$}
					{
						${\mathcal Q}.push(v)$; Remove $v$ from $H$\;
					}
				}
			}
		}
	}
	The ego colorful $\beta$-core $\bar {H} \leftarrow$ the remaining graph of $H$\;
	\For{$u \in \bar G(V)-\bar H(V)$}
	{
	    Remove $u$ from $\bar G(V)$\;
	}
	$\hat G \leftarrow \fcore(\bar G=(U,V,E,A), \alpha, \beta)$\;
	{\bf return} $\hat {G}$;
\end{algorithm}
\vspace*{0.05cm}

With \lamref{lem:eccorekwfc}, we can construct a 2-hop graph $H$ based on the fair side $V$ and prune the vertices in $G(V)$ that cannot form a \nonesidebc~by calculating the ego colorful $\beta$-core of $H$. Obviously, the scale of ego colorful $\beta$-core is smaller than that of $H$. That means that some vertices in the lower side can be removed from $G$, and thus we can further apply the \fcore to prune the vertices in both the upper side and lower side of $G$. Based on this idea, we propose a colorful fair $\alpha$-$\beta$ core pruning algorithm, namely, \cfcore, as shown in Algorithm \ref{alg:colorfulprune}. The \cfcore algorithm works as follows. It first performs \fcore (Algorithm \ref{alg:fairalphabetacore}) to calculate the fair $\alpha$-$\beta$ core $\bar G$ according to \lamref{lem:coreprune} (line 1). The \cfcore algorithm then constructs a 2-hop graph $H$ on the fair (lower) side $G(V)$ (Algorithm \ref{alg:2hopgraph}), and deletes the vertices whose degree is less than $A^V_n \times \beta -1$ as such vertices clearly cannot form a \nonesidebc~(lines 3-5). After that, \cfcore uses the greedy coloring for $H$ which colors vertices based on the order of degree \cite{csDS0310049, MatulaB83}, and computes the ego colorful $\beta$-core $\bar H$ by iteratively peeling vertices from the remaining graph based
on their ego colorful degrees (lines 6-24). According to \lamref{lem:eccorekwfc}, the \cfcore safely removes the vertices that are not contained in the ego colorful $\beta$-core $\bar H$ from $\bar G$. It further performs \fcore (Algorithm \ref{alg:fairalphabetacore}) again to reduce the vertices for both the upper side and lower side of $\bar G$ (lines 25-27). Finally, \cfcore returns the pruned graph $\hat{G}$ which contains all {\nonesidebc}s. Algorithm~\ref{alg:colorfulprune} consumes $O(E + V + \sum_{u \in U}d(u,G)^2 + \sum_{v \in V}d(v,G)^2)$ time using $O(V\times A^{V}_n \times color)$ space.

\begin{algorithm}[t]
    \scriptsize
	\caption{\buildhopgraph}
	\label{alg:2hopgraph}
	\KwIn{$G = (U,V,E,A)$, a integer $\alpha$, the fair side $V$}
	\KwOut{The 2-hop graph $H$ based on the fair side $V$}
	Let $H=(V=G(V),E=\emptyset,A=A_V)$ be an attributed graph\;
	\For{$v \in G(V)$}
	{
	    Initialize an array $C$ with $C[i]=0, 1 \le i \le |G(V)|$\;
	    \For{$u \in N(v, G)$}{
	        \For{$w \in N(u, G)$}{
	            {\bf {if}} $w \neq v$ {\bf {then}} $C[w] \leftarrow {\mathcal C[w]+1}$\;
	        }
	    }
	    \For{$u \in G(V)$}
        {
            {\bf {if}} $C[u] \ge \alpha$ \emph{and} $u<v$ {\bf {then}} $E(H) \leftarrow E(H) \cup (u, v)$\;
    	} 
	}
	{\bf return} ${H}$;
\end{algorithm}

\begin{example}
	Consider the bipartite graph $G = (U, V, E, A)$ in \figref{fig:expgraph1}. Suppose that we set $\alpha=2, \beta=2$. The \cfcore first performs \fcore to calculate fair $\alpha$-$\beta$ core denoted by $\bar G$ as shown in \figref{fig:expgraph2}. Then it constructs 2-hop graph $H$ for the fair side $V$ of $\bar G$ (i.e., the vertices in circle), which is illustrated in \figref{fig:expgraph3}. The vertex $v_3$ in \figref{fig:expgraph3} with two neighbors cannot form a \nonesidebc, and we remove it from $H$. This is because a \nonesidebc~$C$ contains at least $A^V_n \times \beta$ vertices in the lower side $V$, which requires that the vertices in $V(C)$ should have at least $A^V_n \times \beta-1 = 2 \times 2-1=3$ neighbors in the 2-hop graph $H$. Analogously, vertex $v_8$ in \figref{fig:expgraph3} is not included in a \nonesidebc~and we also remove $v_8$ from $H$. After the degree pruning, we color $H$ using a greedy coloring algorithm \cite{14spaacolororder} as shown \figref{fig:expgraph4}, and computes the ego colorful $2$-core $\bar H$. Taking $v_1$ as an example, we derive the ego colorful degrees of $v_1$, i.e., $ED_{a}(v_1, H) = 4$ and $ED_{b}(v_1, H) = 1$. Further, we have $ED_{\min}(v_1, H) = 1 \le \beta=2$. Thus, $v_1$ can be safely removed, since it is not in the ego colorful $2$-core and also not in a \nonesidebc~by \lamref{lem:eccorekwfc}. \figref{fig:expgraph5} shows the ego colorful $2$-core $\bar H$. We use $\bar H$ to prune the bipartite graph $\bar G$. The remaining graph is illustrated in \figref{fig:expgraph6}. Clearly, in the lower side, the pruned $\bar G$ only has 5 vertices while the previous $\bar G$ in \figref{fig:expgraph2} has 9 vertices. Further, \cfcore performs \fcore again to remove the vertices in $\bar G$ as depicted in \figref{fig:expgraph7} and \figref{fig:expgraph8}. The final graph pruned by \cfcore is shown in \figref{fig:expgraph8}, which is significantly small than the original graph in \figref{fig:expgraph1}. \eop
\end{example}




\subsection{The \onesideFBCEM algorithm} \label{sec:onesidealg}	
Before introducing the \onesideFBCEM algorithm, we first give two important definitions, i.e., \emph{fair set} and \emph{maximal fair subset}.

\begin{defn}\label{def:fairset}
	(\kw{Fair~set}) Given an attributed set $S$ with attribute values in $A$ and two integers $k, \delta$, we call $S$ is a fair set if (1) $\forall a_i \in A, |S_{a_i}| \geq k$; (2) $\forall a_i, a_j \in A, ||S_{a_i}|-|S_{a_j}|| \leq \delta$.
\end{defn}

\begin{defn}\label{def:maximalfairsubset}
	(\kw{Maximal~fair~subset}) Given an attributed set $S$ with attribute values in $A$ and two integers $k, \delta$, $\hat{S} \subseteq S$ is a maximal fair subset if (1) $\hat{S}$ is a fair set based on $k, \delta$; (2) there is no fair set $\bar S \subset S$ satisfying $\hat{S} \subset \bar S$.
\end{defn}

\comment{Consider a graph $G = (U,V,E,A)$ in \figref{fig:expgraph1}. Suppose that $k=2$ and $\delta=0$. By Definition~\ref{def:maximalfairsubset}, we can easily derive that the set $\lbrace v_1, v_2, v_3, v_4, v_6, v_7, v_8, v_9 \rbrace$ is a maximal fair subset of $\lbrace v_1, v_2, v_3, v_4, v_5, v_6, v_7, v_8, v_9 \rbrace$.}

Here we propose an efficient algorithm to identify whether a set $\hat{S}$ is the maximal fair subset of the set $S$ as shown in Algorithm \ref{alg:ismaximalfairsubset}. Clearly, $\hat{S}$ is a maximal fair subset when it satisfies there is no subset of $S-\{\hat S\}$ could be added into $\hat{S}$ without harming its fairness.

Equipped with \cfcore pruning techniques, we propose the \onesideFBCEM algorithm which enumerates all {\nonesidebc}s based on a branch and bound search method. In \onesideFBCEM, there are four important sets: $L, R, P, Q$ which control the generation of the search tree. 
Specifically, we use $R$ to denote the currently-found vertices in the lower side $V$ which may be extended to a \nonesidebc. $L$ is the vertex set in the upper side $U$ in which every vertex is a neighbor of all vertices in $R$. $P$ is the candidate set in $V$ that can be used to extend $R$ in the search tree. $Q$ is the set of vertices in which every vertex can be used to expand $R$ but has already been visited in previous search paths. Below, we give some observations to explain our \onesideFBCEM algorithm.

\begin{algorithm}[t]
	\scriptsize
	\caption{\maximalfairset}
	\label{alg:ismaximalfairsubset}
	\KwIn{The sets $S, \hat S$, the set of attribute values $A$, two integers $k, \delta$}
	\KwOut{true: $\hat S$ is a maximal fair subset; false: $\hat S$ is not a maximal fair subset}
	{\bf {if}} $\exists a_i \in A, \hat{S}_{a_i} < k$ {\bf {then}} {\bf return false};\\
	$C \leftarrow S-\hat{S}$\;
	{\bf {if}} $\forall a_i \in A, |C_{a_i}| > 0$ {\bf {then}} {\bf return false};\\
	\For{$a_i \in A$}
	{
		\If{$|C_{a_i}| > 0$}{
	    	{\bf {if}} $\exists u \in C_{a_i}, \hat{S} \cup \lbrace u \rbrace$ is a fair set {\bf {then}} {\bf return false};
		}
	}
	{\bf return true};
\end{algorithm}

\begin{obser}\label{obs:obs1_maximal}
If $\forall a^V_i \in A(V)$, we can find that at least one vertex $v \in Q$ with $v.val = a^V_i$ satisfying $\forall u \in L, (u,v) \in E$, $R$ is not a maximal and thus we can end the current search and all deeper searches.
\end{obser}

\comment{
\begin{proof}
    With this condition, we can add a node in $Q$ of every attribute value into $R$. Thus, the currently-found $R$ is not maximal. As  and $Q$. 
    Hence we can not find any fair one-side biclique due to the maximal. In the following search process, $L$ will shrink by the search tree so that the above argument will always hold on.
\end{proof}	

}

\begin{obser}\label{obs:obs2_maximalfairset}
Given a fair set $R$, if there is no vertex set $S \subseteq P \cup Q$ which is fully connected to $L$ and could be added into $R$ without breaking the fairness, then $(L, R)$ is a \nonesidebc.
\end{obser}


\begin{obser}\label{obs:obs3_fulladdp}
If all nodes in $P$ are fully connected to $R$, and $R \cup P$ is a fair set, then we can add all vertices in $P$ into $R$ without losing solution. 
\end{obser}

\comment{
\begin{proof}
    With this condition, $(L, R \cup P)$ is the maximum \nonesidebc~in this search branch. Any possible \nonesidebc~must be contained in $(L, R \cup P)$, which means it is  not maximal. thus no \nonesidebc~is lost adn we can saadding all vertices in $P$ into $R$ does not lose e .
\end{proof}	
\vspace*{0.05cm}
}

\begin{obser}\label{obs:obs4_alphabeta}
If $|L| < \alpha$ or $\exists a^V_i \in A(V), |R_{a^V_i}|+|P_{a^V_i}|<\beta$, we can terminate the current search branch.
\end{obser}

\comment{
\begin{proof}
    since with the search tree proceed, the size of $L$ and $R\cup P$ shrink, when the search branch is less than the size bound, we end this search branch.
\end{proof}	
\vspace*{0.05cm}
}

Based on above observations, the \onesideFBCEM algorithm for \nonesidebc~enumeration is outlined in Algorithm \ref{alg:FairBCEM}. It first employs the \cfcore pruning to remove vertices that cannot be in a \nonesidebc~and initializes four sets $L, R, P, Q$, and then invokes the \inonesideFBCEM procedure to find all {\nonesidebc}s with the branch-and-bound technique. In \inonesideFBCEM, each vertex $x$ in $P$ is used to extend the current-found $R$. With the adding of $x$, $L$ must be updated to keep out those vertices that are not adjacent to $x$, as each vertex in $L$ is a neighbor of all vertices in $R$ (lines 7-8). A variable $flag$, initialized as true, indicates that whether there is a \nonesidebc~in the current branch. We denote $Q^{FC}$ and $P^{FC}$ the vertices in $Q$ and $P$ that are fully connected to $L$ respectively, which are used to check the maximality of $R$. Clearly, if $|L'|<\alpha$, we cannot find a \nonesidebc~because it violates the restriction on the number of vertices in the upper side in Definition \ref{def:faironesidebiclique}, and thus we set $flag$ to false (line 9). Then, the \inonesideFBCEM procedure identifies whether $R$ is maximal with the set $Q$ based on Observation \ref{obs:obs1_maximal} and maintains the value of $flag$ and the set $Q'$ (lines 10-15). Once $flag$ equals false, there is no \nonesidebc~in the current branch and we move $x$ from $P$ to $Q$ to indicate that $x$ has been searched (lines 29-30). Otherwise, the \inonesideFBCEM computes the sets $P'$ and $P^{FC}$ with the candidate set $P$ (lines 17-20). If $P'=P^{FC}$, all vertices in $P$ are fully connected to $R'$ and we can directly check if $(L', R' \cup P^{FC})$ is a \nonesidebc~according to Observation \ref{obs:obs3_fulladdp}. If so, \inonesideFBCEM adds the biclique $(L', R' \cup P^{FC})$ into the result set $Res$ and updates $P'$ and $P^{FC}$ as empty sets (lines 21-23). After that, the procedure identifies whether $R'$ is a maximal fair set of $R' \cup P^{FC} \cup Q^{FC}$ by Algorithm \ref{alg:ismaximalfairsubset} and adds $(L', R')$ into $Res$ by Observation \ref{obs:obs2_maximalfairset} (lines 24-26). Subsequently, If $P' \neq \emptyset$ and $\forall a^V_i \in A(V), |R'_{a^V_i}|+|P'_{a^V_i}| \geq \beta$, \inonesideFBCEM performs the next backtracking with the new $L', R', P', Q'$ (lines 27-28). The final set $Res$ maintains all {\nonesidebc}s in $G$ (line 4).

\begin{algorithm}[t]
	\scriptsize
	\caption{\onesideFBCEM}
	\label{alg:FairBCEM}
	\KwIn{A bipartite graph $G = (U, V, E, A)$, three integers $\alpha,\beta,\delta$}
	\KwOut{The set of all {\nonesidebc}s $Res$}
	$\hat G=(\hat U,\hat V, \hat E, A) \leftarrow \cfcore(G, \alpha, \beta)$\;
	$L \leftarrow \hat U$; $R \leftarrow \emptyset$; $P \leftarrow \hat {V}$; $Q \leftarrow \emptyset$\;
	$\inonesideFBCEM(L,R,P,Q)$\;
	{\bf return} ${Res}$\;
    \vspace*{0.1cm}
	{\bf Procedure} $\inonesideFBCEM(L, R, P, Q)$\\
    \While{$P \neq \emptyset$}
    {
	    $x \leftarrow$ a vertex in $P$; $flag \leftarrow true$\;
	    $R' \leftarrow R \cup \lbrace x \rbrace$; $L' \leftarrow \lbrace u \in L | (u,x) \in {\hat E} \rbrace$\;
	    {\bf {if}} $|L'| < \alpha$ {\bf {then}} $flag \leftarrow false$\;
    	\For{$u \in Q$}{
    	    $N(u) = \lbrace v \in L'| (u,v) \in {\hat E} \rbrace$\;
    	    {\bf {if}} $|N(u)| = |L'|$ {\bf {then}} $Q^{FC} \leftarrow Q^{FC} \cup \lbrace u \rbrace$\;
    	    {\bf {if}} $|N(u)| \geq \alpha$ {\bf {then}} $Q' \leftarrow Q' \cup \lbrace u\rbrace$\;
    	}
    	\If{$\forall a^V_i \in A(V), Q^{FC}_{a^V_i} > 0$}{
    	    $flag \leftarrow false$; 
    	}
    	\If{flag}{
    		\For{$v \in P, v \neq x $}{
    		    $N(v) = \lbrace u \in L'|(u,v) \in {\hat E} \rbrace$\;
    		    {\bf {if}} $|N(v)|=|L'|$ {\bf {then}} $P^{FC}\leftarrow P^{FC} \cup \lbrace v \rbrace$\;
    	        {\bf {if}} $|N(v)| \geq \alpha$ {\bf {then}} $P'\leftarrow P' \cup \lbrace v \rbrace$\;
    		}
    		\If{$P^{FC}=P'$}{
    		    \If{$(L',R'\cup P^{FC})$ is a fair one-side biclique}{
    		    $R' \leftarrow R' \cup P^{FC}$; $P^{FC}\leftarrow \emptyset;P' \leftarrow \emptyset$\;
    		    }
    		}
    		\If{$R'$ is a fair set}{
    		    \If{$R'$ is maximal fair subset of $R'\cup P^{FC} \cup Q^{FC}$}{
    		        $Res \leftarrow Res \cup (L',R')$\;
    		    }
    		 }
    	    \If{$P' \neq \emptyset$ and $\forall a^V_i \in A(V), |R'_{a^V_i}|+|P'_{a^V_i}| \geq \beta$}
    	    {
    		        $\inonesideFBCEM(L',R',P',Q')$\;
    	    }
        }
    $P \leftarrow P-\{x\}$\;
    $Q \leftarrow Q \cup \{x\}$\;
    }
\end{algorithm}

\stitle{Correctness analysis}. Clearly, we enumerate all possible $R$ based on the sets $P, Q$ and all {\nonesidebc}s~lie in the enumeration tree, thus the completeness of our algorithm is satisfied. The fairness and maximality of a biclique are satisfied at line 22 and line 25 of Algorithm \ref{alg:FairBCEM}. Besides, the set $Q$ can guarantee that each {\nonesidebc} only be enumerated once, thus our algorithm also satisfy the non-redundancy property. In conclusion, our \onesideFBCEM algorithm can correctly output all {\nonesidebc}s.

\comment{
\begin{example}
	Consider the bipartite graph $G = (U, V, E, A)$ in \figref{example9}. Assume that we set $\alpha=2,\beta=1,\delta=0$, we have the search tree generated by \onesideFBCEM in \figref{example10}, the red content means it is a \nonesidebc. When $L= \lbrace v_{1}, v_{2}, v_{3} \rbrace$ and $R=\lbrace v_{6},v_{7} \rbrace$, according to Observation \ref{obs:obs2_maximalfairset}, we have that we find out a fair one-side bilique, the same as the situation for $L=\lbrace v_1,v_3 \rbrace$ and $R=\lbrace v_{7},v_{8} \rbrace$. When the $L=\lbrace v_{1},v_{3} \rbrace$, $R=\lbrace v_8 \rbrace$ and $Q=\lbrace v_{6},v_{7} \rbrace$, we end this search branch according to observation \ref{obs:obs1}. When $L=\lbrace v_{4},v_{5} \rbrace and R=\lbrace v_{9} \rbrace, P=\lbrace v_{10} \rbrace$, according to observation \ref{obs:obs3}, we add $P$ into $R$, we have another fair one-side biclique whose $L=\lbrace v_{4},v_{5} \rbrace,R=\lbrace v_{9},v_{10} \rbrace$.
	\eop
\end{example}
\vspace*{0.05cm}
}

\subsection{The \onesideFBCEMPP~algorithm} \label{sec:onesidealgplus}	
The \onesideFBCEM algorithm may suffer from large search space due to enormous {\nonesidebc}s. To further improve the efficiency, we propose a new algorithm, called \onesideFBCEMPP, which first enumerates all maximal bicliques and then uses a combinatorial enumeration technique to find all {\nonesidebc}s in the set of all maximal bicliques. Our algorithm is based on the key observation that any \nonesidebc~must be contained in a biclique.

\begin{algorithm}[t]
	\scriptsize
	\caption{\onesideFBCEMPP}
	\label{alg:FairBCEMplus}
	\KwIn{A bipartite graph $G = (U, V, E, A)$, three integers $\alpha,\beta,\delta$}
	\KwOut{The set of all {\nonesidebc}s $Res$}
	$\hat G=(\hat U,\hat V, \hat E, A) \leftarrow \cfcore(G, \alpha, \beta)$\;
	$L \leftarrow \hat U$; $R \leftarrow \emptyset$; $P \leftarrow \hat {V}$; $Q \leftarrow \emptyset$\;
	$\inonesideFBCEMPP(L,R,P,Q)$\;
	{\bf return} ${Res}$\;
	\vspace*{0.1cm}
	{\bf Procedure} $\inonesideFBCEMPP(L, R, P, Q)$\\
	\While{$P \neq \emptyset$}{
		$x \leftarrow$ a vertex in $P$; $flag \leftarrow true$\;
	    $R' \leftarrow R \cup \lbrace x \rbrace$; $L' \leftarrow \lbrace u \in L | (u,x) \in {\hat E} \rbrace$\;
	    {\bf {if}} $|L'| < \alpha$ {\bf {then}} $flag \leftarrow false$\;
		\For{$u \in Q$}{
			$N(u)=\lbrace v \in L' | (u,v) \in {\hat E} \rbrace$\;
			{\bf {if}} $|N(u)| = |L'|$ {\bf {then}} $flag \leftarrow false;$ {\bf {break}}\;
			{\bf {if}} $|N(u)| > 0$ {\bf {then}} $Q'\leftarrow Q'\cup\lbrace u\rbrace$\;
		}
		$C \leftarrow C \cup \lbrace u \rbrace$\;
		\If{flag}{
			\For{$v \in P, v \neq x $}{
				$N(v)= \lbrace u \in L'|(u,v) \in {\hat E} \rbrace$\;
				\If{$|N(v)|=|L'|$}{
					$R'\leftarrow R' \cup \lbrace v \rbrace$\;
					$N^{lap}(v)=\lbrace u| u\in L/L',(u,v) \in {\hat E} \rbrace$\;
					{\bf {if}} $|N^{lap}(v)|=0$ {\bf {then}} $C \leftarrow C \cup \lbrace v \rbrace$\;
				}
				{\bf {if}} $|N(v)| \geq \alpha$ {\bf {then}} $P'\leftarrow P' \cup \lbrace v \rbrace$\;
			}
			\If{$(L', R')$ is a \nonesidebc}{
				$Res \leftarrow Res \cup (L',R')$\;
			}\Else{
				${\cal R}' \leftarrow \combination(R', A(V), \beta, \delta)$\; 
				\For{$r' \in {\cal R}'$}{
					{\bf {if}} $N(r')=L$ {\bf {then}} $Res \leftarrow Res \cup (L', r')$\;
				}	
			}
			\If{$P' \neq \emptyset$ and $\forall a^V_i \in A(V), |R'_{a^V_i}|+|P'_{a^V_i}| \geq \beta$}
    	    {
    		        $\inonesideFBCEMPP(L',R',P',Q')$\;
    	    }
		}
		$P=P-C$\;
		$Q=Q \cup C$\;		
	}
\end{algorithm}

\comment{
\begin{algorithm}[t]
	\scriptsize
	\caption{Combination($S$,$k$,$\delta$)}
	\label{alg:Combination}
	\If{$\hat{S}_{a_i} \textless k, \exists a_i \in A$}{
		{\bf return $\emptyset$};
	}
	minsize=$min(S_{a_i}),a_i \in A$\;
	\For{$a_i \in A$}{
		contraintsize=$min(S_{a_i},minsize+\delta)$\;
		$res[a_i] \leftarrow$ all subsets of $S_{a_i}$ that size is contraintsize\;
	}
	candidates $\leftarrow res[0]$\;
	\For{$a_i \in A,a_i \neq 0$}{
		candidates=candidates $\times$ $res[a_i]$\;
	} 
	{\bf return candidates};
\end{algorithm}
}

More specifically, \onesideFBCEMPP~first find all maximal bicliques satisfying $|L| \geq \alpha$ and $R_{a^V_i} \geq \beta, \forall a^V_i \in A(V)$, and then enumerates all {\nonesidebc}s among them. The pseudo-code of \onesideFBCEMPP~is depicted in Algorithm \ref{alg:FairBCEMplus}. Similar to \onesideFBCEM, \onesideFBCEMPP~uses the \cfcore pruning to remove unpromising vertices and then performs the \inonesideFBCEMPP~procedure to find all {\nonesidebc}s (lines 1-3). In each iteration of \inonesideFBCEMPP, we find all maximal bicliques based on the idea of the MBEA++ algorithm \cite{ZhangPRBCL14} which adds a set of vertices (i.e., the set $C$) into $R$ once. Specifically, it first extends $R$ by adding $x$ and obtain the set $L'$ in which vertices are linked to $x$ (lines 7-8). Then, it determines whether $(L', R')$ is a maximal biclique by trying to add each vertex $u$ in $Q$ to the current biclique. Clearly, if not, we can terminate the current search as any \nonesidebc~must be in a biclique (lines 10-13). Otherwise, we move the vertices connected to all vertices in $L'$ from $P$ to $R'$ once and update the sets $C$ and $P'$ (lines 16-22). 
We consider two cases for $(L', R')$: (1) $R'$ is a fair set then $(L',R')$ is a \nonesidebc~(lines 23-24); (2) $R'$ is not a fair set then we calculate all maximal fair subsets of $R'$ to further enumerate {\nonesidebc}s (lines 25-28). The maximal fair subsets can be obtained by a combinatorial enumeration method as illustrated in Algorithm \ref{alg:Combination}. Let $r' \in {\cal R}'$ be a maximal fair subset of $R'$. If $N(r')$ equals $L$, we obtain a \nonesidebc and the \inonesideFBCEMPP~procedure adds $(L', r')$ into the result set $Res$ (line 28). Similar to \onesideFBCEM, \inonesideFBCEMPP~invokes the next backtracking procedure if $P' \neq \emptyset$ and $\forall a^V_i \in A(V), |R'_{a^V_i}|+|P'_{a^V_i}| \geq \beta$ hold (lines 29-30). Finally, the set $Res$ maintains all {\nonesidebc}s in $G$ (line 4).

\stitle{Correctness analysis.} The bicliques with $|L| \geq \alpha, |R_{a_i}| \geq \beta$ are enumerated due to the correctness of MBEA++ \cite{ZhangPRBCL14}. For any maximal biclique $B(L,R')$, the algorithm enumerates all {\nonesidebc}s in $B$. Since every {\nonesidebc} is contained in a maximal bilcique, \onesideFBCEMPP~satisfies completeness. In line 26, we find all maximal fair subsets of $R'$ by the \combination~algorithm and identify whether they form a biclique with $L$. Thus, the fairness constraint is satisfied. As $L$ is shrinking during the search process, the maximality is also met due to the line 28. Meanwhile, each {\nonesidebc}~$B'(L,R')$'s $L$ is the $L$ of a maximal biclique $B(L,R')$ and every maximal biclique has different $L$, thus every {\nonesidebc} only be enumerated in one maximal biclique, which avoids repeated enumeration.

\begin{algorithm}[t]
	\scriptsize
	\caption{\combination}
	\label{alg:Combination}
	\KwIn{A set $S$, the set of attribute value $A$, two integers $k, \delta$}
	\KwOut{The set of all combinations ${\cal C}an{\cal S}et$}
	\If{$\exists a_i \in A, S_{a_i} < k$}{
		{\bf return} $\emptyset$;
	}
	$msize=\mathop{\min}_{a_i \in A} S_{a_i}$\;
	\For{$a_i \in A$}{
		$csize=\mathop{\min} (S_{a_i}, msize+\delta)$\;
		${\cal R}es(a_i) \leftarrow$ all subsets of $S_{a_i}$ that with size equals $csize$\;
	}
	${\cal C}an{\cal S}et \leftarrow {\cal R}es(a_0)$\;
	\For{$a_i \in A, i \neq 0$}{
		${\cal C}an{\cal S}et={\cal C}an{\cal S}et \times {\cal R}es(a_i)$\;
	} 
	{\bf return} ${\cal C}an{\cal S}et$;
\end{algorithm}

\stitle{Extending to finding all {\osbcp}s.} We propose an algorithm, called \onesideFBCEMPPPRO, to enumerate all {\osbcp}s by slightly modifying \onesideFBCEMPP~(Algorithm \ref{alg:FairBCEMplus}). Specifically, in line 23 of Algorithm \ref{alg:FairBCEMplus}, \onesideFBCEMPPPRO~replaces the inspection for a \nonesidebc~with the inspection for a \nonesidebcpro~which can be easily implemented. Additionally, in line 26 of Algorithm \ref{alg:FairBCEMplus}, we use a different algorithm, called \combinationpro, instead of \combination, to enumerate {\nonesidebcpro}s. The workflow of \combinationpro~is similar to that of \combination, and the difference is that \combinationpro~calculates $csize$ by $min(S_{a_{i}},msize+\delta,msize*\frac{(1-\theta)}{\theta})$ (line 5 in Algorithm \ref{alg:Combination}). The third item comes from the proportion constraint which can be easily derived by the inequality $\frac{msize}{msize+csize} \geq \theta$. Due to the space limit, we omit the pseudo-codes of \onesideFBCEMPPPRO~and \combinationpro.

\vspace{-1.0mm}
\section{Bi-side fair biclique enumeration} \label{sec:twosidebc}
This section first revises the pruning techniques for solving the \nonesidebc~enumeration problem to fit into our \ntwosidebc~enumeration problem. Then, we propose an algorithm, called \twosideFBCEM, by extending \onesideFBCEM~to enumerate all fair {\ntwosidebc}s. Similarly, we also propose an algorithm called \twosideFBCEMPP~by extending the \onesideFBCEMPP~algorithm. Finally, we present the \twosideFBCEMPPPRO~algorithm to solve {\tsbcp} enumeration problem by adapting the \twosideFBCEMPP~algorithm.


\subsection{The pruning techniques} \label{sec:TwoFairalphabetacorepruning}
In \nonesidebc~enumeration, we derive two pruning techniques by considering the attribute degrees of vertices on the fair side (i.e., the lower side $V$). In the \ntwosidebc~model, the attribute constraint is expanded to both the upper side and lower side, thus a natural idea is to employ the attribute degrees of vertices in $U$ and $V$ to design the pruning methods. Below, we give two pruning techniques, namely, \bfcore and \bcfcore, which are variants of \fcore and \cfcore, respectively.

\stitle{Bi-fair $\alpha$-$\beta$ core pruning} (\bfcore). Similar to \fcore, we introduce the concept of bi-fair~$\alpha$-$\beta$~core as Definition \ref{def:bifairalphabetacore} and derive the \lamref{lem:bicoreprune} to prune vertices in both $U$ and $V$ that are definitely not in any \ntwosidebc.

\begin{defn}
	\label{def:bifairalphabetacore}
	(\kw{Bi\text{-}fair}~$\alpha$-$\beta$~\kw{core}) Given an attributed bipartite graph $G= (U,V,E,A)$, a subgraph $H=(L,R,E,A)$ is a bi-fair $\alpha$-$\beta$ core if (1) $D_{a_i}(u, H) \ge \beta, u \in L, a_i \in A(V)$; (2) $D_{a_i}(v, H) \ge \alpha, v \in R, a_i \in A(U)$; (3) there is no subgraph $H' \supset H$ that satisfies (1) and (2) in $G$.
\end{defn}

\begin{Lemma}
	\label{lem:bicoreprune}
	Given an attributed bipartite graph $G = (U, V, E, A)$ and two integers $\alpha, \beta$, any \ntwosidebc~must be contained in a bi-fair $\alpha$-$\beta$ core.
\end{Lemma}

With \lamref{lem:bicoreprune}, a question is how to calculate the bi-fair $\alpha$-$\beta$ core of a bipartite graph $G$. We devise a peeling algorithm, called \bfcore, by slightly modifying \fcore (Algorithm \ref{alg:fairalphabetacore}), as Definition \ref{def:bifairalphabetacore} is also a variant of the classic $k$-core \cite{csDS0310049,MatulaB83}. Specifically, for each vertex $v$ in $V$, \bfcore calculates the attribute degree $D_{a^U_i}(v)$ instead of the degree $D(v)$ (lines 2-6). When a vertex $u$ is removed, the algorithm updates the attribute degrees for its neighbors and maintains the priority queue $Q$. If a neighbor $v$ is in the lower side $V$, \bfcore calculates the new attribute degree $D_{a^U_i}(v)$ as it is in the upper side $U$ (lines 16-19). The other steps of \bfcore are similar to those of \fcore and thus we omit the pseudo-code of \bfcore.

\begin{algorithm}[t]
    \scriptsize
	\caption{\bbuildhopgraph}
	\label{alg:22hopgraph}
	\KwIn{$G = (U,V,E,A)$, a integer $\alpha$, the fair side $V$}
	\KwOut{The 2-hop graph $H$ based on the fair side $V$}
	Let $H=(V=G(V),E=\emptyset,A=A_V)$ be an attributed graph\;
	\For{$v \in G(V)$}
	{
	    $C$ is an array with $C[i][j]=0, 1 \le i \le |G(V)|, 1 \le j \le |A(U)|$\;
	    \For{$u \in N(v, G)$}{
	        \For{$w \in N(u, G)$}{
	            {\bf {if}} $w \neq v$ {\bf {then}} $C[w][w.val] \leftarrow {\mathcal C[w][w.val]+1}$\;
	        }
	    }
	    \For{$u \in G(V)$}
        {
            \If{$\forall a^U_i \in A(U), C[u][a^U_i] \ge \alpha$ and $u<v$}{
	            $E(H) \leftarrow E(H) \cup (u, v)$\;
	        }
    	} 
	}
	{\bf return} ${H}$;
\end{algorithm}

\begin{algorithm}[t]
	\scriptsize
	\caption{\twosideFBCEM}
	\label{alg:BFairBCEM}
	\KwIn{A bipartite graph $G = (U, V, E, A)$, three integers $\alpha, \beta, \delta$}
	\KwOut{The set of all {\ntwosidebc}s $Res$}
	$\hat G=(\hat U,\hat V, \hat E, A) \leftarrow \bcfcore(G, \alpha, \beta)$\;
	$L \leftarrow \hat U$; $R \leftarrow \emptyset$; $P \leftarrow \hat {V}$; $Q \leftarrow \emptyset$\;
	Enumerate all {\nonesidebc}s by $\onesideFBCEM(\hat G, \alpha, \beta, \delta)$\;
	\For{each \nonesidebc~$B(L', R')$}{
	    ${\cal L'} \leftarrow \combination(L', A(U), \alpha, \delta)$\; 
    	\For{$l' \in {\cal L'}$}{
    		\If{$R'$ is a maximal fair subset of $N(l')$}{
    			$Res \leftarrow Res \cup(l', R')$\;
    		}
    	}
	}
	{\bf return} ${Res}$\;
\end{algorithm}

\stitle{Bi-colorful fair $\alpha$-$\beta$ core pruning} (\bcfcore). In \cfcore, we construct the 2-hop graph on the fair side $V$ by adding an edge for two vertices with at least $\alpha$ common neighbors (i.e., the condition (1) in Definition \ref{def:faironesidebiclique}). While the \ntwosidebc~model considers the fairness on both $U$ and $V$. Thus, when building the 2-hop graph on $V$, we only add an edge for two vertices if they share at least $\alpha$ common neighbors for each attribute value $a^U_i \in A(U)$ (i.e., the condition (1) in Definition \ref{def:fairtwosidebiclique}). Here, we revise the 2-hop graph algorithm to fit the \ntwosidebc~enumeration problem, which is outlined in Algorithm \ref{alg:22hopgraph}. In the graph constructed by \bbuildhopgraph, we can still calculate the ego colorful $\beta$-core to prune the unpromising vertices in $V$. 

In addition, the \ntwosidebc~model also requires fairness on the upper side $U$, and thus we can prune the vertices in $U$ like handling the lower side $V$. Based on this idea, we propose the \bcfcore algorithm which is similar to \fcore and we only make the following minor changes. In particular, for the lower side $V$, \bcfcore constructs the 2-hop graph by \bbuildhopgraph instead of \buildhopgraph (line 3 in Algorithm \ref{alg:colorfulprune}), and computes the ego colorful $\beta$-core to prune the vertices in $V$. And for the upper side $U$, \bcfcore again builds the 2-hop graph by \bbuildhopgraph with parameters $(G, \beta, U)$, and calculates the ego colorful $\alpha$-core to prune the unpromising vertices in $U$. Due to the space limitation, we omit the pseudo-code of \bcfcore.

\subsection{The \twosideFBCEM algorithm}
\label{sec:twosidealg}

Before introducing our \twosideFBCEM algorithm, we first give the following observation. 

\begin{obser}\label{obs:obs6}
	A \ntwosidebc~must be contained in {\nonesidebc}s.
\end{obser}

With Observation \ref{obs:obs6}, we present the \twosideFBCEM algorithm as shown in Algorithm \ref{alg:BFairBCEM}. We first search all {\nonesidebc}s and then enumerate all {\ntwosidebc}s by combination of the upper side. Specifically, \twosideFBCEM  invokes \onesideFBCEM to search all {\nonesidebc}s (line 3). Given a \nonesidebc~$B(L',R')$, it satisfies the fairness restriction on the lower side, and we enumerate all maximal fair subsets of $L'$ in the upper side to ensure fairness by the \combination~algorithm (line 5). For a maximal fair subset of $l'$ in ${\cal L}'$, the \twosideFBCEM algorithm determines whether $R'$ is a maximal subset of $N(l')$ (line 7). Clearly, if yes, $(l', R')$ is a \ntwosidebc~and we add it into $Res$. As all {\ntwosidebc}s are contained in all {\nonesidebc}s based on Observation \ref{obs:obs6}. The \twosideFBCEM algorithm correctly returns all {\ntwosidebc}s.

\stitle{Correctness analysis.} All {\nonesidebc}s are correctly enumerated by \onesideFBCEM and any \ntwosidebc~must be included in a {\nonesidebc}, so the completeness is satisfied. The maximality is met by the line 7 of Algorithm \ref{alg:BFairBCEM}, since $l'$ is a maximal fair subset of $N(R)$ and $R'$ is a maximal fair subset of $N(l')$, which also verifies the fairness restriction. For non-redundancy, it is obviously that any {\ntwosidebc} enumerated in a {\nonesidebc} has the same $R'$, and there is no two different {\nonesidebc}s has the same $R$, thus any {\ntwosidebc} is enumerated once.

\subsection{The \twosideFBCEMPP~algorithm}
\label{sec:twosidealgplus}
Based on Observation \ref{obs:obs6}, we can also invoke the \onesideFBCEMPP~algorithm to search all {\nonesidebc}s and then enumerate all {\ntwosidebc}s by the combinatoral enumeration method. Hence, we propose the \twosideFBCEMPP~algorithm which can be easily devised by slightly modifying  Algorithm \ref{alg:BFairBCEM}. That is, we use \onesideFBCEMPP~instead of \onesideFBCEM~in line 3 to find all {\nonesidebc}s. Due to the space limitation, we omit the pseudo-code of \twosideFBCEMPP.

\stitle{Extending to finding all {\tsbcp}s.} We can slightly adapt the \twosideFBCEMPP~algorithm to solve {\tsbcp} enumeration problem, which is called \twosideFBCEMPPPRO. That is, we replace \combination~with \combinationpro~(line 5 in Algorithm \ref{alg:BFairBCEM}), and use the inspection for a \tsbcp instead of that for a \tsbc~(lines 3-4 in Algorithm \ref{alg:BFairBCEM}). It is worth noting that we also need to check whether the ratio constraint is satisfied for maximal fair subset checking (line 7 in Algorithm \ref{alg:BFairBCEM}). We omit the details of \twosideFBCEMPPPRO~due to the space limit.

\vspace{-1.0mm}
\section{Experiments} \label{sec:experiments}

\subsection{Experimental setup} \label{sec:setup}
For \nonesidebc~enumeration problem, we implement \onesideFBCEM (Algorithm \ref{alg:FairBCEM}) and \onesideFBCEMPP~(Algorithm \ref{alg:FairBCEMplus}) equipped with the pruning techniques \fcore (Algorithm \ref{alg:fairalphabetacore}) and \cfcore (Algorithm \ref{alg:colorfulprune}). To enumerate all {\ntwosidebc}s, the \twosideFBCEM (Algorithm \ref{alg:BFairBCEM}) and \twosideFBCEMPP~are implemented armed with the \bfcore and \bcfcore pruning techniques. For comparison, we implement two naive search algorithms, i.e., \naivesearchtree and \tsnaivesearchtree, to find all {\osbc}s and {\tsbc}s, which reserve the pruning techniques such as Algorithm \ref{alg:fairalphabetacore} and Algorithm \ref{alg:colorfulprune} and drop off all pruning techniques in the search process such as Observation \ref{obs:obs1_maximal}, Observation \ref{obs:obs3_fulladdp} and Observation \ref{obs:obs4_alphabeta}. We also implement the above enumeration algorithms with two different vertex selection orderings, i.e., \degreeorder and \idorder, which are obtained by sorting the vertices based on a non-increasing manner of their degrees and IDs respectively. All algorithms are implemented in C++. We conduct all experiments on a PC with a 2.10GHz Inter Xeon CPU and 256GB memory. We set the time limit for all algorithms to $24$ hours, and use the symbol ``INF'' to denote that the algorithm cannot terminate within $24$ hours.

\stitle{Datasets.} We evaluate the efficiency of the proposed algorithms in five real-world graphs. Specifically, \wiki is a feature network. \youtube, \imdb are affiliation networks, \twi is an interaction network and \dblp is an authorship network. All datasets can be downloaded from \url{http://konect.cc/}. Note that all these datasets are non-attributed bipartite graphs, thus we randomly assign an attribute to each vertex to generate attributed graphs for evaluating the efficiency of all algorithms. 

{\comment{There are four parameters in our algorithms: $\alpha$, $\beta$, $\delta$ and $\theta$. Since different datasets have various scales, the parameter $\alpha$ and $\beta$ is set within different integers. For \osbc~(\osbcp)~and \tsbc~(\tsbcp)~enumeration problems, we also set parameters within different integers. The detailed parameter settings are illustrated in Table \ref{tab:paras}. In the experiments, we study the performance of our algorithms with varying $\alpha$, $\beta$, $\delta$ and $\theta$. Unless otherwise specified, the value of a parameter is set to its default value when varying another parameter.
}}

\stitle{Parameters.} There are four parameters in our algorithms: $\alpha$, $\beta$, $\delta$ and $\theta$. $\alpha$ and $\beta$ are used to restrict the size of fair bicliques. If $\alpha$ and $\beta$ are too small, we will obtain too many small bicliques which are not meaningful. When $\alpha$ and $\beta$ are too large, most of the vertices will be pruned during the pruning processing and the remaining graph will miss much structural information, resulting in few bicliques being outputted. We carefully fine-tune them to extract meaningful fair bicliques based on the biclique numbers in real-life datasets. $\delta$ represents the maximum difference between the number of vertices of every attribute. With $\delta$ increases, the fairness between different attributes in vertex set decreases. Therefore, $\delta$ should not be set to be too large or the problem will degenerate to the maximal biclique enumeration problem. The parameter $\theta$ is the fairness ratio threshold and we can easily derive that $\theta$ is no larger than $0.5$. Thus, $\theta$ also should not be set to be too large. Since different datasets have various scales, the parameter $\alpha$ and $\beta$ is set within different integers. For \osbc~(\osbcp)~and \tsbc~(\tsbcp)~enumeration problems, we also set parameters within different integers. The detailed parameter settings can be found on the website \url{https://github.com/Heisenberg-Yin/fairnesss-biclique}. \comment{We study the performance of our algorithms with varying $\alpha$, $\beta$, $\delta$ and $\theta$. Unless otherwise specified, when varying another parameter, the value of a parameter is set to its default value which is listed in Table I.}

{}

\begin{table}[t!]\vspace*{-0.3cm}
\scriptsize
\setlength{\tabcolsep}{0.3mm}
\caption{Datasets and Parameters}
\label{tab:datasets}
\vspace*{-0.5cm}
\begin{center}
{
    \begin{tabular}{c|c|c|c|c|c|c|c|c|c|c} \hline
	{\bf Dataset} & $|U|$ &$|V|$ &$|E|$& Density &$\alpha^{*s}$&$\beta^{*s}$&$\alpha^{*b}$&$\beta^{*b}$&$\delta^*$&$\theta^*$\\ \hline
	{\bf \youtube} &$94,238$&$30,087$ &$293,360$&$1.0 \times 10^{-4}$&8 &8  &5  &5  &2  &0.4\\
    {\bf \twi} &$175,214$&$530,418$&$1,890,661$&$2.0\times 10^{-5}$&8   &8  &6  &7  &2  &0.4\\
	{\bf \imdb} &$303,617$&$896,302$&$3,782,463$&$1.4 \times 10^{-5}$&10   &10  &6  &6  &2  &0.4\\
    {\bf \wiki} &$1,853,493$&$182,947$&$3,795,796$&$1.1\times 10^{-5}$&7   &7  &6  &6  &2  &0.4\\
    {\bf \dblp} &$1,953,085$&$5,624,219$&$12,282,059$&$1.1\times 10^{-6}$&7   &7  &4  &4  &2  &0.4\\
	\hline
    \end{tabular}
}
\end{center}
\vspace*{-0.1cm}
Note: $\alpha^{*s}, \beta^{*s}$ and $\alpha^{*b}, \beta^{*b}$ are the default values of $\alpha, \beta$ for \osbc (\osbcp) and \tsbc~(\tsbcp) models respectively, $\delta^*, \theta^*$ are the default values of $\delta$ and $\theta$.
\vspace*{-0.5cm}
\end{table}



\vspace{-1.0mm}
\subsection{Efficiency testing} \label{sec:results}
\vspace{-0.5mm}
\begin{figure*}[t!]\vspace*{-0.5cm}
\centering
    \subfigure[{\scriptsize \youtube (vary $\alpha$)}]{
      \label{fig:exp-oneside-alg-time-youtube-alpha}
      \begin{minipage}{3.2cm}
      \centering
      \includegraphics[width=\textwidth]{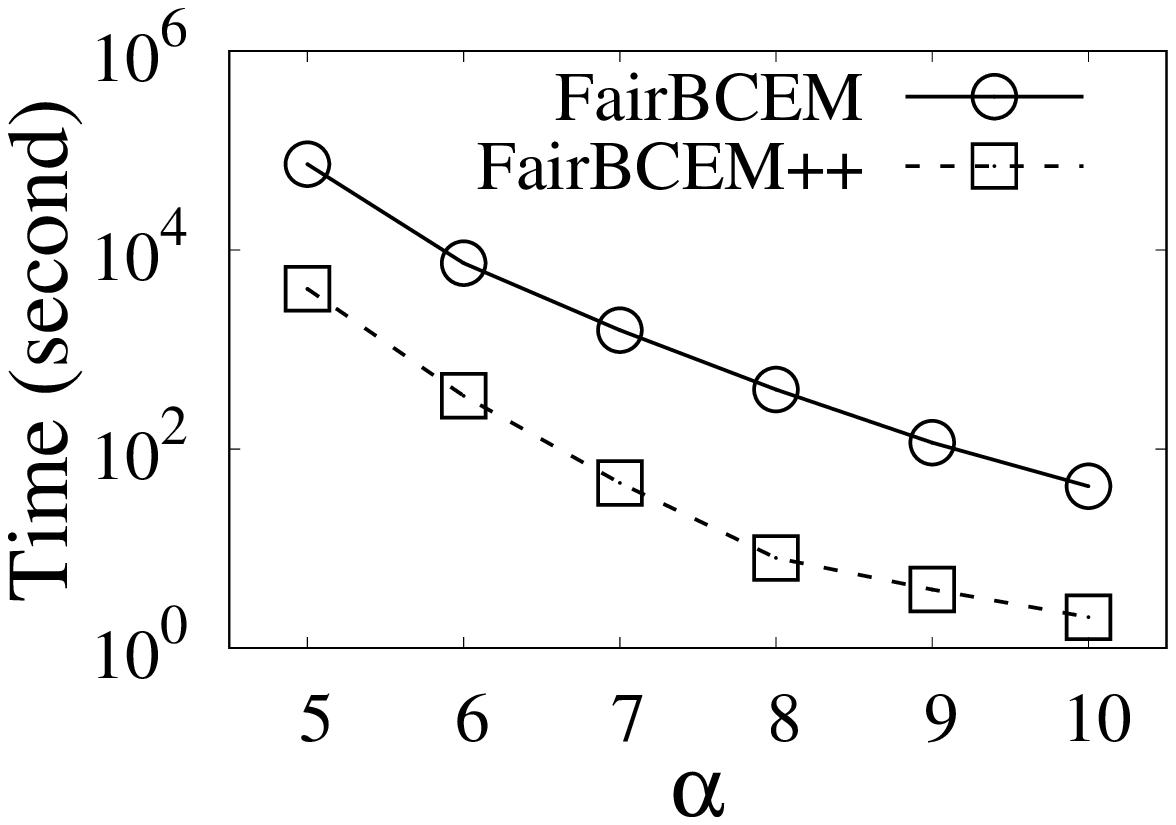}
      \end{minipage}
    }
    \subfigure[{\scriptsize \twi (vary $\alpha$)}]{
      \label{fig:exp-oneside-alg-time-twi-alpha}
      \begin{minipage}{3.2cm}
      \centering
      \includegraphics[width=\textwidth]{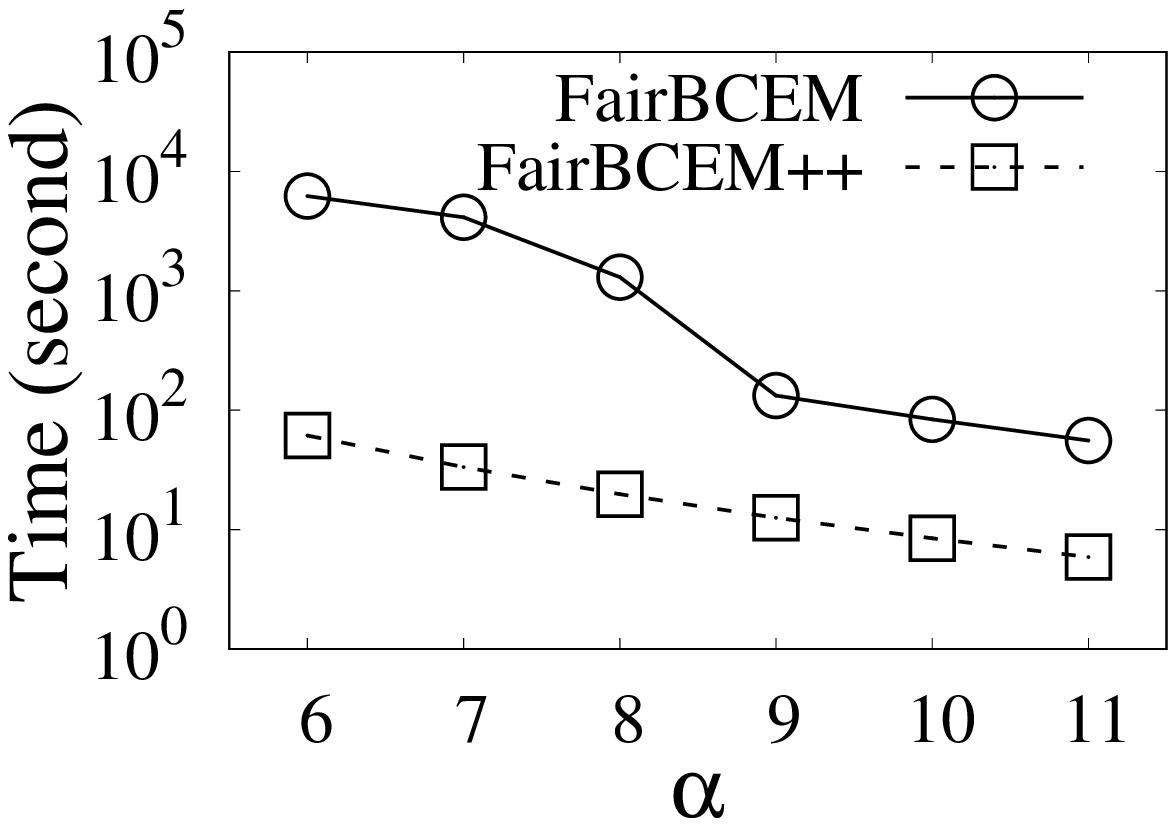}
      \end{minipage}
    }
    \subfigure[{\scriptsize \imdb (vary $\alpha$)}]{
      \label{fig:exp-oneside-alg-time-imdb-alpha}
      \begin{minipage}{3.2cm}
      \centering
      \includegraphics[width=\textwidth]{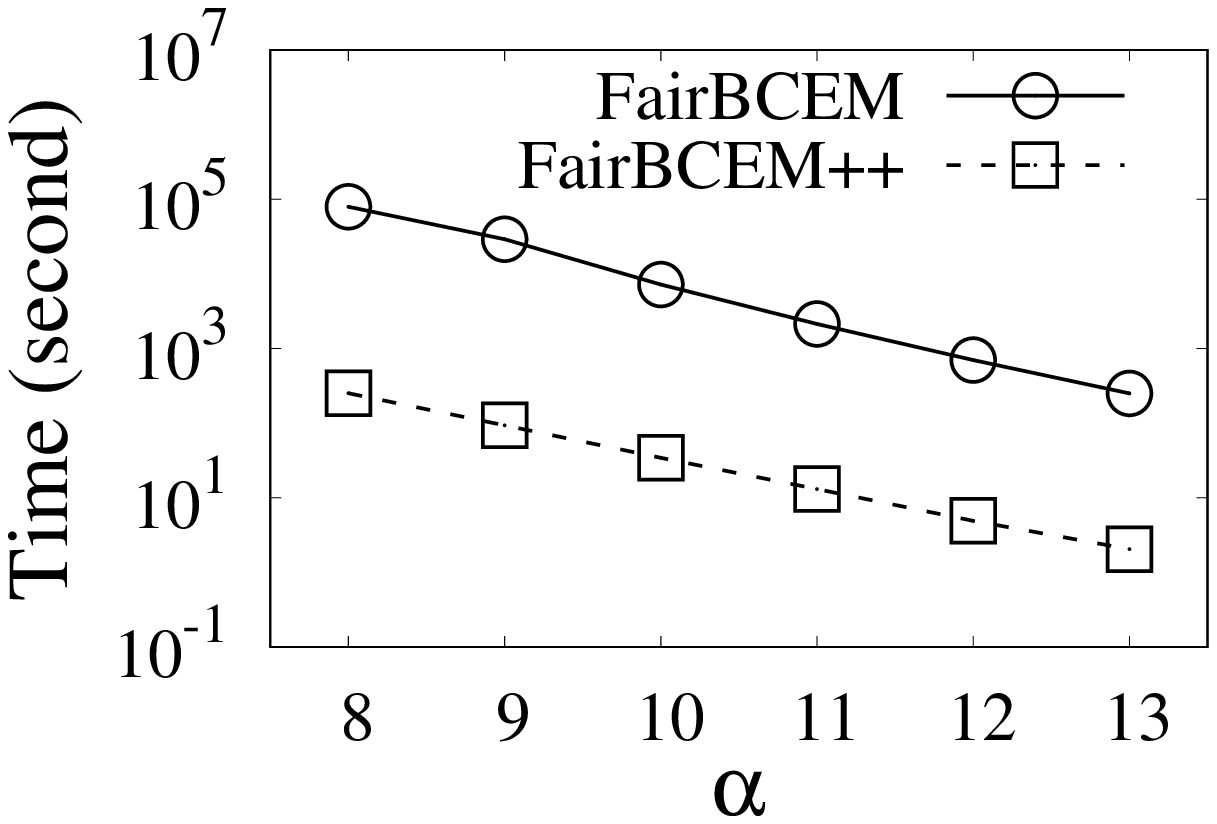}
      \end{minipage}
    }
    \subfigure[{\scriptsize \wiki~(vary $\alpha$)}]{
      \label{fig:exp-oneside-alg-time-wiki-alpha}
      \begin{minipage}{3.2cm}
      \centering
      \includegraphics[width=\textwidth]{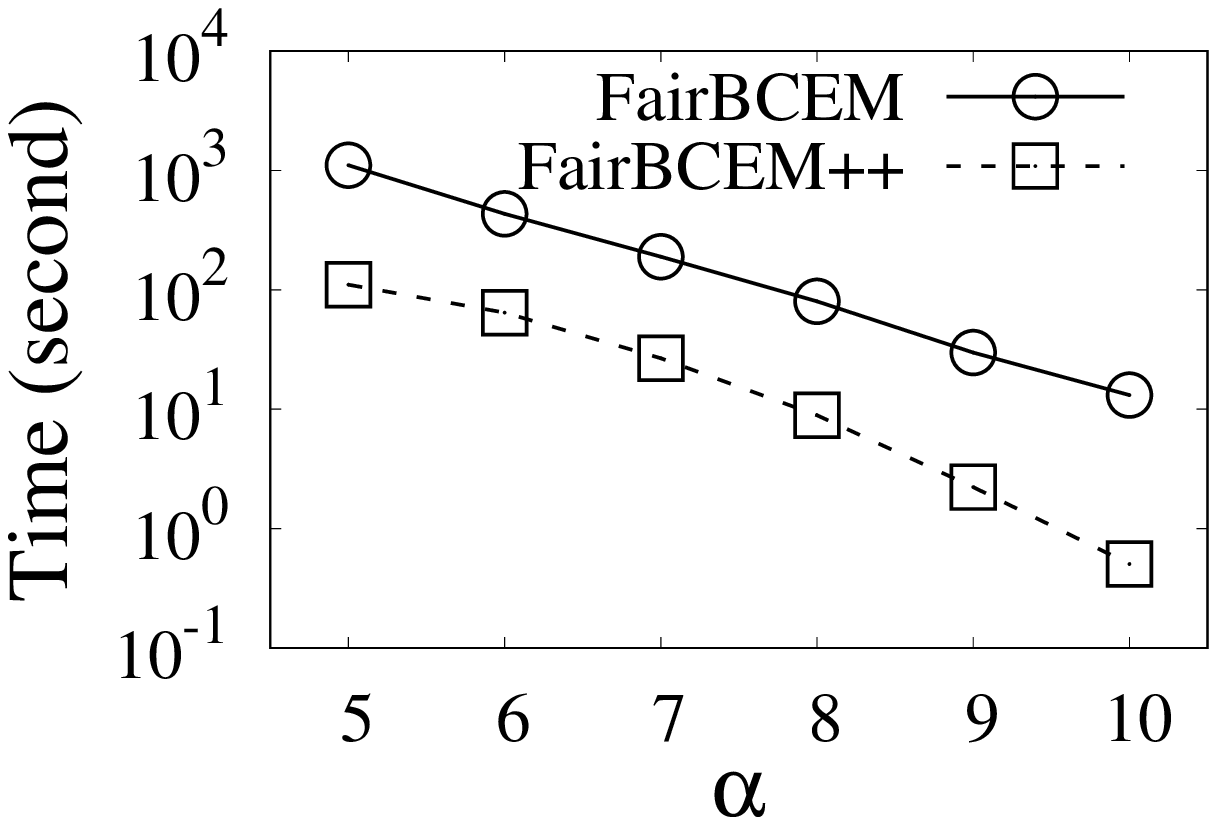}
      \end{minipage}
    }
    \subfigure[{\scriptsize \dblp (vary $\alpha$)}]{
      \label{fig:exp-oneside-alg-time-dblp-alpha}
      \begin{minipage}{3.2cm}
      \centering
      \includegraphics[width=\textwidth]{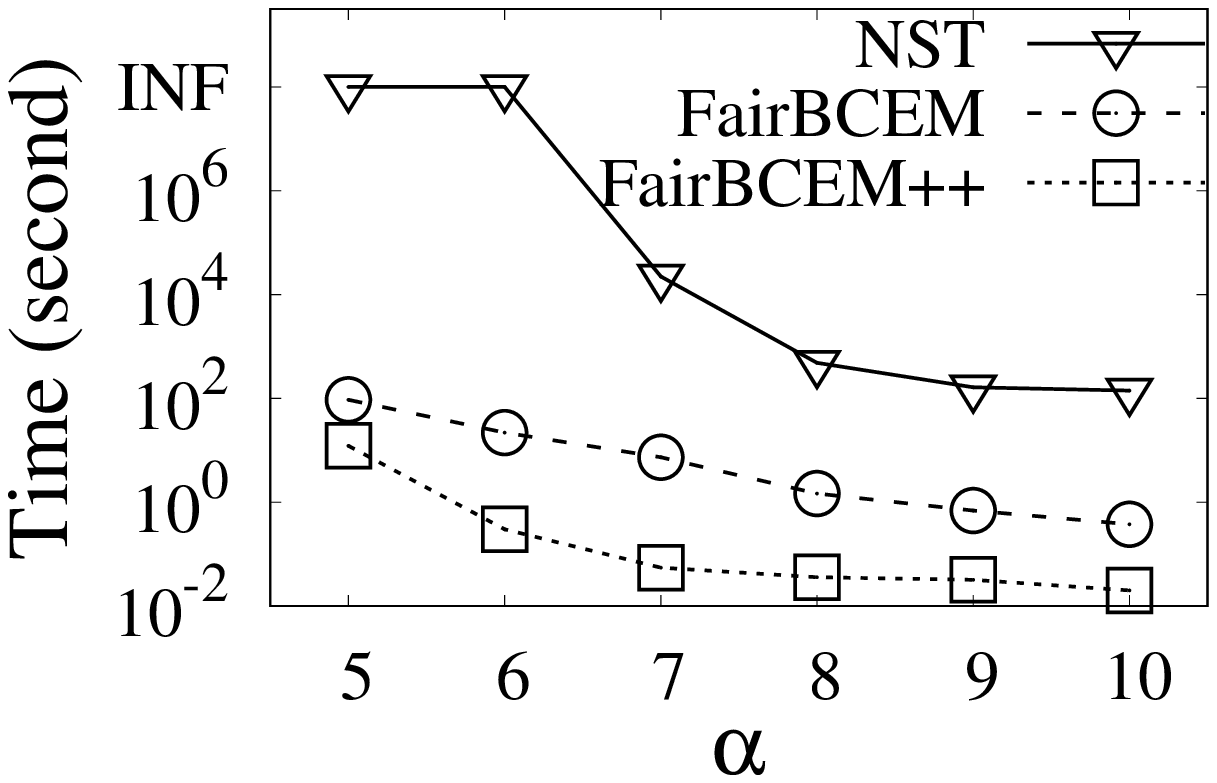}
      \end{minipage}
    }
    \vspace*{-0.3cm}
    
    \subfigure[{\scriptsize \youtube (vary $\beta$)}]{
      \label{fig:exp-oneside-alg-time-youtube-beta}
      \begin{minipage}{3.2cm}
      \centering
      \includegraphics[width=\textwidth]{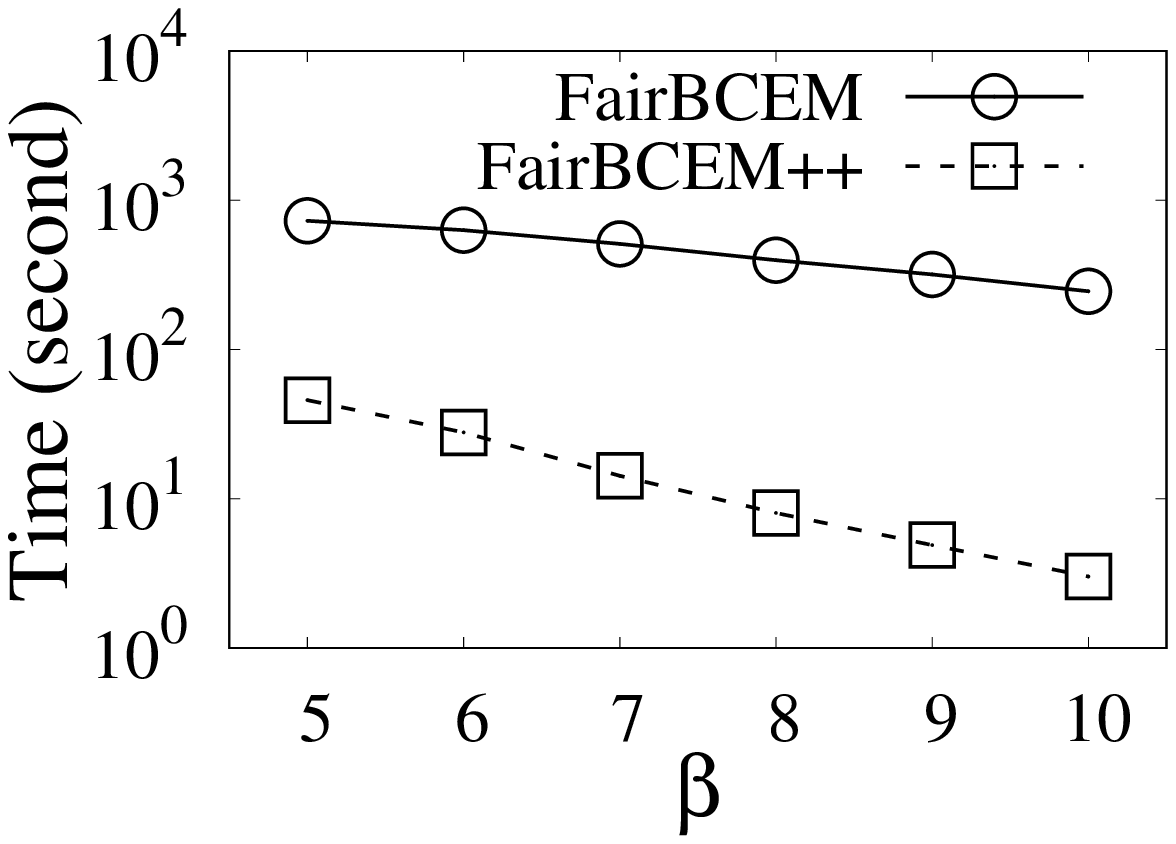}
      \end{minipage}
    }
    \subfigure[{\scriptsize \twi (vary $\beta$)}]{
      \label{fig:exp-oneside-alg-time-twi-beta}
      \begin{minipage}{3.2cm}
      \centering
      \includegraphics[width=\textwidth]{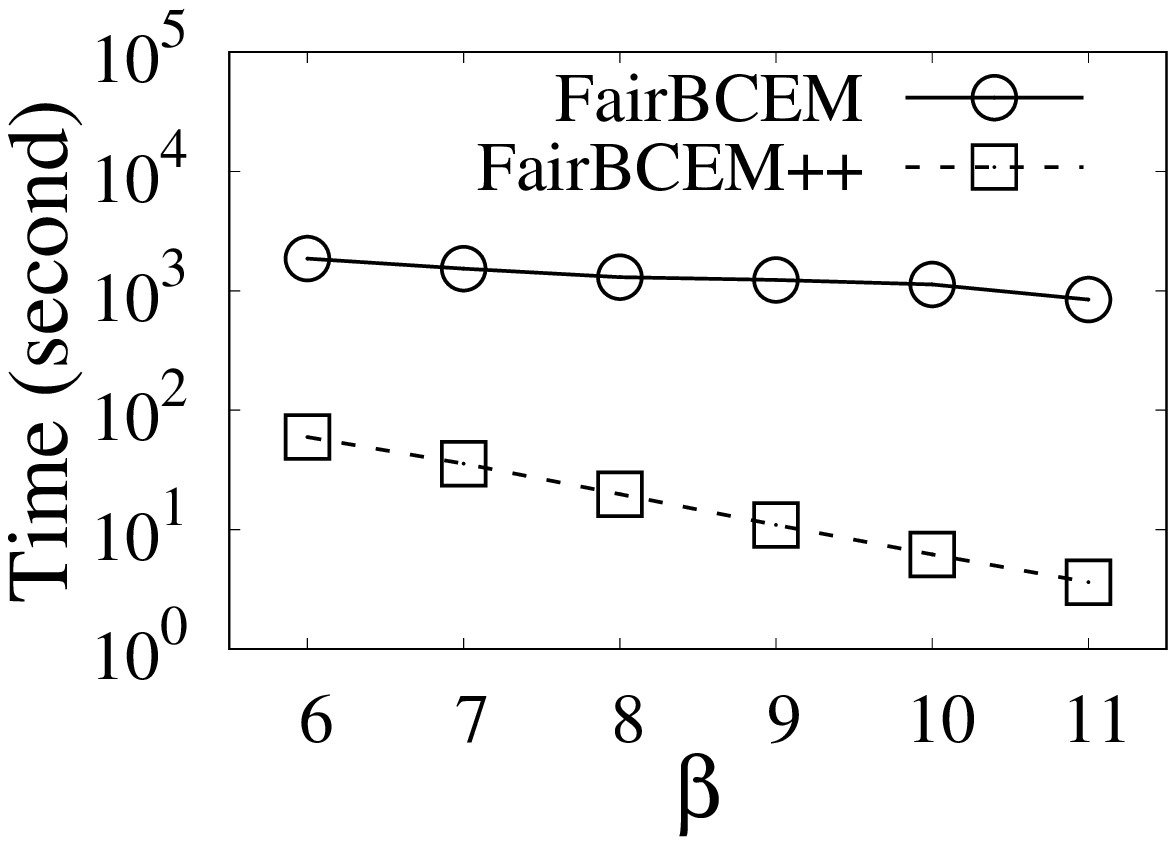}
      \end{minipage}
    }
    \subfigure[{\scriptsize \imdb (vary $\beta$)}]{
      \label{fig:exp-oneside-alg-time-imdb-beta}
      \begin{minipage}{3.2cm}
      \centering
      \includegraphics[width=\textwidth]{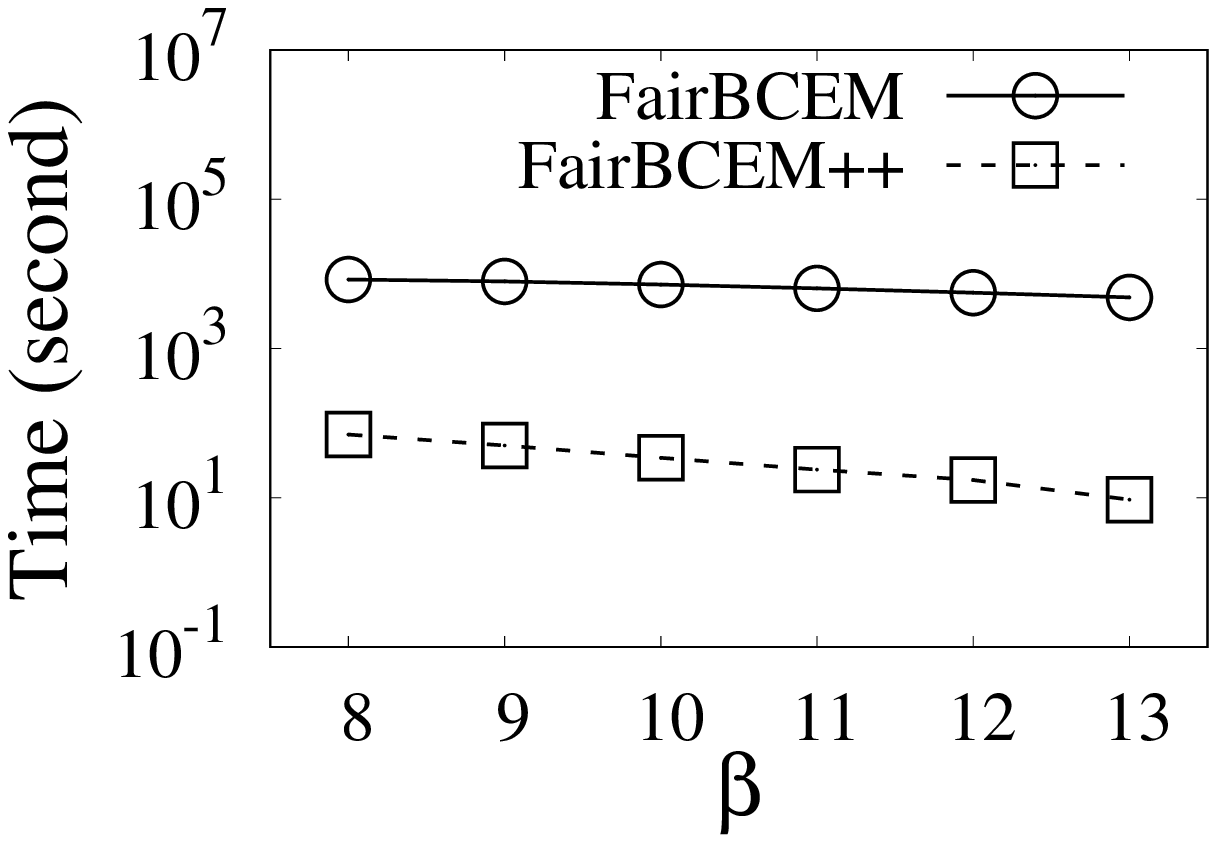}
      \end{minipage}
    }
    \subfigure[{\scriptsize \wiki~(vary $\beta$)}]{
      \label{fig:exp-oneside-alg-time-wiki-beta}
      \begin{minipage}{3.2cm}
      \centering
      \includegraphics[width=\textwidth]{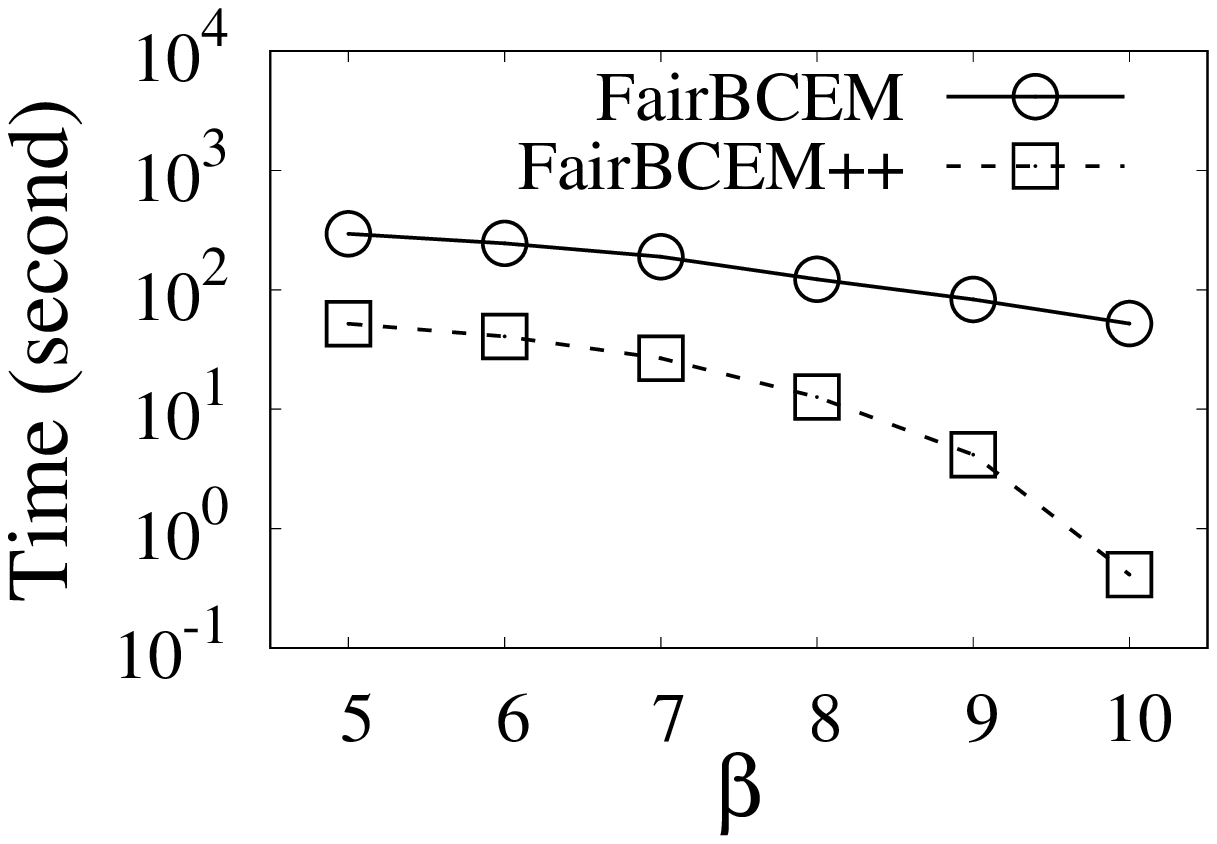}
      \end{minipage}
    }
    \subfigure[{\scriptsize \dblp (vary $\beta$)}]{
      \label{fig:exp-oneside-alg-time-dblp-beta}
      \begin{minipage}{3.2cm}
      \centering
      \includegraphics[width=\textwidth]{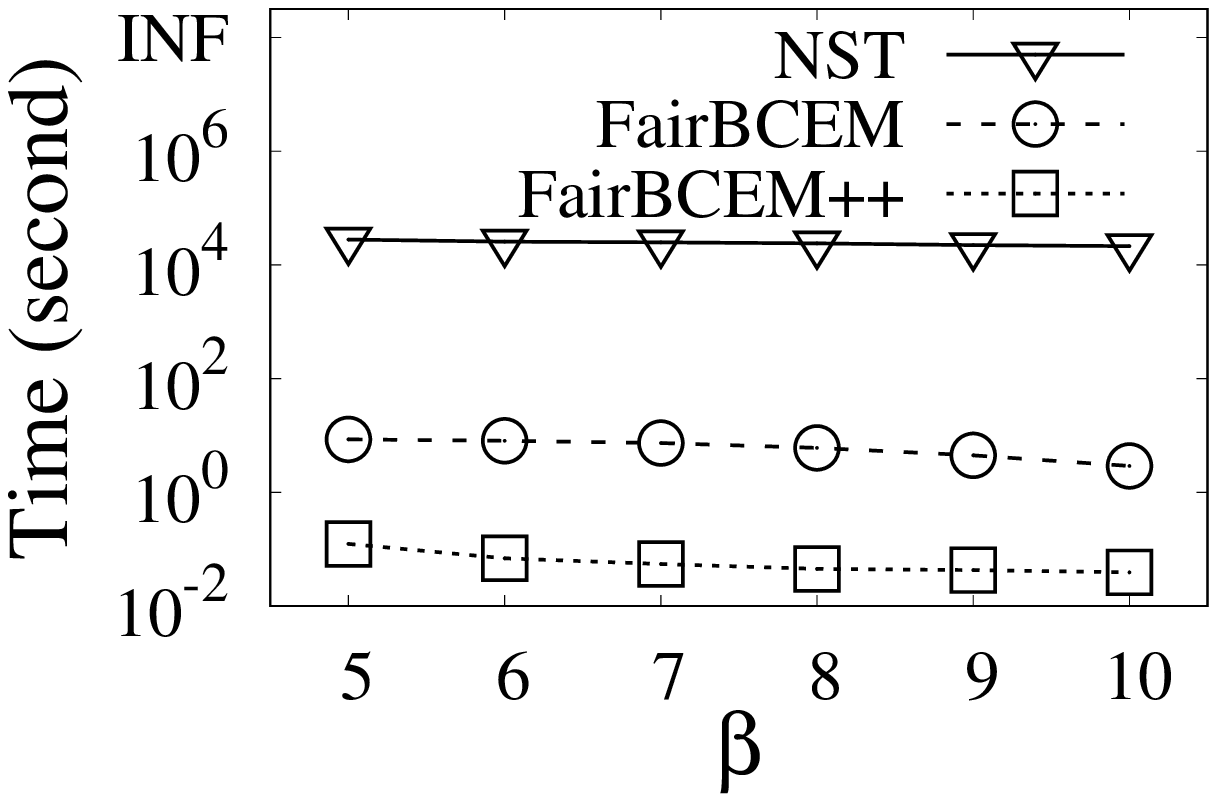}
      \end{minipage}
    }
    \vspace*{-0.3cm}
    
    \subfigure[{\scriptsize \youtube (vary $\delta$)}]{
      \label{fig:exp-oneside-alg-time-youtube-delta}
      \begin{minipage}{3.2cm}
      \centering
      \includegraphics[width=\textwidth]{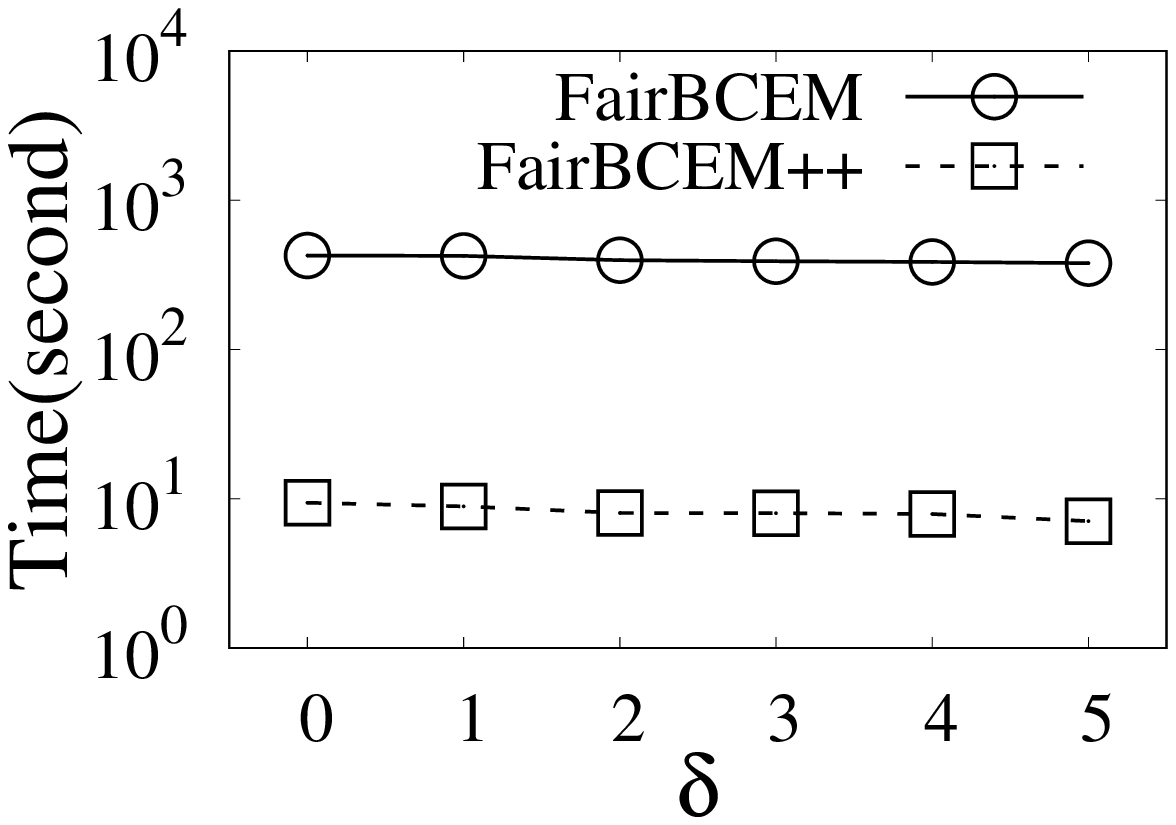}
      \end{minipage}
    }
    \subfigure[{\scriptsize \twi (vary $\delta$)}]{
      \label{fig:exp-oneside-alg-time-twi-delta}
      \begin{minipage}{3.2cm}
      \centering
      \includegraphics[width=\textwidth]{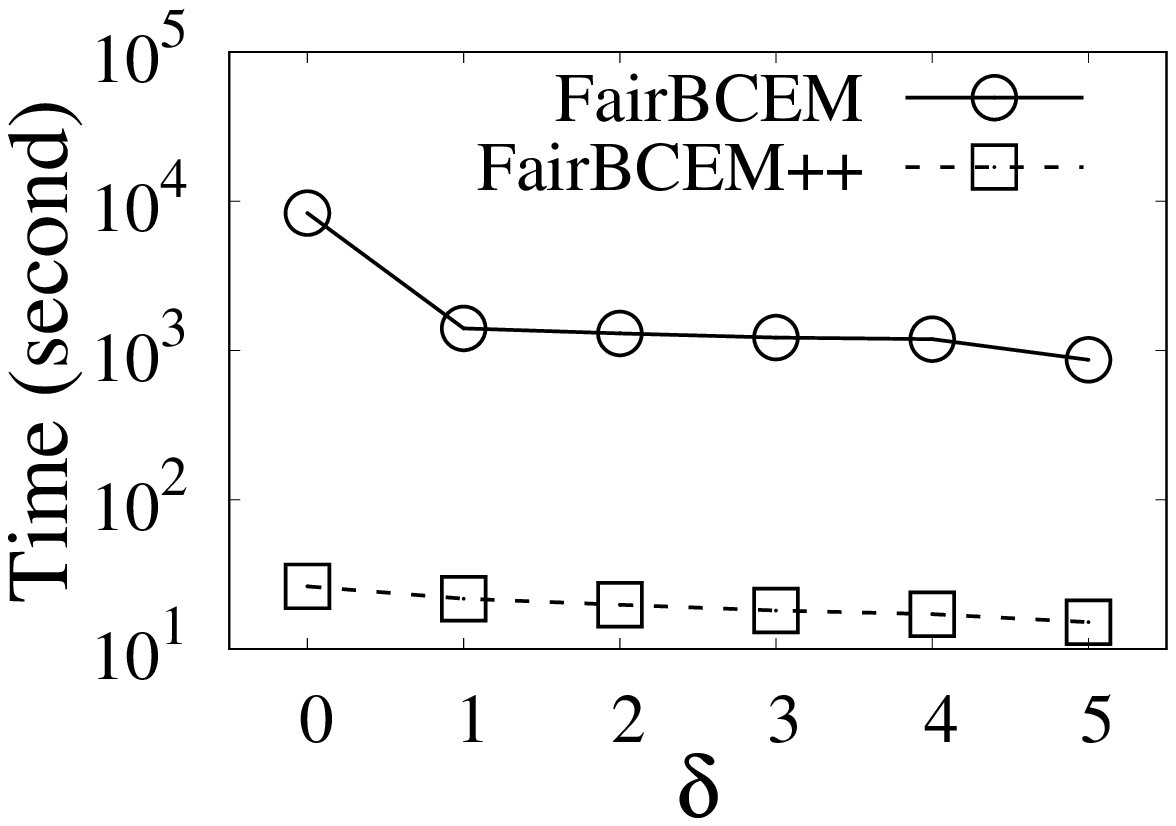}
      \end{minipage}
    }
    \subfigure[{\scriptsize \imdb (vary $\delta$)}]{
      \label{fig:exp-oneside-alg-time-imdb-delta}
      \begin{minipage}{3.2cm}
      \centering
      \includegraphics[width=\textwidth]{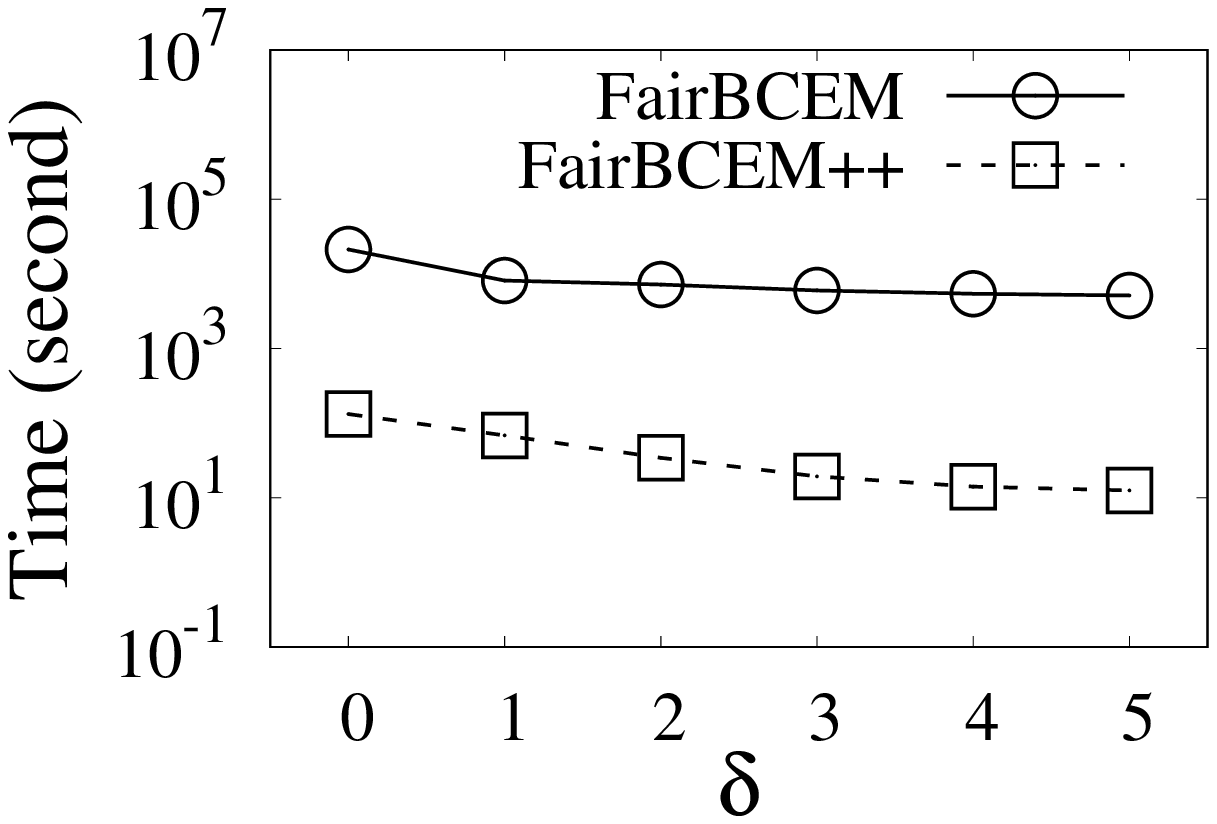}
      \end{minipage}
    }
    \subfigure[{\scriptsize \wiki~(vary $\delta$)}]{
      \label{fig:exp-oneside-alg-time-wiki-delta}
      \begin{minipage}{3.2cm}
      \centering
      \includegraphics[width=\textwidth]{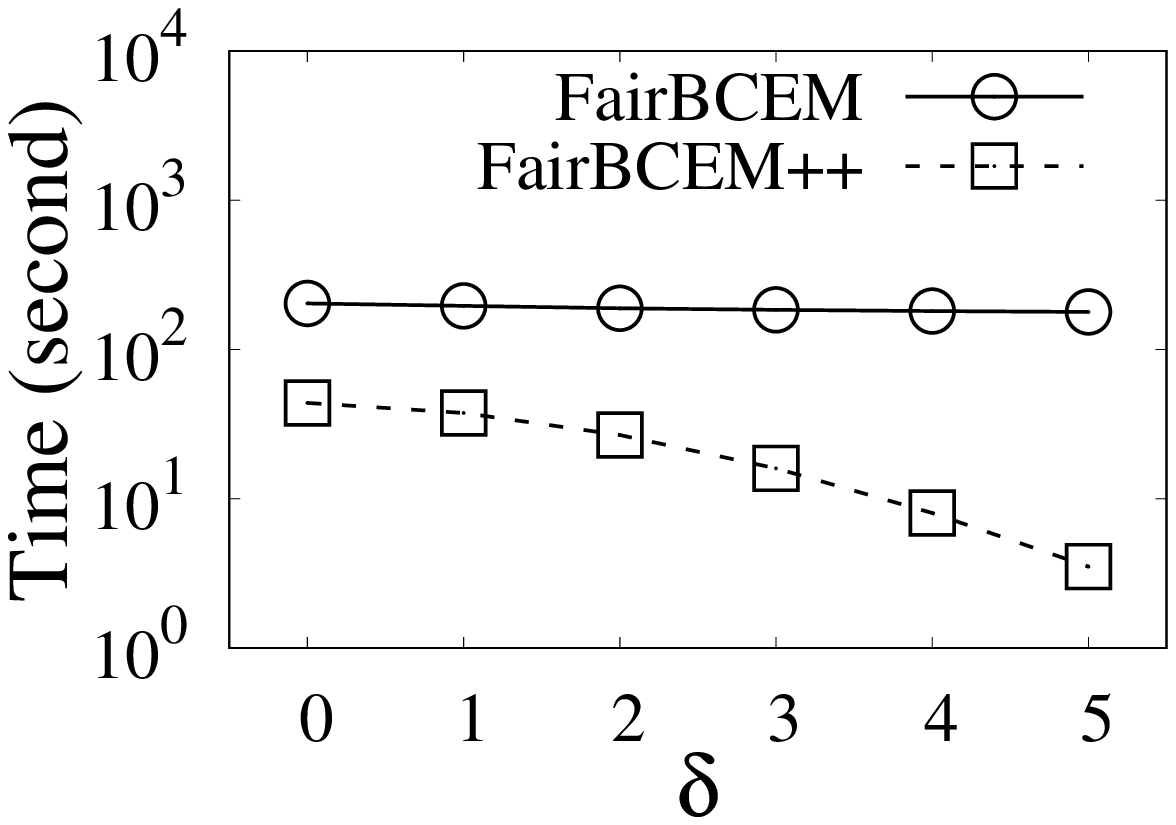}
      \end{minipage}
    }
    \subfigure[{\scriptsize \dblp (vary $\delta$)}]{
      \label{fig:exp-oneside-alg-time-dblp-delta}
      \begin{minipage}{3.2cm}
      \centering
      \includegraphics[width=\textwidth]{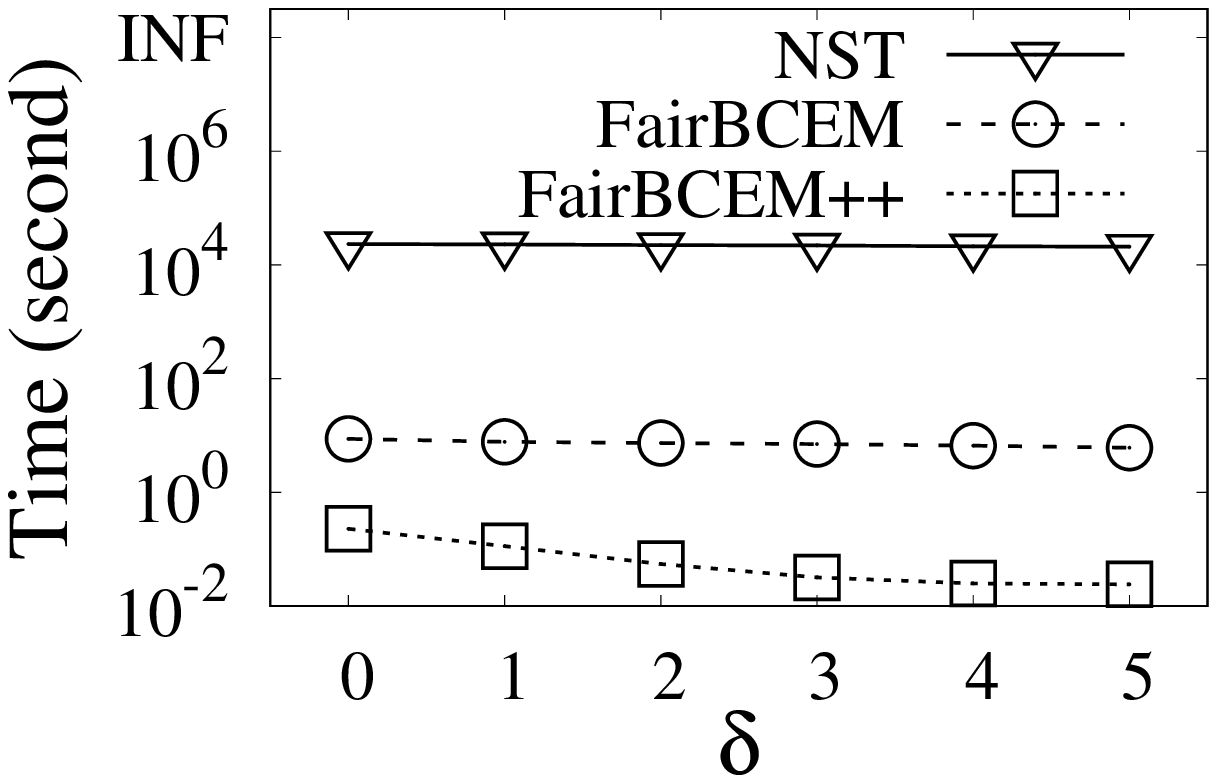}
      \end{minipage}
    }
	\vspace*{-0.3cm}
	\caption{{The running time of the \naivesearchtree, \onesideFBCEM and \onesideFBCEMPP~algorithms in different datasets.}}
	\vspace*{-0.2cm}
	\label{fig:exp-oneside-alg-time}
\end{figure*}

\begin{figure}[t!]\vspace*{-0.2cm}
\centering
    \subfigure[{\scriptsize \imdb (vary $\alpha$)}]{
      \label{fig:exp-oneside-pruning-node-imdb-alpha}
      \begin{minipage}{3.2cm}
      \centering
      \includegraphics[width=\textwidth]{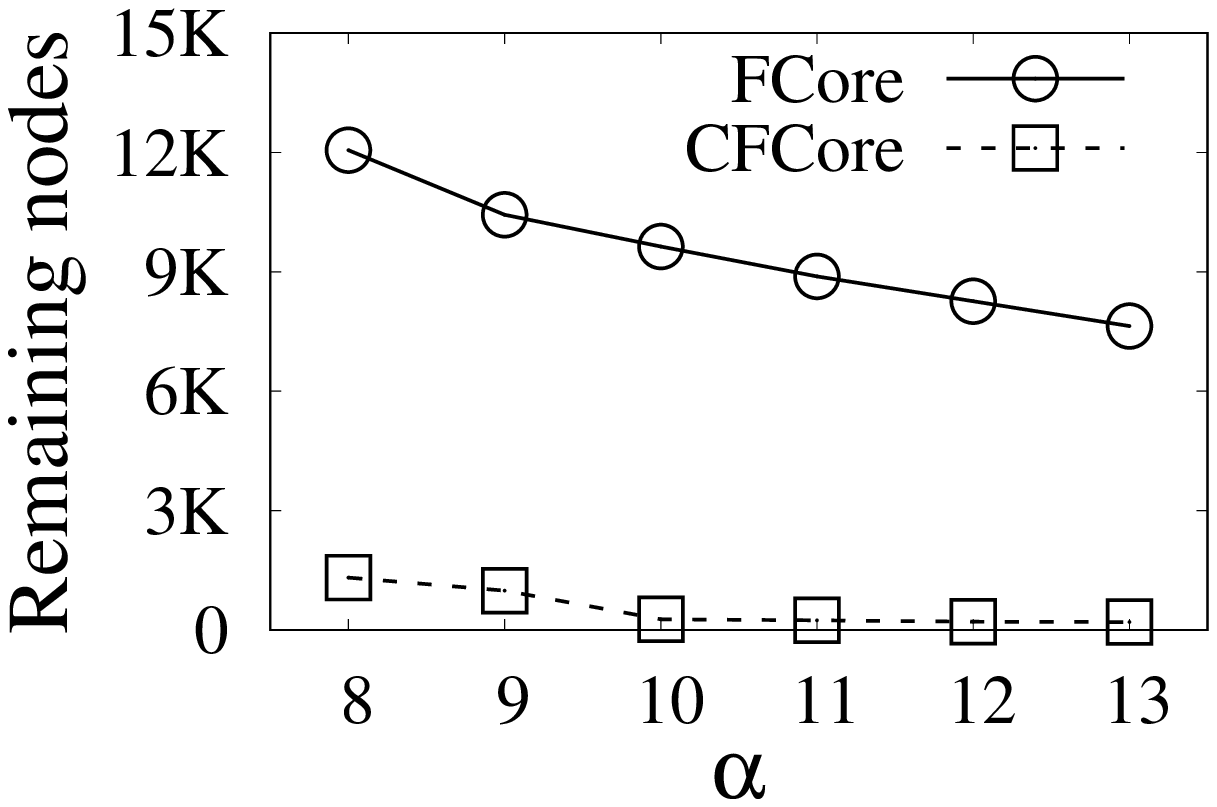}
      \end{minipage}
    }
    \subfigure[{\scriptsize \imdb (vary $\beta$)}]{
      \label{fig:exp-oneside-pruning-node-imdb-beta}
      \begin{minipage}{3.2cm}
      \centering
      \includegraphics[width=\textwidth]{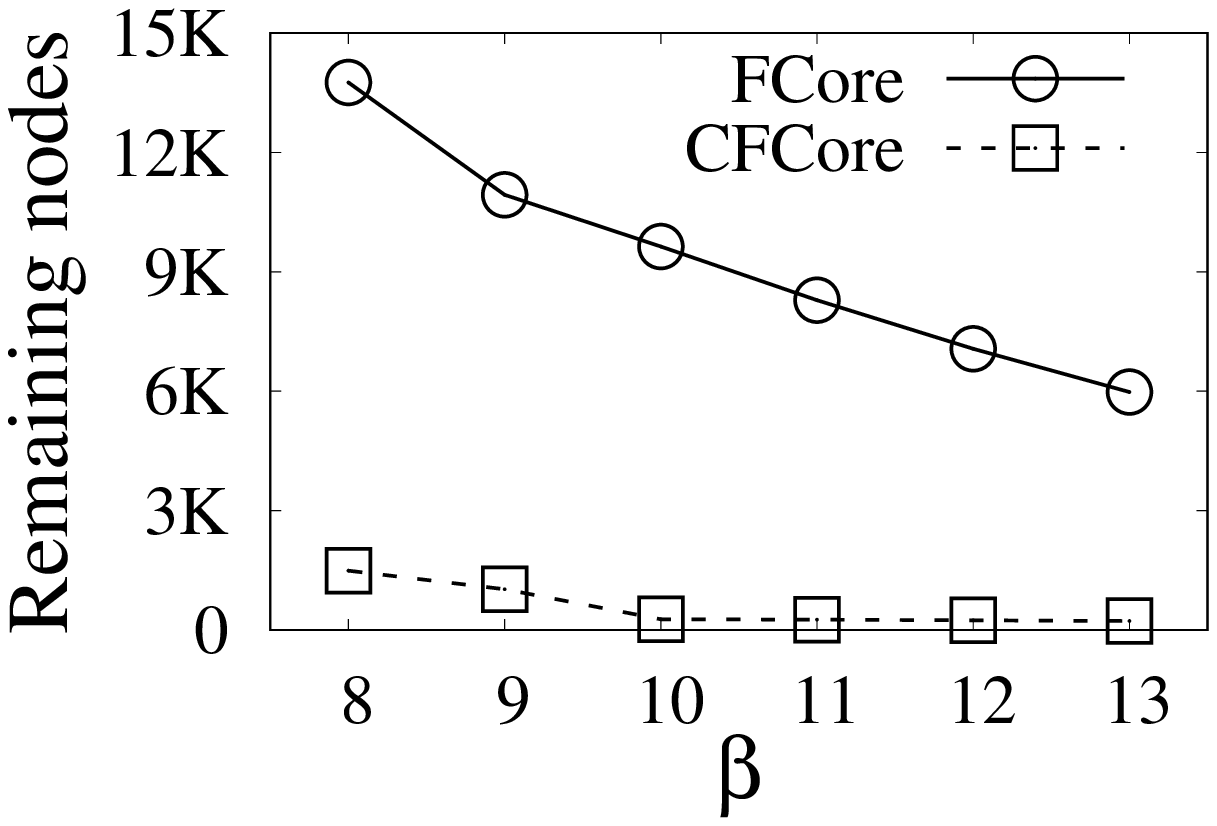}
      \end{minipage}
    }
    \vspace*{-0.3cm}
    	
    \subfigure[{\scriptsize \imdb (vary $\alpha$)}]{
      \label{fig:exp-oneside-pruning-time-imdb-alpha}
      \begin{minipage}{3.2cm}
      \centering
      \includegraphics[width=\textwidth]{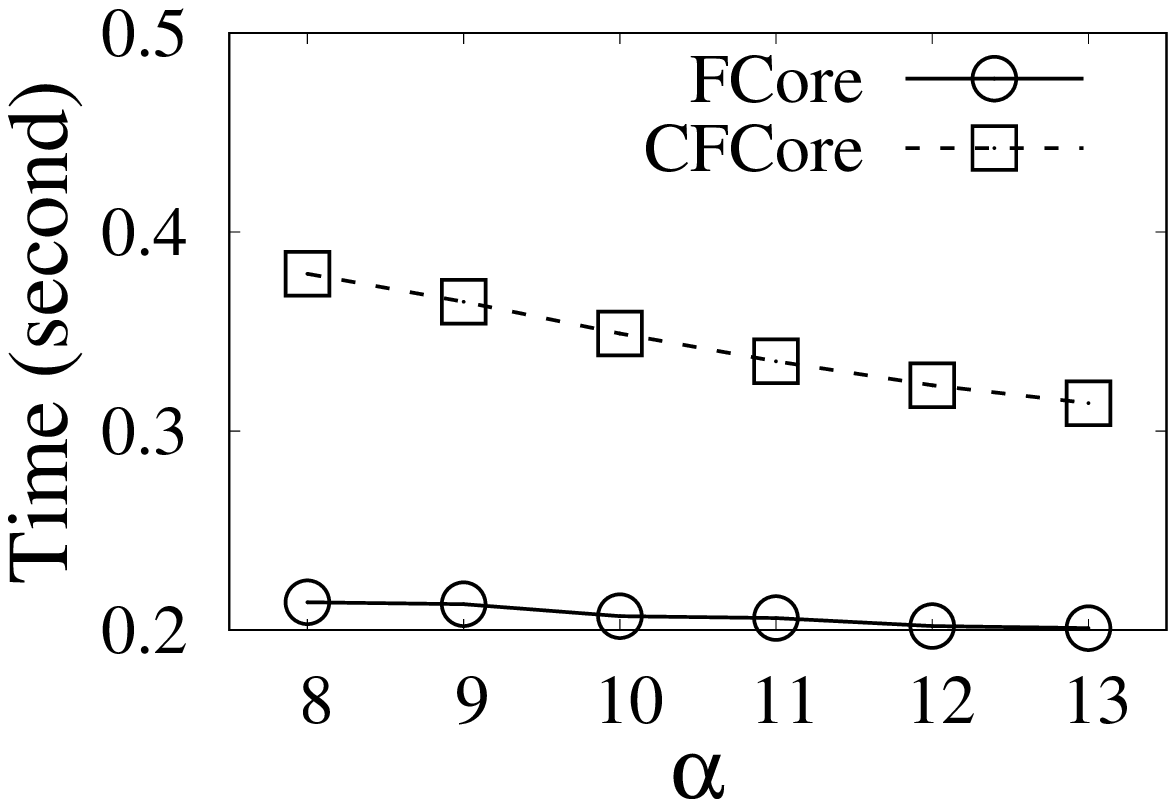}
      \end{minipage}
    }
    \subfigure[{\scriptsize \imdb (vary $\beta$)}]{
      \label{fig:exp-oneside-pruning-time-imdb-beta}
      \begin{minipage}{3.2cm}
      \centering
      \includegraphics[width=\textwidth]{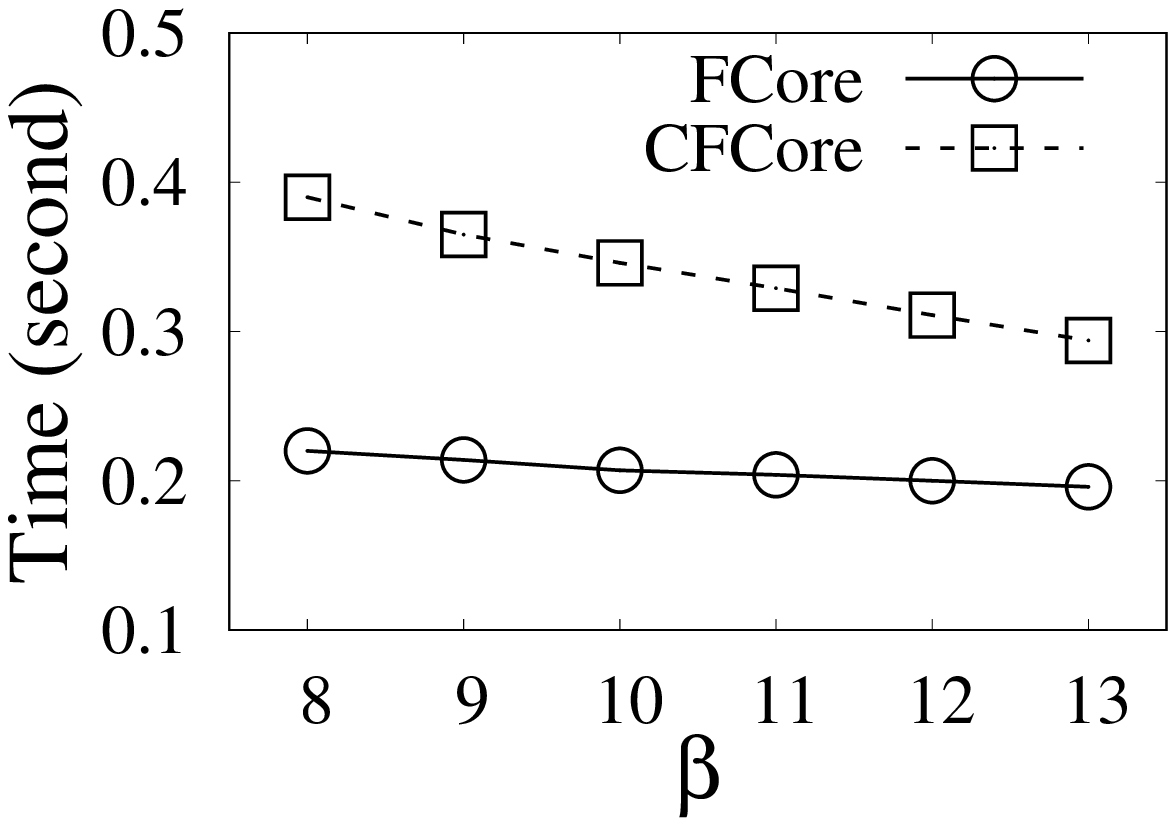}
      \end{minipage}
    }
	\vspace*{-0.3cm}
	\caption{The pruning time and remaining nodes of \fcore and \cfcore.}
	\vspace*{-0.2cm}
	\label{fig:exp-oneside-pruning-time-node}
\end{figure}

\begin{figure}[t!]\vspace*{-0.2cm}
\centering
    \subfigure[{\scriptsize \twi (vary $\alpha$)}]{
      \label{fig:exp-twoside-pruning-node-twi-alpha}
      \begin{minipage}{3.2cm}
      \centering
      \includegraphics[width=\textwidth]{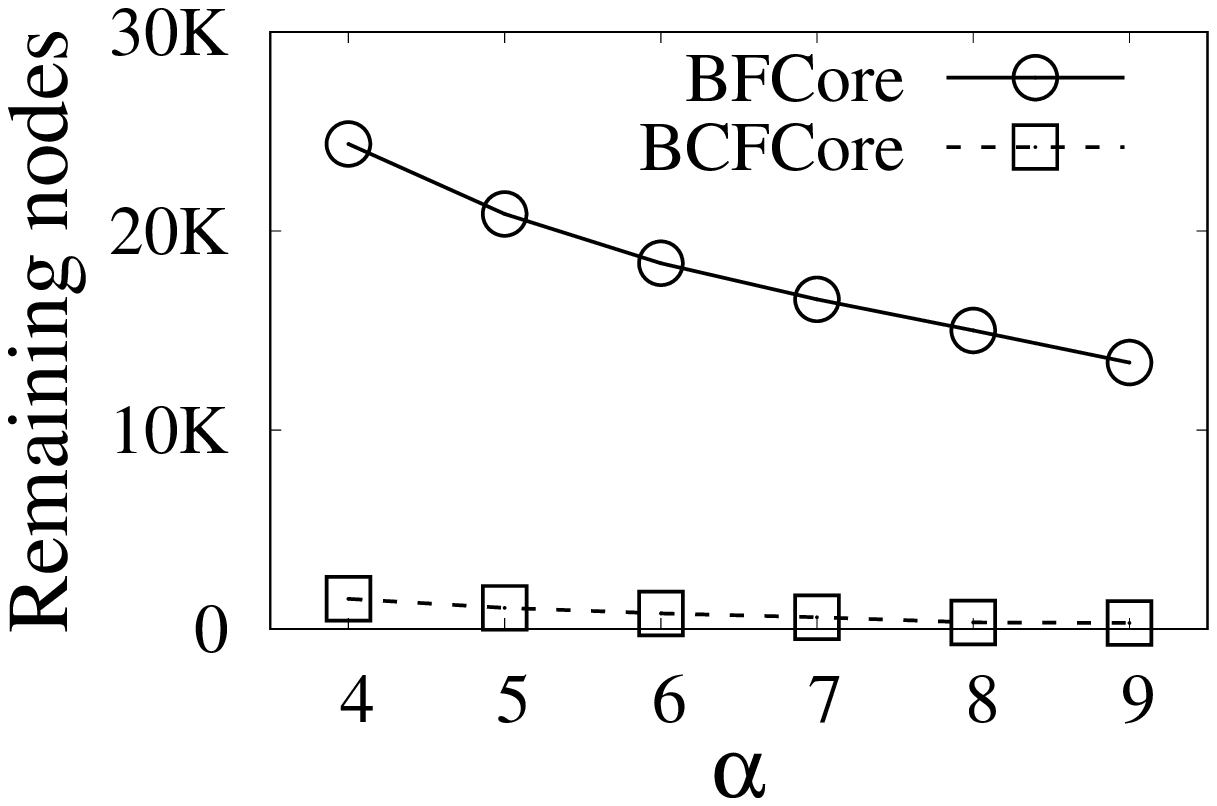}
      \end{minipage}
    }
    \subfigure[{\scriptsize \twi (vary $\beta$)}]{
      \label{fig:exp-twoside-pruning-node-twi-beta}
      \begin{minipage}{3.2cm}
      \centering
      \includegraphics[width=\textwidth]{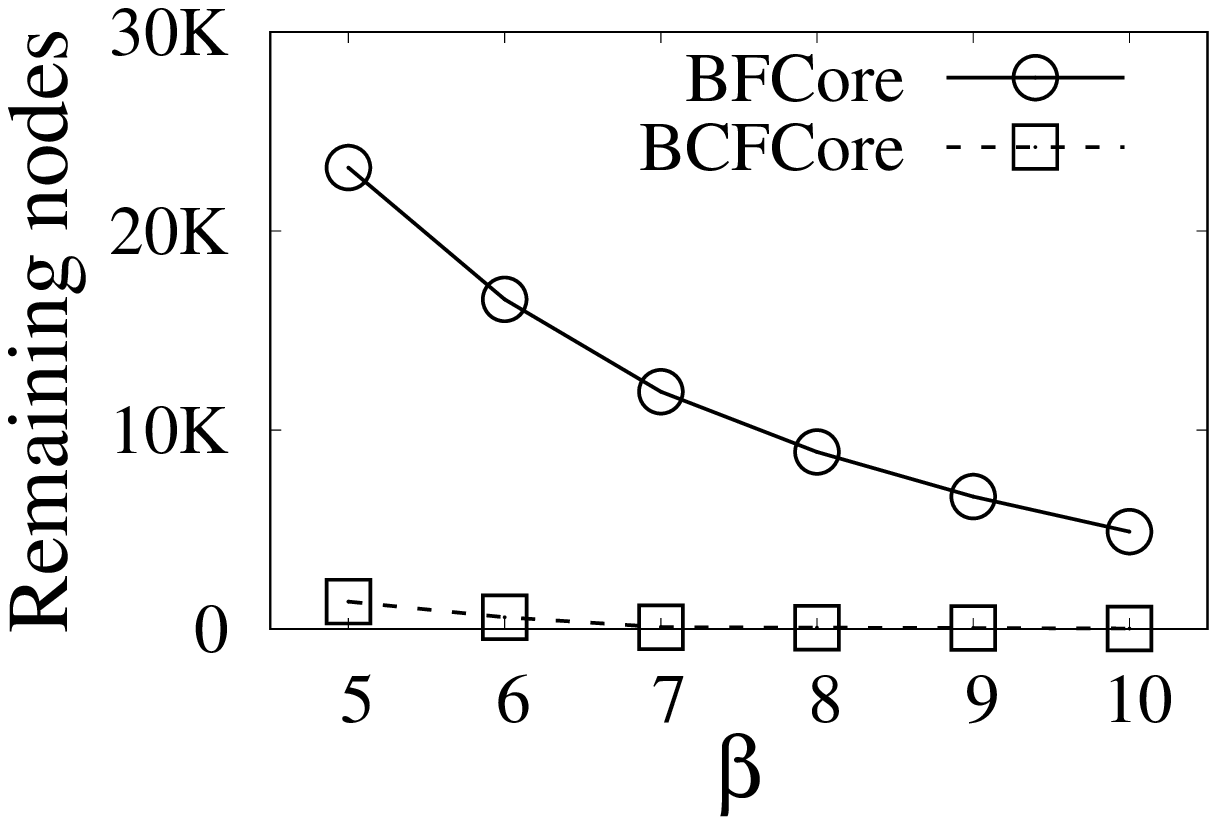}
      \end{minipage}
    }
    \vspace*{-0.3cm}
    	
    \subfigure[{\scriptsize \twi (vary $\alpha$)}]{
      \label{fig:exp-twoside-pruning-time-twi-alpha}
      \begin{minipage}{3.2cm}
      \centering
      \includegraphics[width=\textwidth]{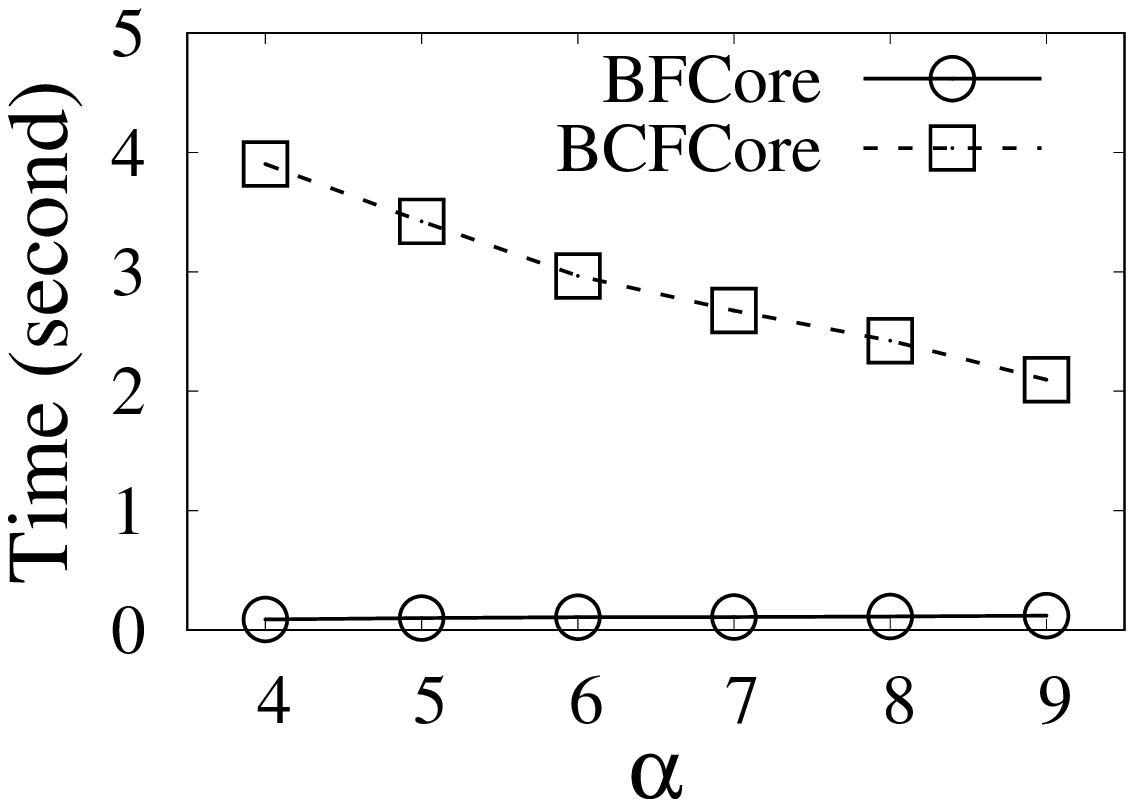}
      \end{minipage}
    }
    \subfigure[{\scriptsize \twi (vary $\beta$)}]{
      \label{fig:exp-twoside-pruning-time-twi-beta}
      \begin{minipage}{3.2cm}
      \centering
      \includegraphics[width=\textwidth]{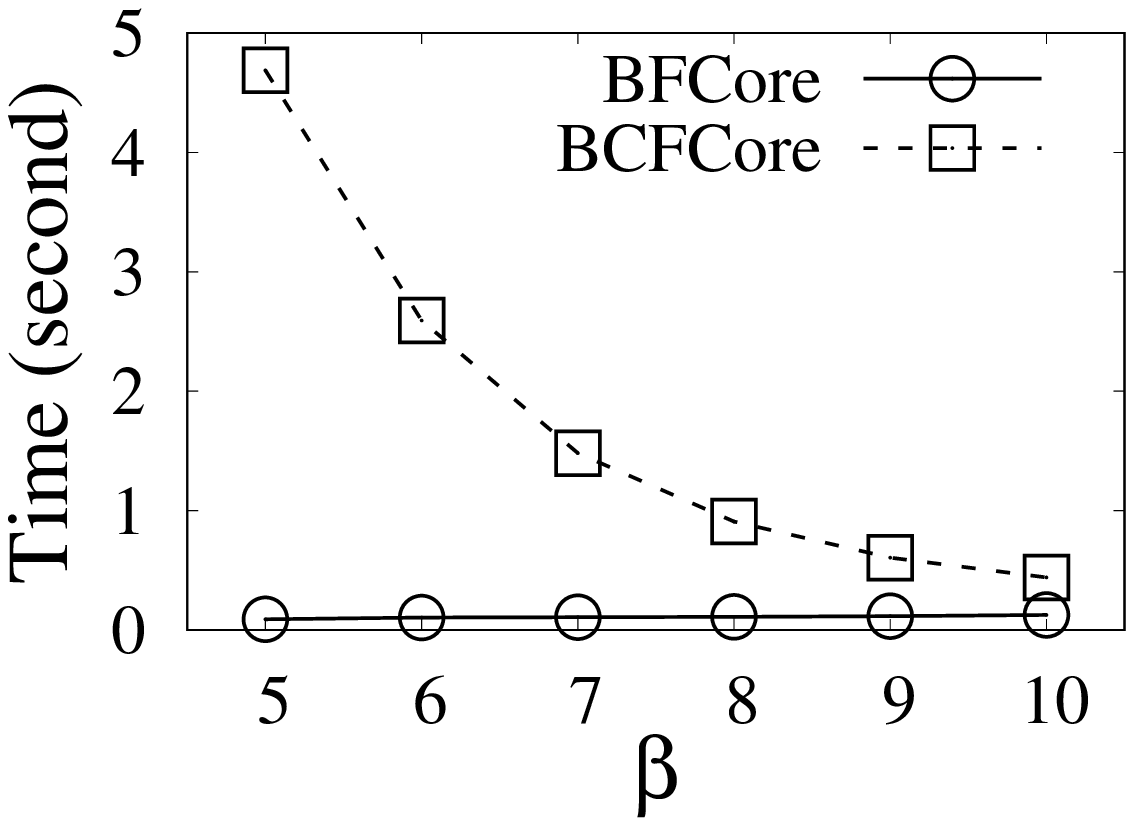}
      \end{minipage}
    }
	\vspace*{-0.3cm}
	\caption{The pruning time and remaining nodes of \bfcore and \bcfcore.}
	\vspace*{-0.2cm}
	\label{fig:exp-twoside-pruning-time-node}
\end{figure}

\begin{figure*}[t!]\vspace*{-0.5cm}
\centering
    \subfigure[{\scriptsize \youtube (vary $\alpha$)}]{
      \label{fig:exp-twoside-alg-time-youtube-alpha}
      \begin{minipage}{3.2cm}
      \centering
      \includegraphics[width=\textwidth]{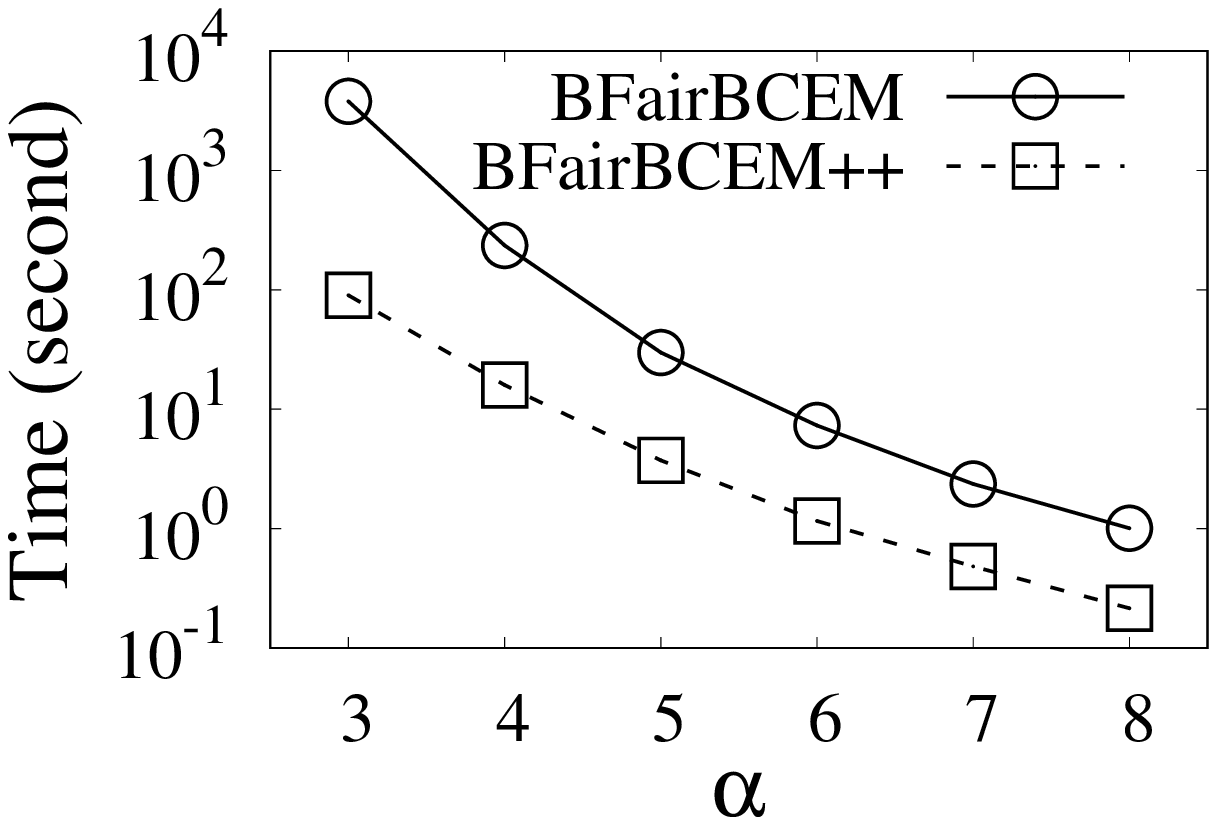}
      \end{minipage}
    }
    \subfigure[{\scriptsize \twi (vary $\alpha$)}]{
      \label{fig:exp-twoside-alg-time-twi-alpha}
      \begin{minipage}{3.2cm}
      \centering
      \includegraphics[width=\textwidth]{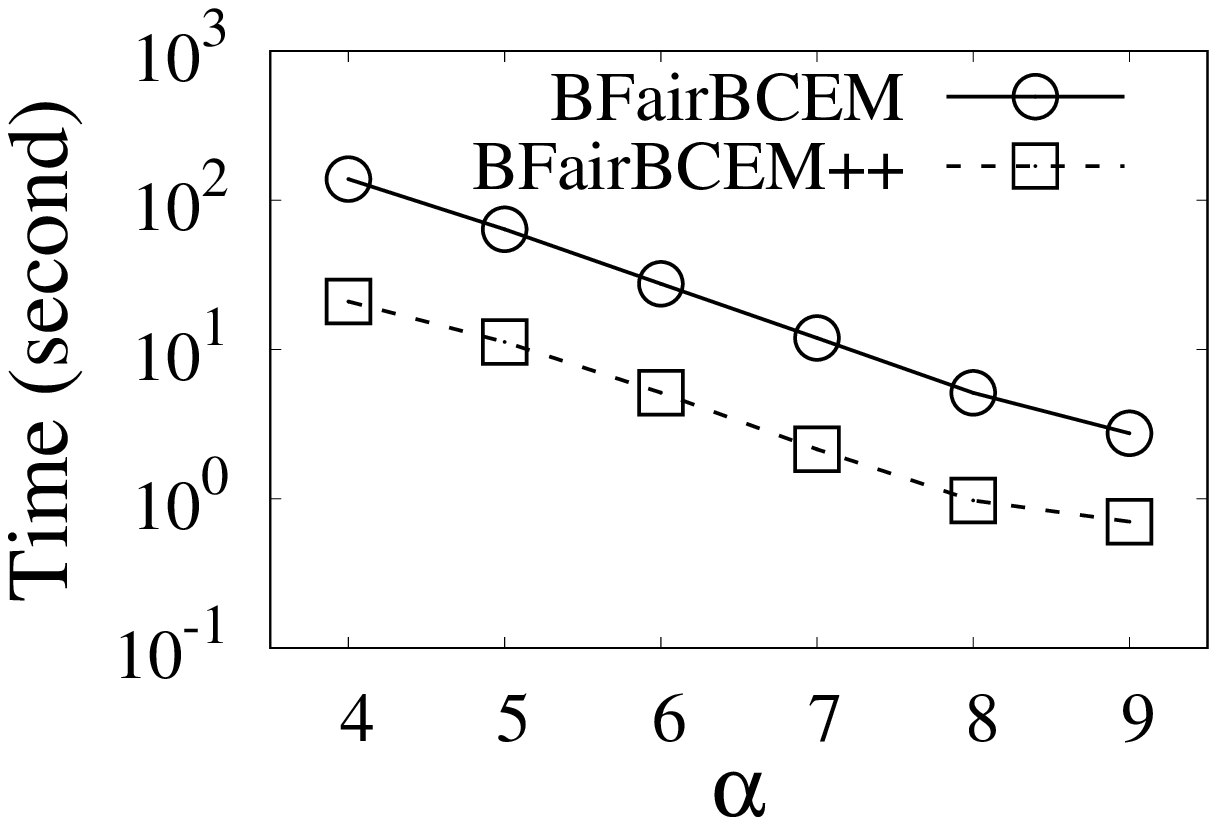}
      \end{minipage}
    }
    \subfigure[{\scriptsize \imdb (vary $\alpha$)}]{
      \label{fig:exp-twoside-alg-time-imdb-alpha}
      \begin{minipage}{3.2cm}
      \centering
      \includegraphics[width=\textwidth]{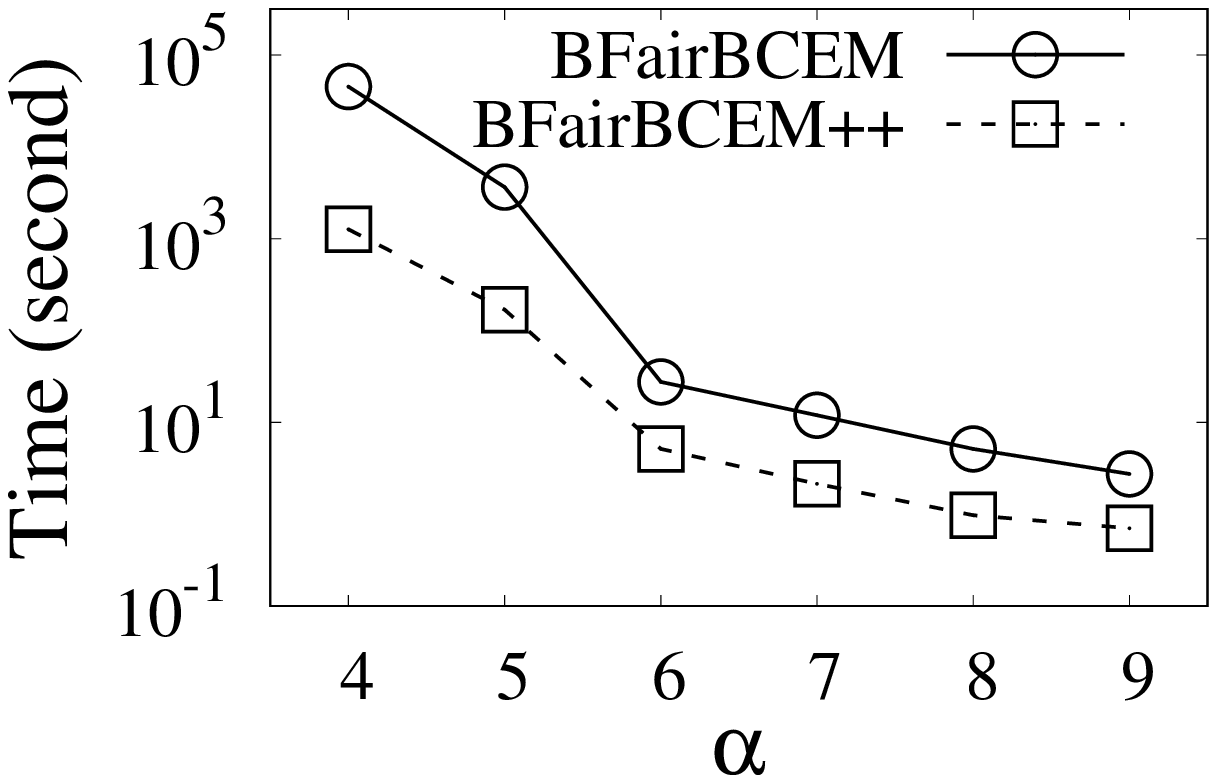}
      \end{minipage}
    }
    \subfigure[{\scriptsize \wiki~(vary $\alpha$)}]{
      \label{fig:exp-twoside-alg-time-wiki-alpha}
      \begin{minipage}{3.2cm}
      \centering
      \includegraphics[width=\textwidth]{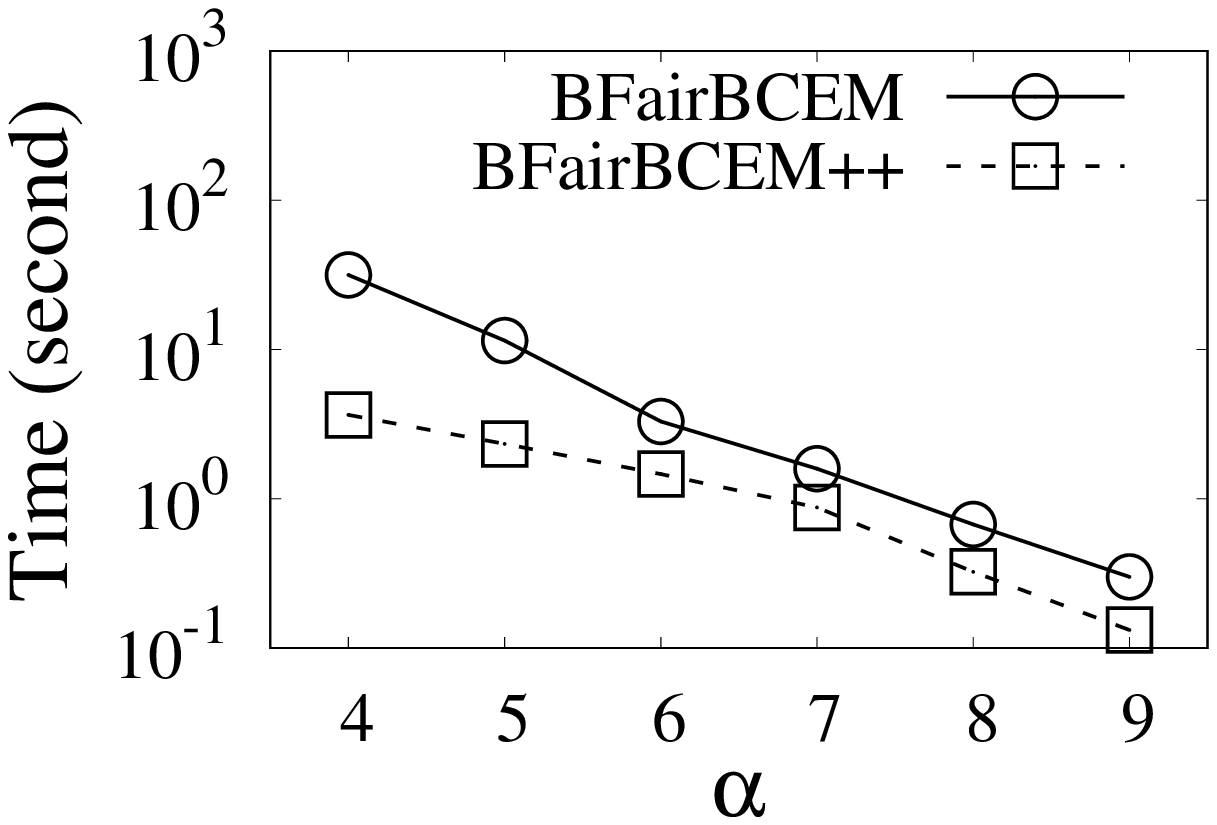}
      \end{minipage}
    }
    \subfigure[{\scriptsize \dblp (vary $\alpha$)}]{
      \label{fig:exp-twoside-alg-time-dblp-alpha}
      \begin{minipage}{3.2cm}
      \centering
      \includegraphics[width=\textwidth]{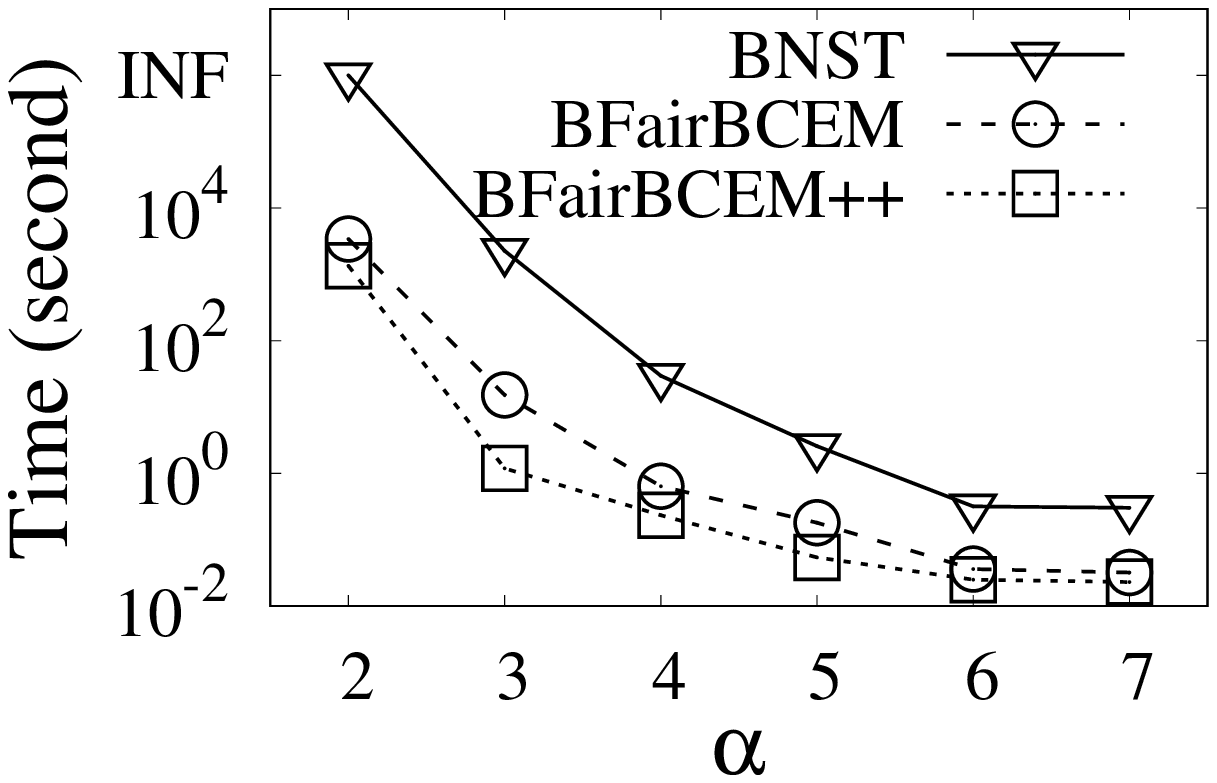}
      \end{minipage}
    }
    \vspace*{-0.3cm}
    
    \subfigure[{\scriptsize \youtube (vary $\beta$)}]{
      \label{fig:exp-twoside-alg-time-youtube-beta}
      \begin{minipage}{3.2cm}
      \centering
      \includegraphics[width=\textwidth]{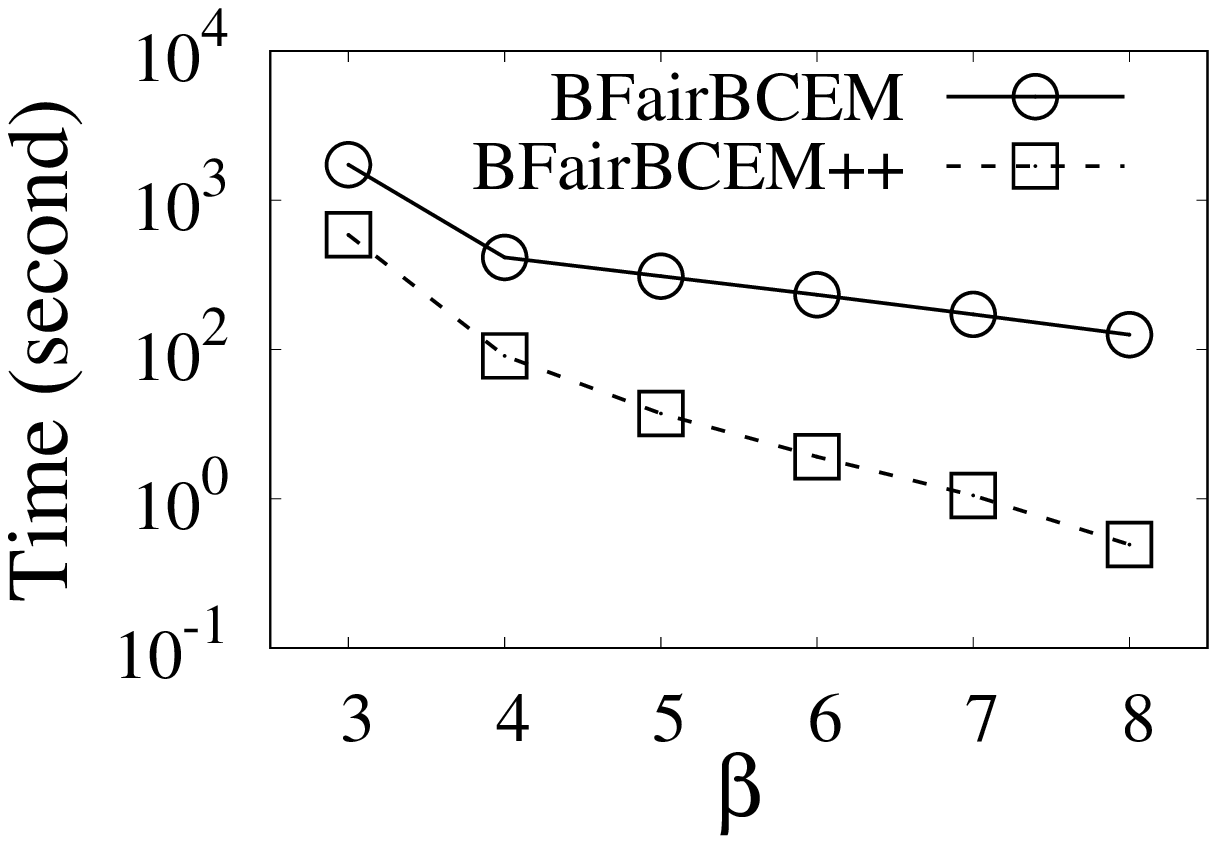}
      \end{minipage}
    }
    \subfigure[{\scriptsize \twi (vary $\beta$)}]{
      \label{fig:exp-twoside-alg-time-twi-beta}
      \begin{minipage}{3.2cm}
      \centering
      \includegraphics[width=\textwidth]{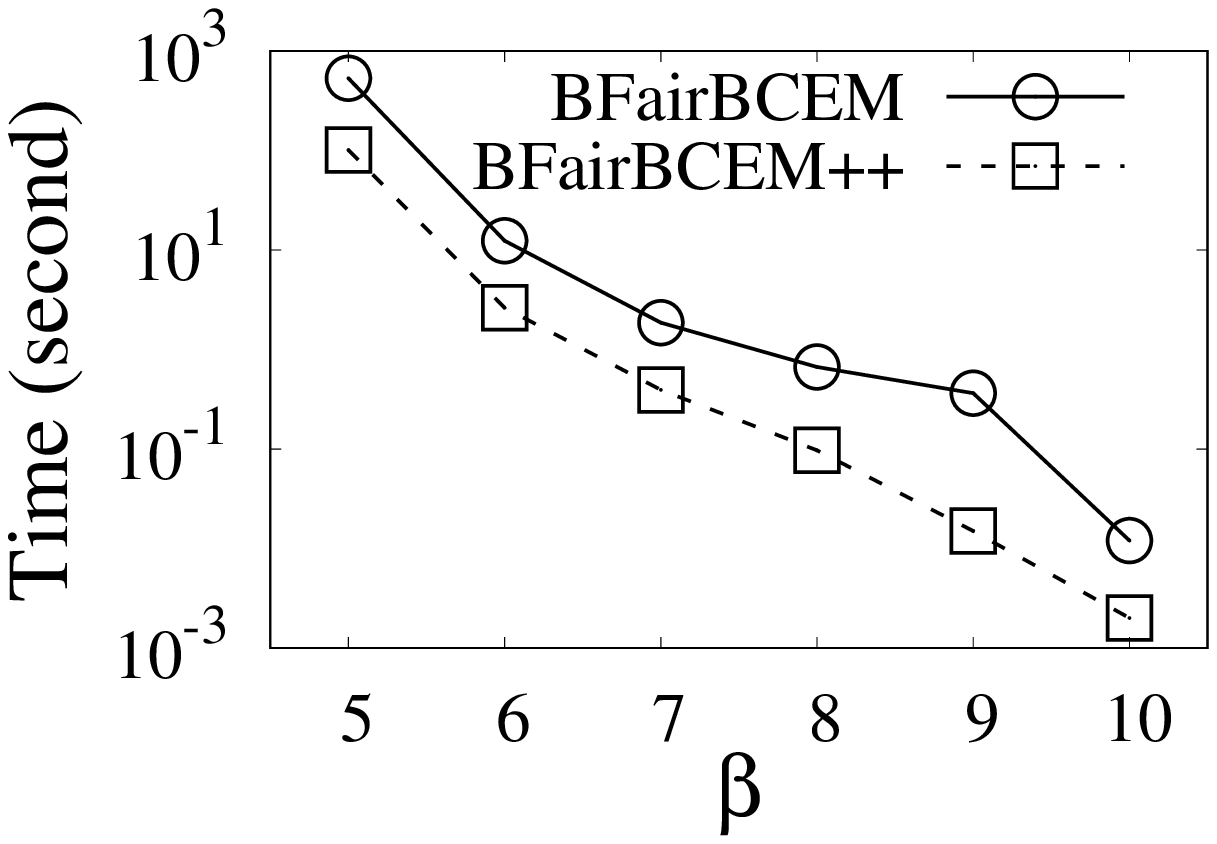}
      \end{minipage}
    }
    \subfigure[{\scriptsize \imdb (vary $\beta$)}]{
      \label{fig:exp-twoside-alg-time-imdb-beta}
      \begin{minipage}{3.2cm}
      \centering
      \includegraphics[width=\textwidth]{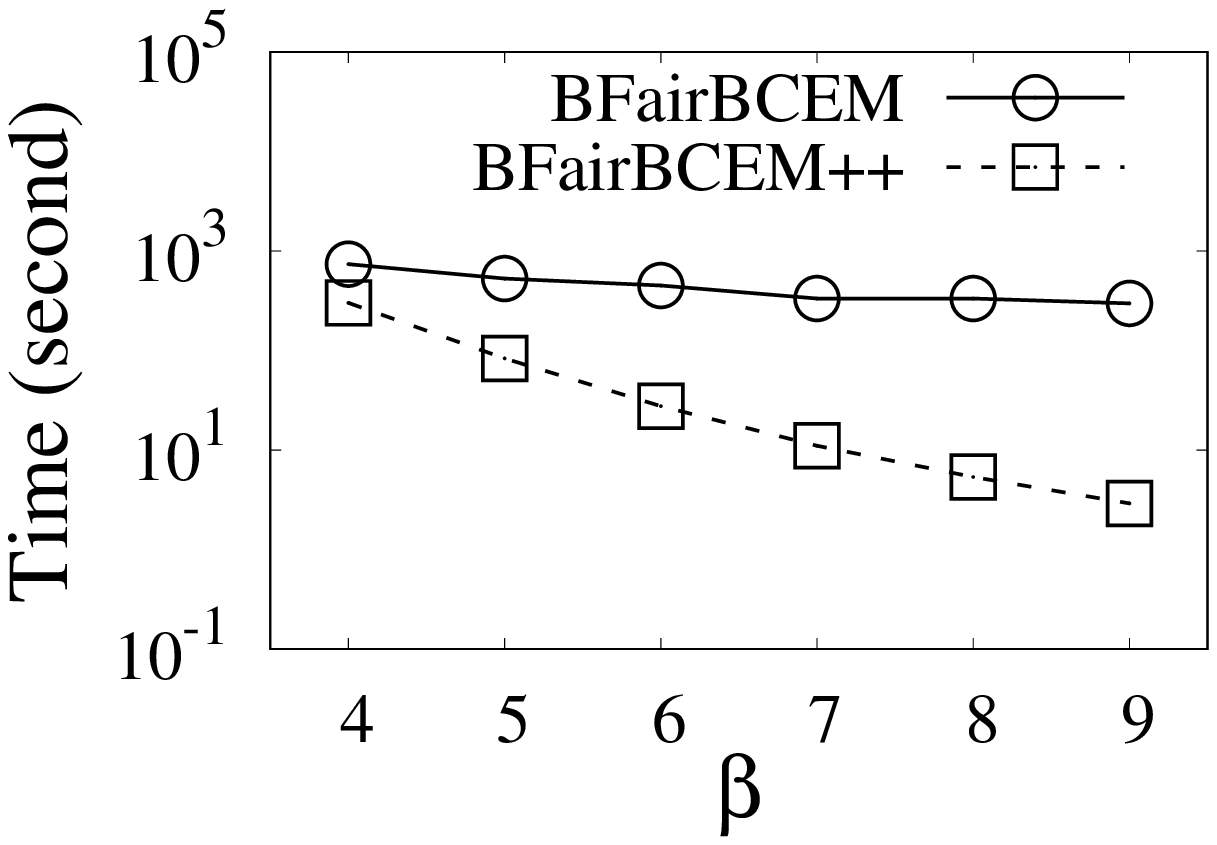}
      \end{minipage}
    }
    \subfigure[{\scriptsize \wiki~(vary $\beta$)}]{
      \label{fig:exp-twoside-alg-time-wiki-beta}
      \begin{minipage}{3.2cm}
      \centering
      \includegraphics[width=\textwidth]{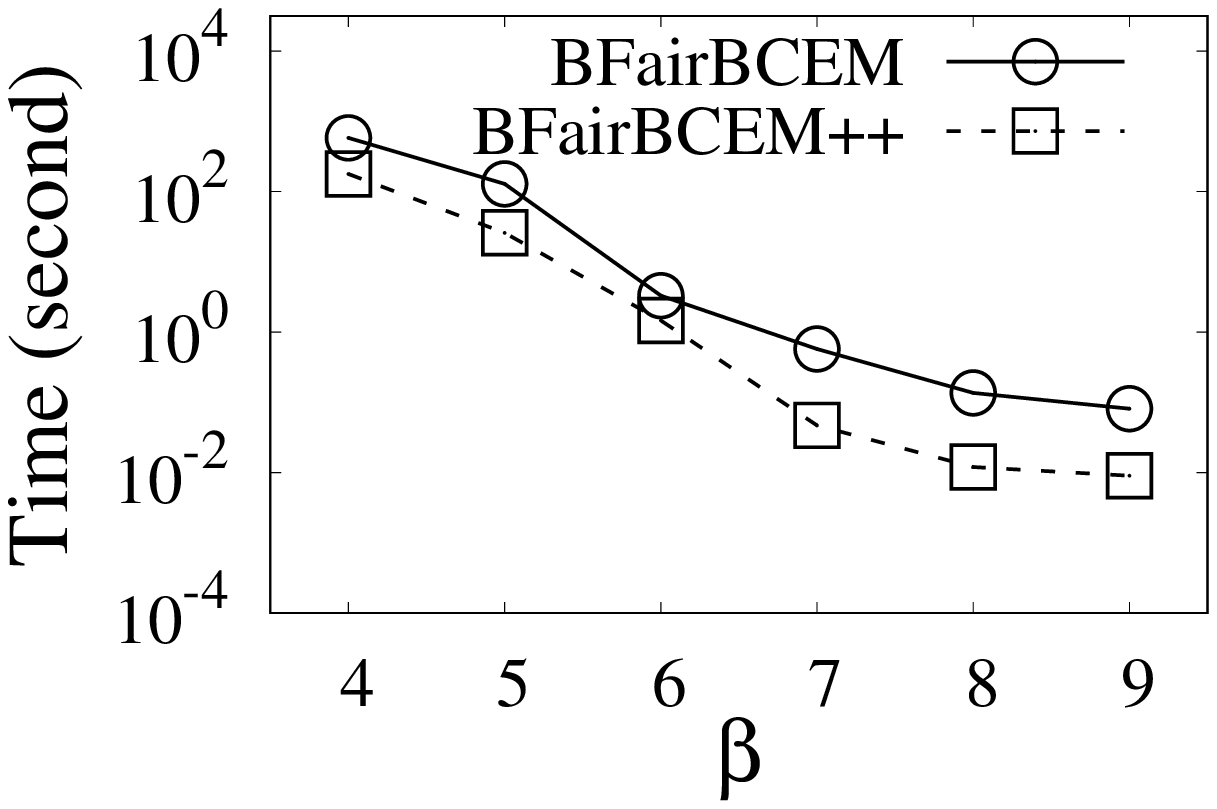}
      \end{minipage}
    }
    \subfigure[{\scriptsize \dblp (vary $\beta$)}]{
      \label{fig:exp-twoside-alg-time-dblp-beta}
      \begin{minipage}{3.2cm}
      \centering
      \includegraphics[width=\textwidth]{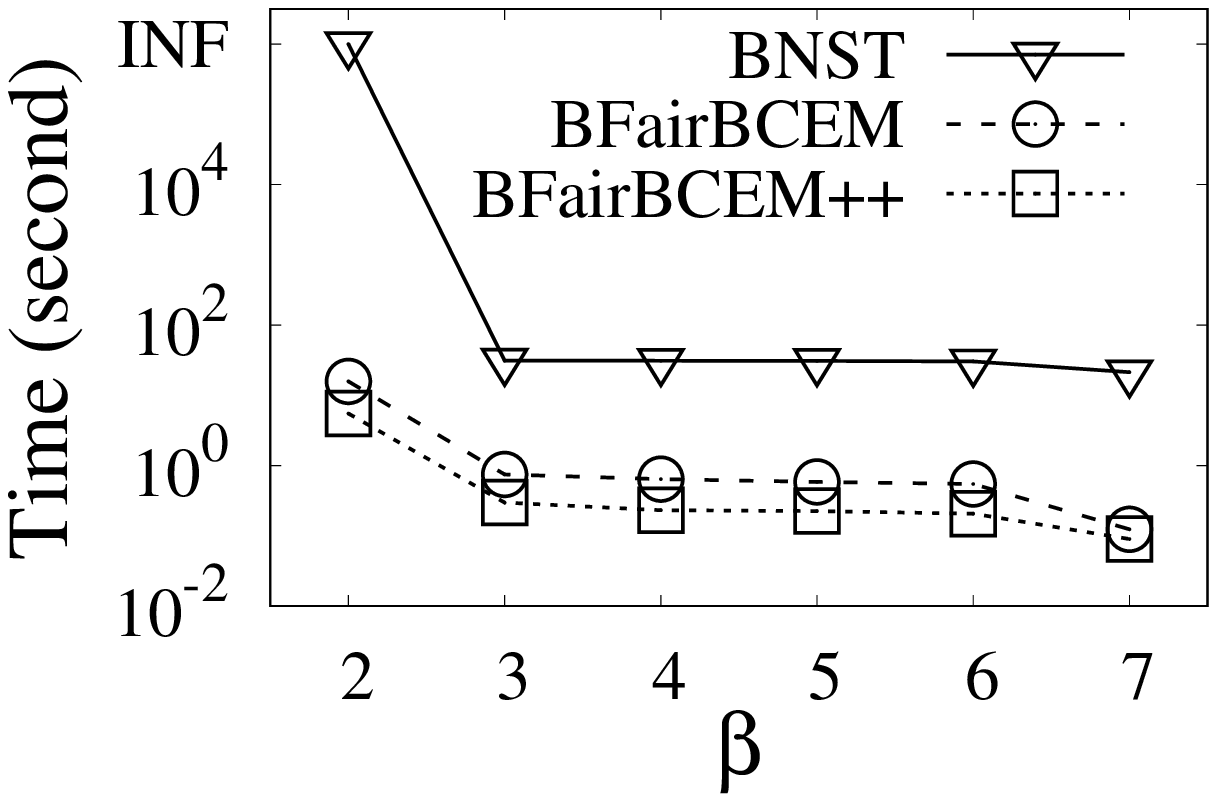}
      \end{minipage}
    }
    \vspace*{-0.3cm}
    
    \subfigure[{\scriptsize \youtube (vary $\delta$)}]{
      \label{fig:exp-twoside-alg-time-youtube-delta}
      \begin{minipage}{3.2cm}
      \centering
      \includegraphics[width=\textwidth]{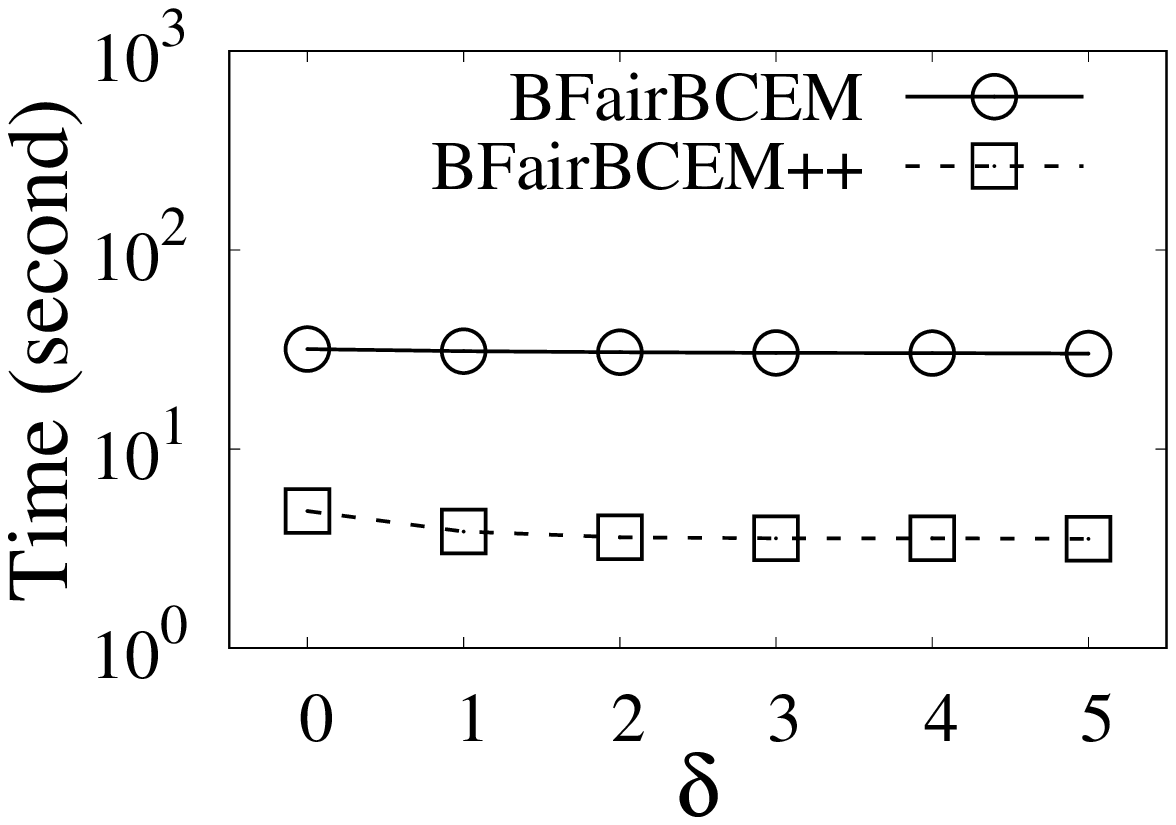}
      \end{minipage}
    }
    \subfigure[{\scriptsize \twi (vary $\delta$)}]{
      \label{fig:exp-twoside-alg-time-twi-delta}
      \begin{minipage}{3.2cm}
      \centering
      \includegraphics[width=\textwidth]{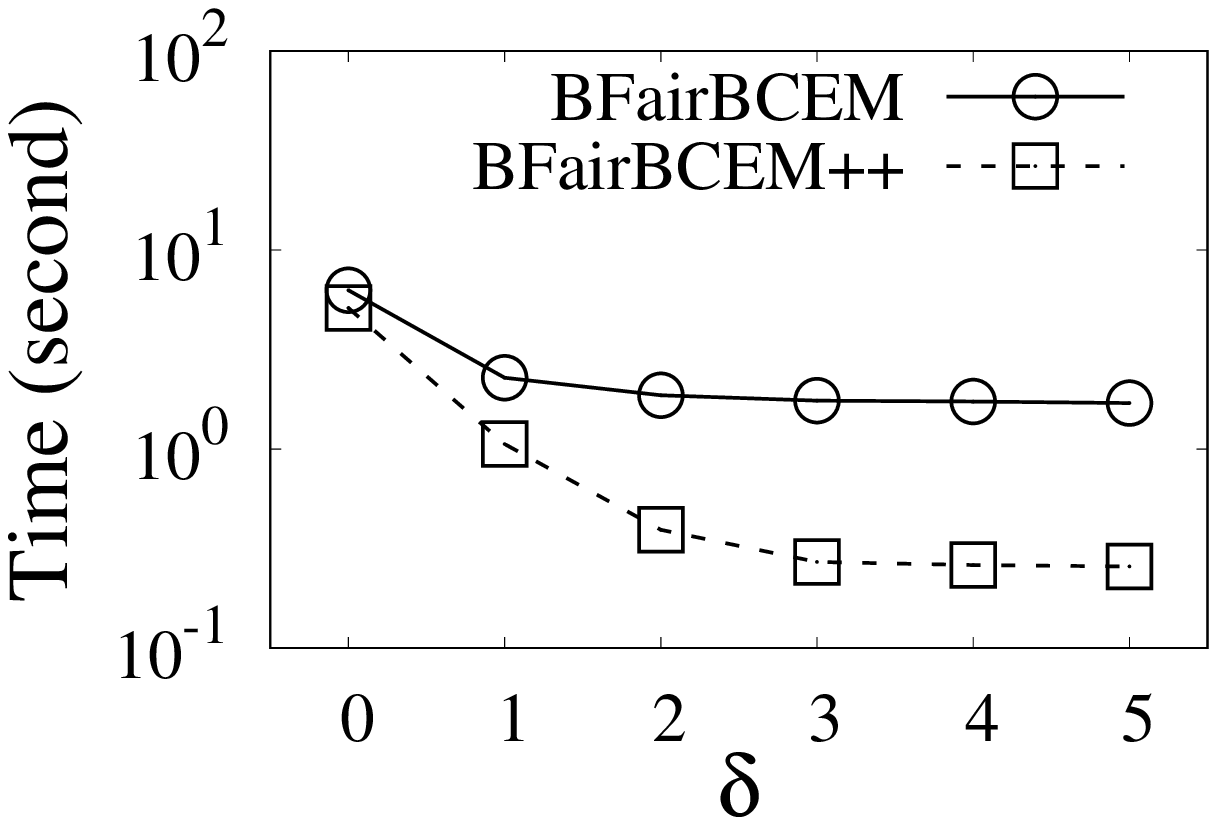}
      \end{minipage}
    }
    \subfigure[{\scriptsize \imdb (vary $\delta$)}]{
      \label{fig:exp-twoside-alg-time-imdb-delta}
      \begin{minipage}{3.2cm}
      \centering
      \includegraphics[width=\textwidth]{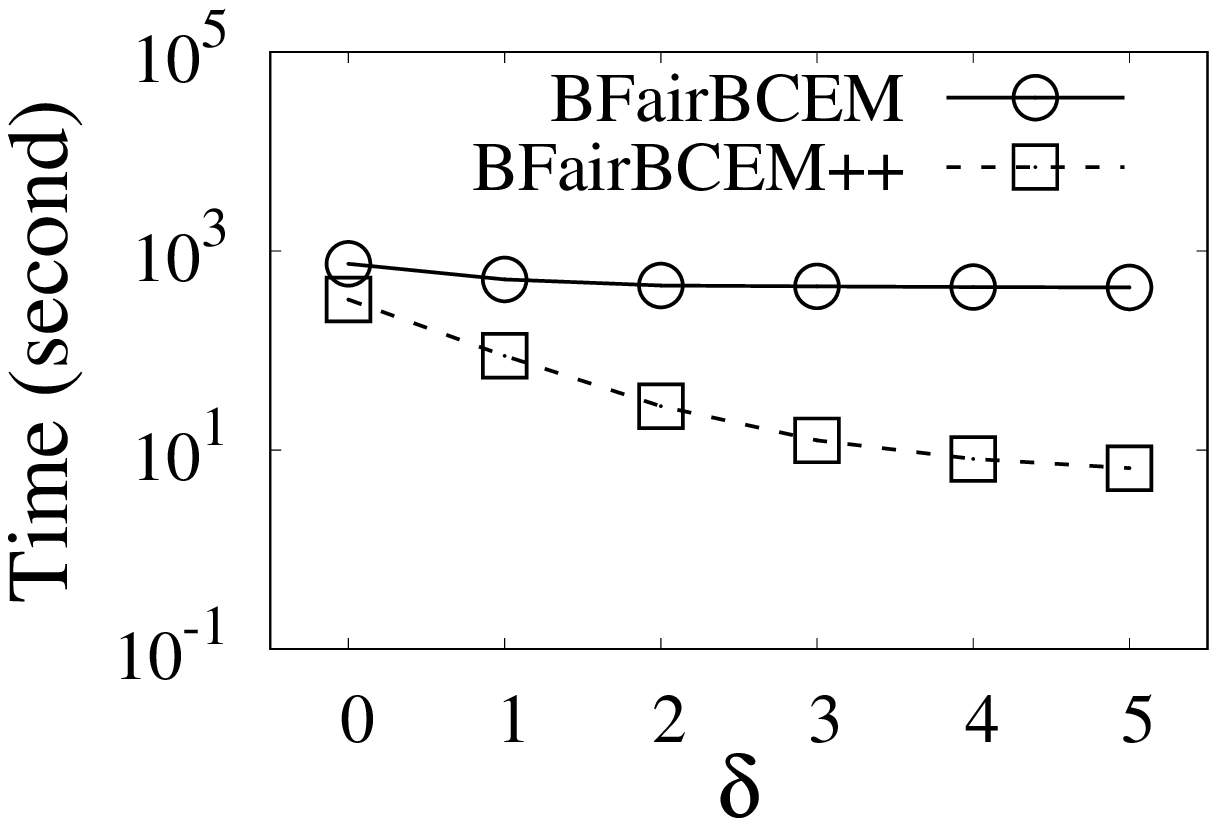}
      \end{minipage}
    }
    \subfigure[{\scriptsize \wiki~(vary $\delta$)}]{
      \label{fig:exp-twoside-alg-time-wiki-delta}
      \begin{minipage}{3.2cm}
      \centering
      \includegraphics[width=\textwidth]{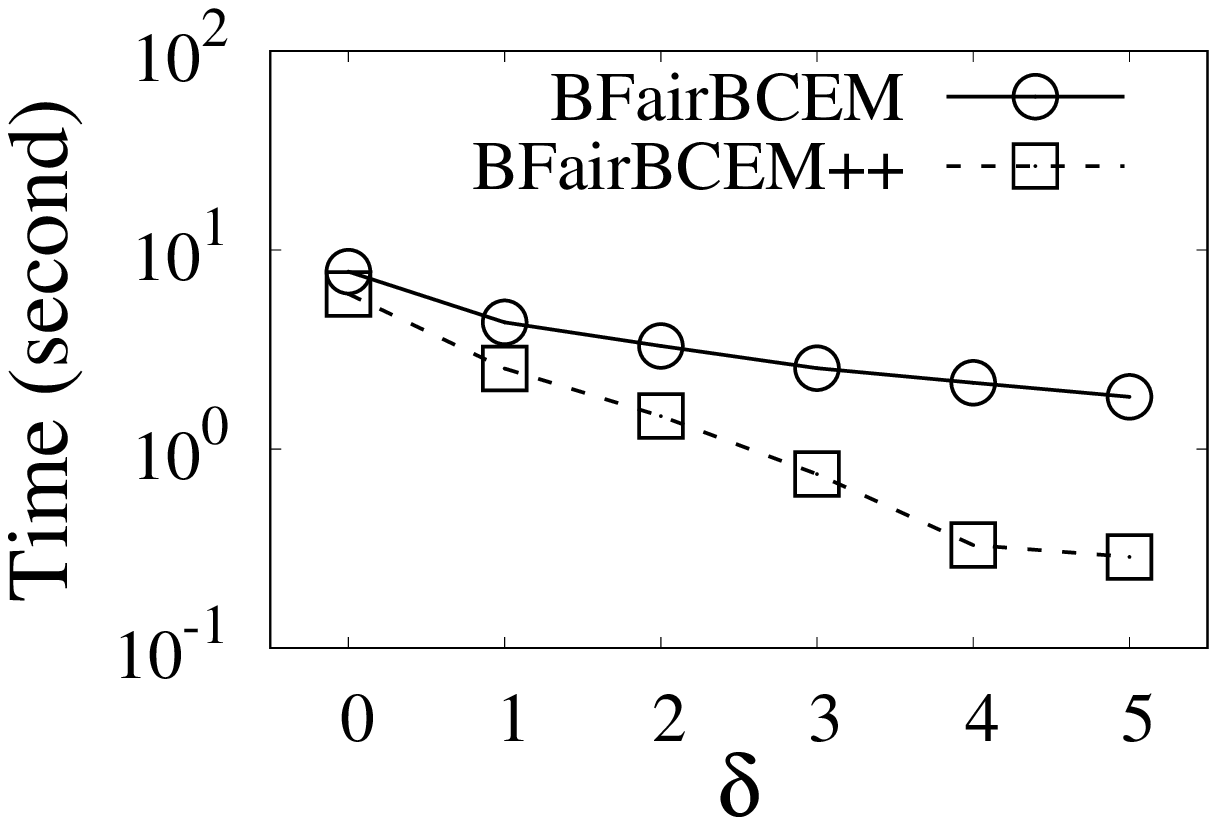}
      \end{minipage}
    }
    \subfigure[{\scriptsize \dblp (vary $\delta$)}]{
      \label{fig:exp-twoside-alg-time-dblp-delta}
      \begin{minipage}{3.2cm}
      \centering
      \includegraphics[width=\textwidth]{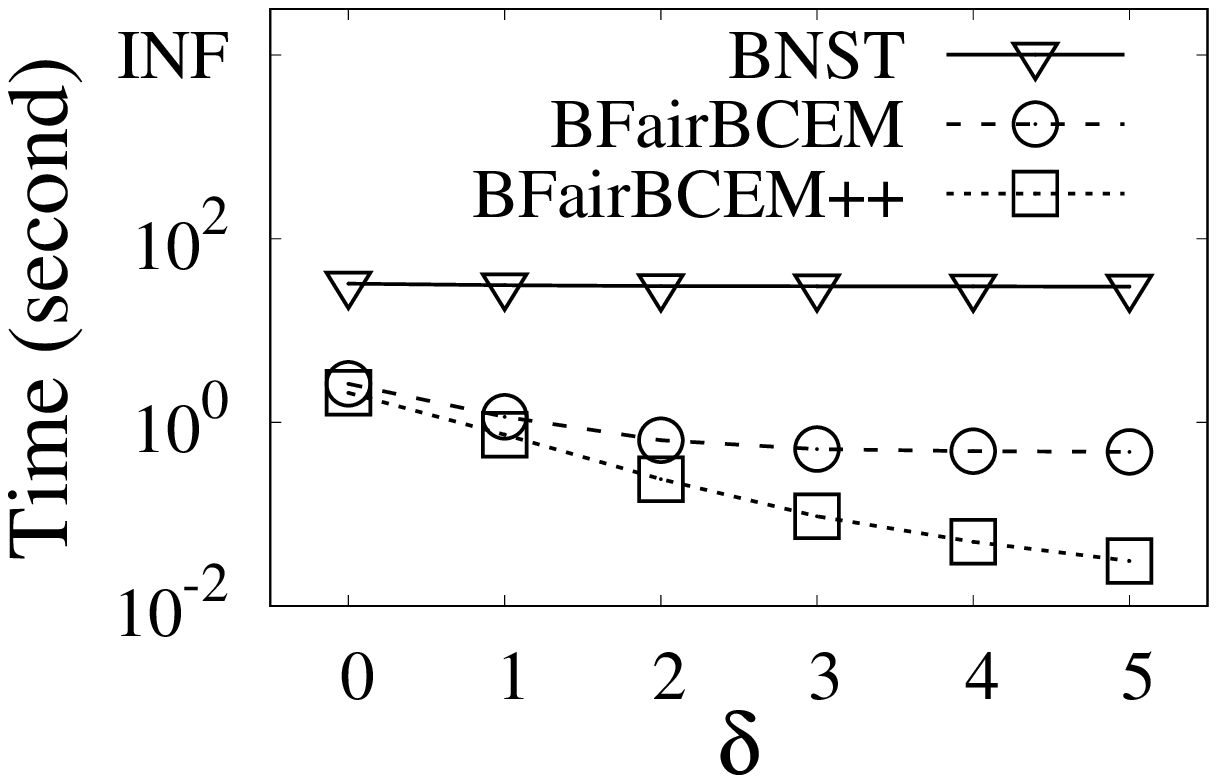}
      \end{minipage}
    }
	\vspace*{-0.3cm}
	\caption{The running time of the \tsnaivesearchtree, \twosideFBCEM and \twosideFBCEMPP~algorithms on different datasets.}
	\vspace*{-0.2cm}
	\label{fig:exp-twoside-alg-time}
\end{figure*}

\stitle{Exp-1: Evaluation of the pruning techniques.} For \nonesidebc~enumeration problem, both \onesideFBCEM and \onesideFBCEMPP~algorithms can use \fcore and \cfcore to prune unpromising nodes. For \ntwosidebc~enumeration problem, the pruning techniques \bfcore and \bcfcore can reduce the graph size in \twosideFBCEM and \twosideFBCEMPP. In this experiment, we evaluate these pruning techniques by comparing the number of remaining vertices after pruning and the consuming time with varying $\alpha$ and $\beta$. \figref{fig:exp-oneside-pruning-time-node} and \figref{fig:exp-twoside-pruning-time-node} illustrate the results for \nonesidebc~and \ntwosidebc~enumeration on \imdb, respectively. The results on the other datasets are consistent. \figref{fig:exp-oneside-pruning-time-node} (a)-(b) show that both \fcore and \cfcore can significantly reduce the number of vertices compared to the original graph as expected. Moreover, the number of remaining vertices decreases with larger $\alpha$ or $\beta$. In general, \cfcore outperforms \fcore in terms of the pruning performance, especially for relatively small $\alpha$ or $\beta$ values. As shown in \figref{fig:exp-oneside-pruning-time-node} (c)-(d), the running time of \fcore and \cfcore decreases as $\alpha$ or $\beta$ increases and \cfcore takes more time than \fcore to prune unpromising vertices. This is because \cfcore performs \fcore first and further reduces the graph by ego fair $\alpha$-$\beta$ core pruning in 2-hop graph (Algorithm \ref{alg:colorfulprune}). For example, in \figref{fig:exp-oneside-pruning-node-imdb-alpha} with $\alpha=8$, \fcore reduces the number of vertices from 9,266,649 to 12,507; and \cfcore further reduces the number of vertices to 1,318. When $\beta$ equals $8$, the number of remaining vertices after \fcore and \cfcore are 13,757 and 1,490 respectively as shown in
\figref{fig:exp-oneside-pruning-node-imdb-beta}. As a result, the \cfcore pruning can achieve superior pruning effect over the \fcore with slightly time consuming. Besides, similar results can also be found in \figref{fig:exp-twoside-pruning-time-node} for \twosidebc~enumeration. To sum up, the above experimental results validate the effectiveness and efficiency of the \fcore, \cfcore, \bfcore and \bcfcore pruning techniques.

\stitle{Exp-2: Evaluation of \osbc enumeration algorithms.} Here we evaluate \onesideFBCEM and \onesideFBCEMPP~algorithms equipped with descending \degreeorder by varying $\alpha, \beta$ and $\delta$. The results are depicted in \figref{fig:exp-oneside-alg-time}. As expected, the runtime of \onesideFBCEM and \onesideFBCEMPP~decreases with increasing $\alpha, \beta, \delta$ on all datasets. This is because for a large $\alpha, \beta$, many vertices can be pruned by the \fcore and \cfcore pruning techniques and the search space can also be correspondingly reduced during the branch and bound procedure. For a large $\delta$, the number of {\onesidebc}s decreases with increasing $\delta$ due to the maximality constraint, thus resulting in a trend of decreasing time. Moreover, we can also see that the runtime of \onesideFBCEMPP~is at least two orders of magnitude lower than that of \onesideFBCEM within all parameter settings over all datasets. For instance, when $\alpha=10$ with default $\beta$ and $\delta$, \onesideFBCEM consumes 29,192 seconds to find all {\onesidebc}s on \imdb, while \onesideFBCEMPP~takes only 91 seconds to output the results, which is almost three orders of magnitude faster than the \onesideFBCEM algorithm. These results validate the efficiency of the proposed \onesideFBCEM and \onesideFBCEMPP~algorithms.

In \onesideFBCEM and \onesideFBCEMPP~algorithms, a vertex is selected from the candidate set to the current biclique for performing a backtracking search procedure. Since the search spaces with various orderings are significantly different, we also evaluate the two algorithms with \idorder and \degreeorder orderings. Table.\ref{table:order} depicts the runtime of \onesideFBCEM and \onesideFBCEMPP~equipped with \idorder and \degreeorder in the case of default $\alpha, \beta, \delta$ over all datasets. As shown in Table.\ref{table:order}, the \onesideFBCEM with \degreeorder is significantly faster than that with \idorder. For example, in \imdb, the \onesideFBCEM algorithms with \idorder and \degreeorder consume 4,378 seconds and 2,098 seconds to output all {\nonesidebc}s. Clearly, the latter is almost 2 times faster than the former. Similar results can also be found for \onesideFBCEMPP~algorithms with \idorder and \degreeorder. Again, the \onesideFBCEMPP~algorithm outperforms \onesideFBCEM on all datasets, which is consistent with our previous founding. The results indicate that the \degreeorder ordering is more efficient that the \idorder ordering during the search procedure.

In addition, We compare \naivesearchtree~with the proposed \onesideFBCEM~and \onesideFBCEMPP~on all datasets. We only show the results on \dblp in \figref{fig:exp-oneside-alg-time} as \naivesearchtree runs out of time on other datasets with most parameter settings. As can be seen, \onesideFBCEM~is at least two orders of magnitude faster than \naivesearchtree. These results confirm that our proposed algorithms significantly outperform the \naivesearchtree algorithm.

\begin{table}[t!]\vspace*{-0.3cm}
\scriptsize
\caption{\textbf{The runtime of different algorithms with \idorder and \degreeorder.}}
\label{table:order}
\centering
\setlength{\tabcolsep}{1mm}
\vspace*{-0.2cm}
\begin{tabular}{cccccccc}
\toprule
Algorithm (s)&Ordering&\imdb&\youtube&\twi&\wiki&\dblp\\
\midrule
\multirow{2}{*}{\onesideFBCEM}
&\idorder&7,022.7&157.1&854.2&90.6&6.3 \\
&\degreeorder&1,612.9&43.6&611.8&45.9&2.6 \\
\midrule
\multirow{2}{*}{\onesideFBCEMPP}
&\idorder&78.6&16.1&72.5&13.2&0.6 \\
&\degreeorder&61.9&8.3&65.1&12.4&0.5 \\
\midrule
\multirow{2}{*}{\twosideFBCEM}
&\idorder&174.2&2.3&76.8&0.9&1.5 \\
&\degreeorder&68.1&1.4&69.1&0.4&1.1 \\
\midrule
\multirow{2}{*}{\twosideFBCEMPP}
&\idorder&19.8&7.4&63.8&0.3&0.7 \\
&\degreeorder&17.2&1.7&59.7&0.2&0.6 \\
\midrule
\end{tabular}
	\vspace*{-0.4cm}
\end{table}

\begin{figure}[t!]\vspace*{-0.5cm}
\centering
    \subfigure[{\scriptsize \wiki~(vary $\alpha$)}]{
      \label{fig:exp-oneside-max-num-wiki-alpha}
      \begin{minipage}{3.2cm}
      \centering
      \includegraphics[width=\textwidth]{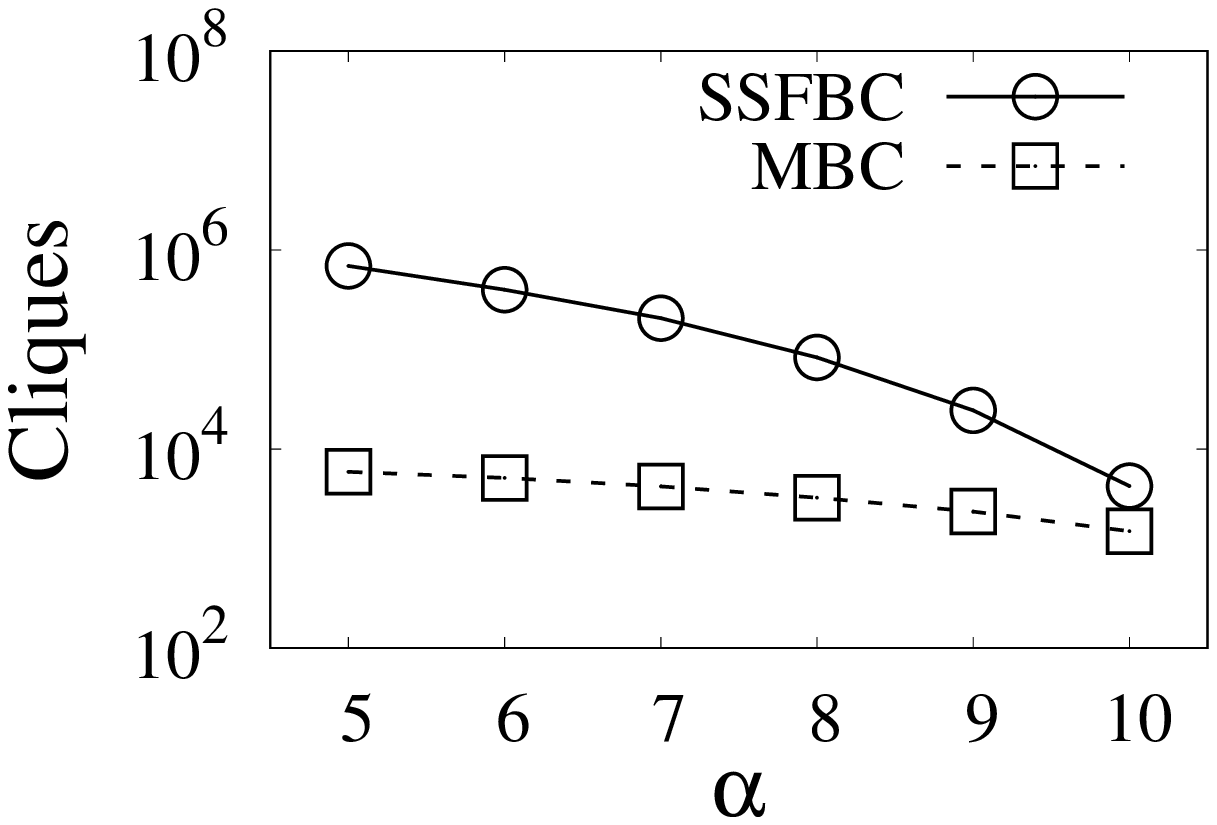}
      \end{minipage}
    }
    \subfigure[{\scriptsize \wiki~(vary $\alpha$)}]{
      \label{fig:exp-twoside-max-num-wiki-}
      \begin{minipage}{3.2cm}
      \centering
      \includegraphics[width=\textwidth]{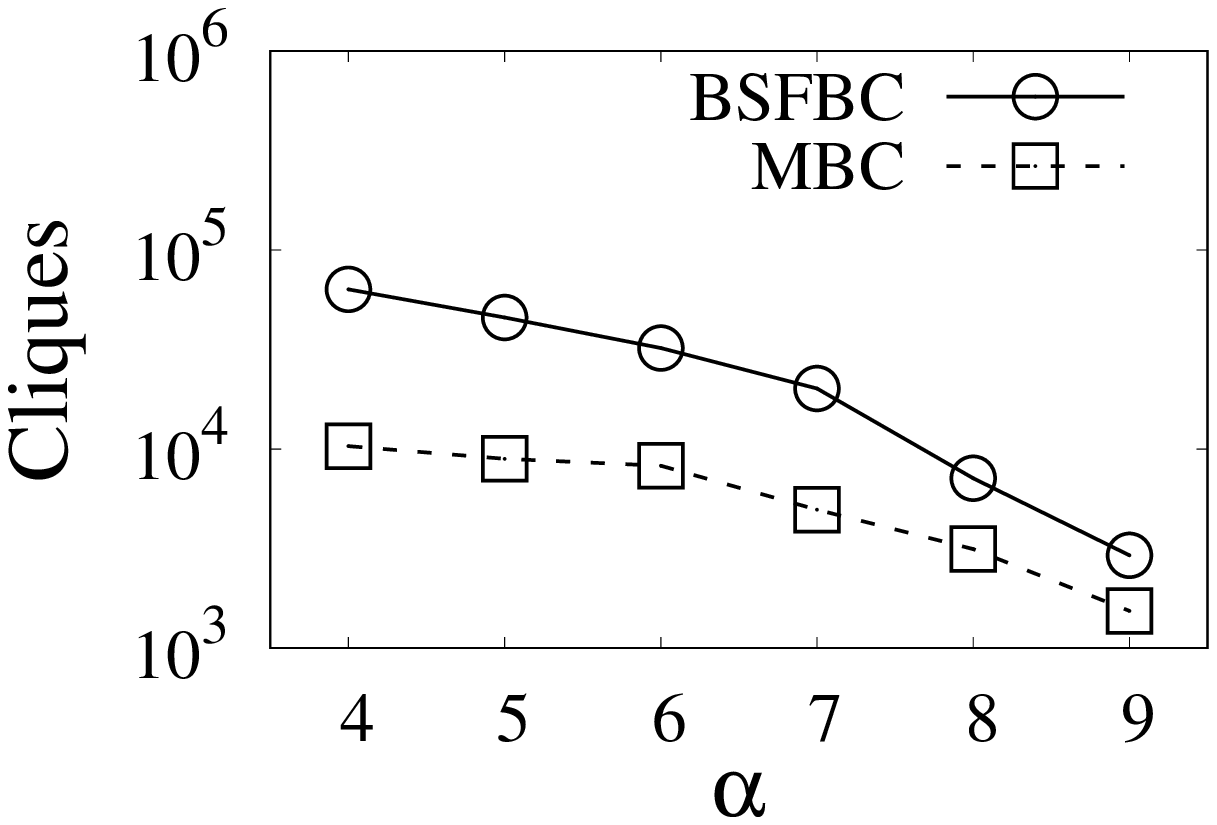}
      \end{minipage}
    }
    \vspace*{-0.3cm}
        
    \subfigure[{\scriptsize \wiki~(vary $\beta$)}]{
      \label{fig:exp-oneside-max-num-wiki-beta}
      \begin{minipage}{3.2cm}
      \centering
      \includegraphics[width=\textwidth]{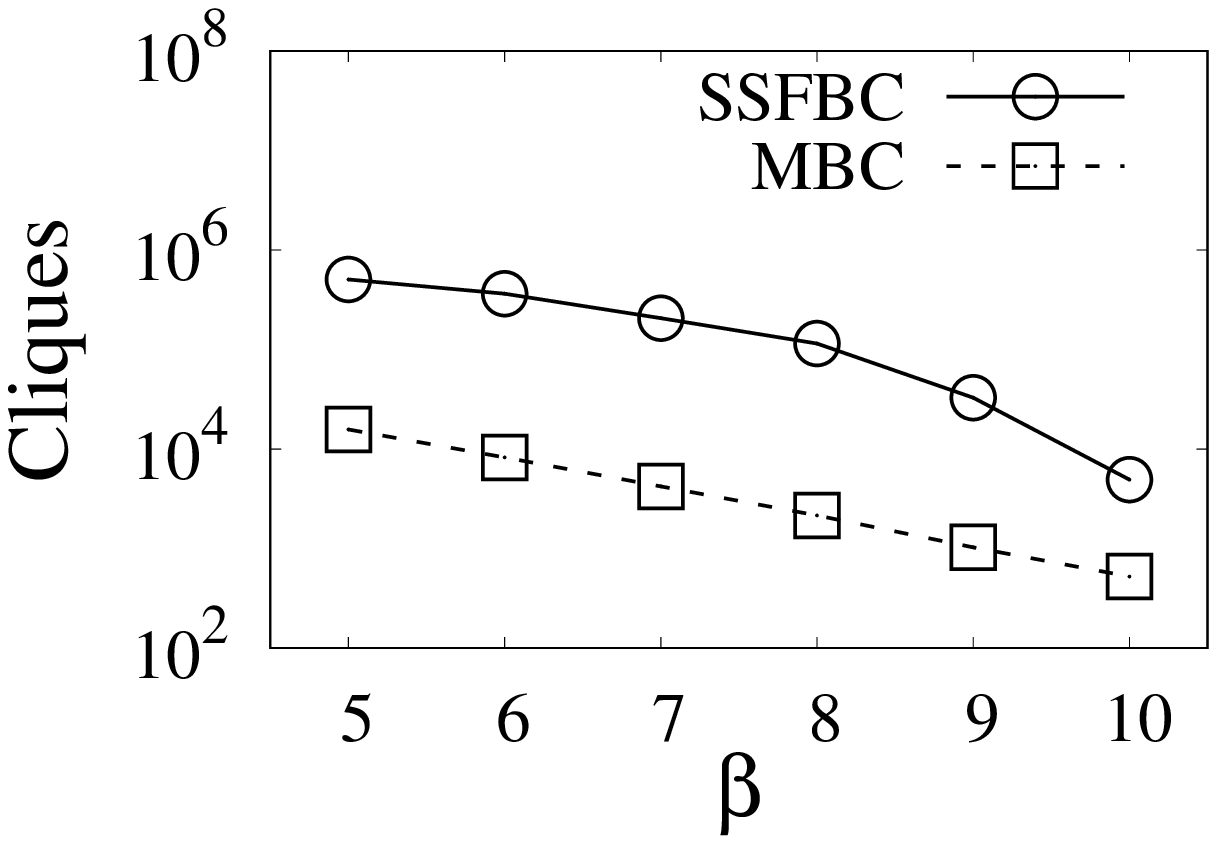}
      \end{minipage}
    }
    \subfigure[{\scriptsize \wiki~(vary $\beta$)}]{
      \label{fig:exp-twoside-max-num-wiki-beta}
      \begin{minipage}{3.2cm}
      \centering
      \includegraphics[width=\textwidth]{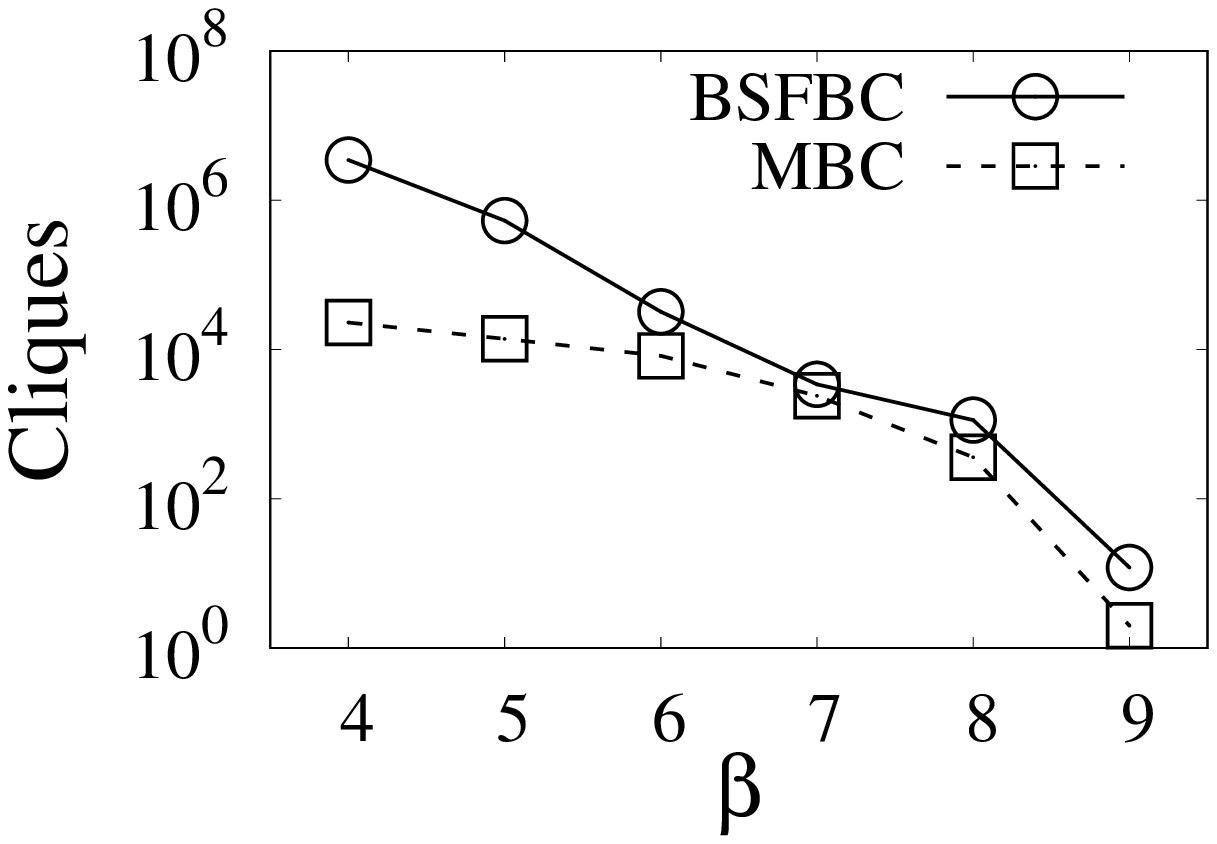}
      \end{minipage}
    }
    \vspace*{-0.3cm}
    
    \subfigure[{\scriptsize \wiki~(vary $\delta$)}]{
      \label{fig:exp-oneside-max-num-wiki-delta}
      \begin{minipage}{3.2cm}
      \centering
      \includegraphics[width=\textwidth]{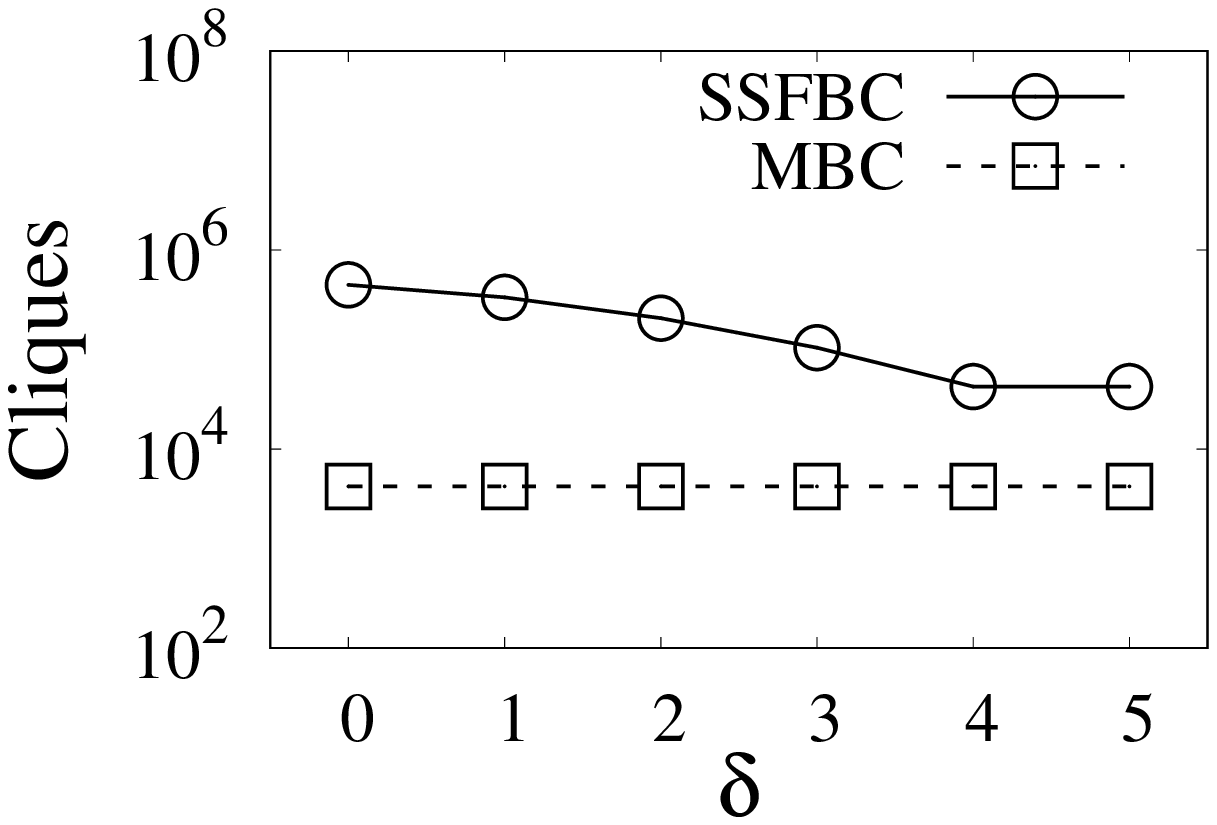}
      \end{minipage}
    }
    \subfigure[{\scriptsize \wiki~(vary $\delta$)}]{
      \label{fig:exp-twoside-max-num-wiki-delta}
      \begin{minipage}{3.2cm}
      \centering
      \includegraphics[width=\textwidth]{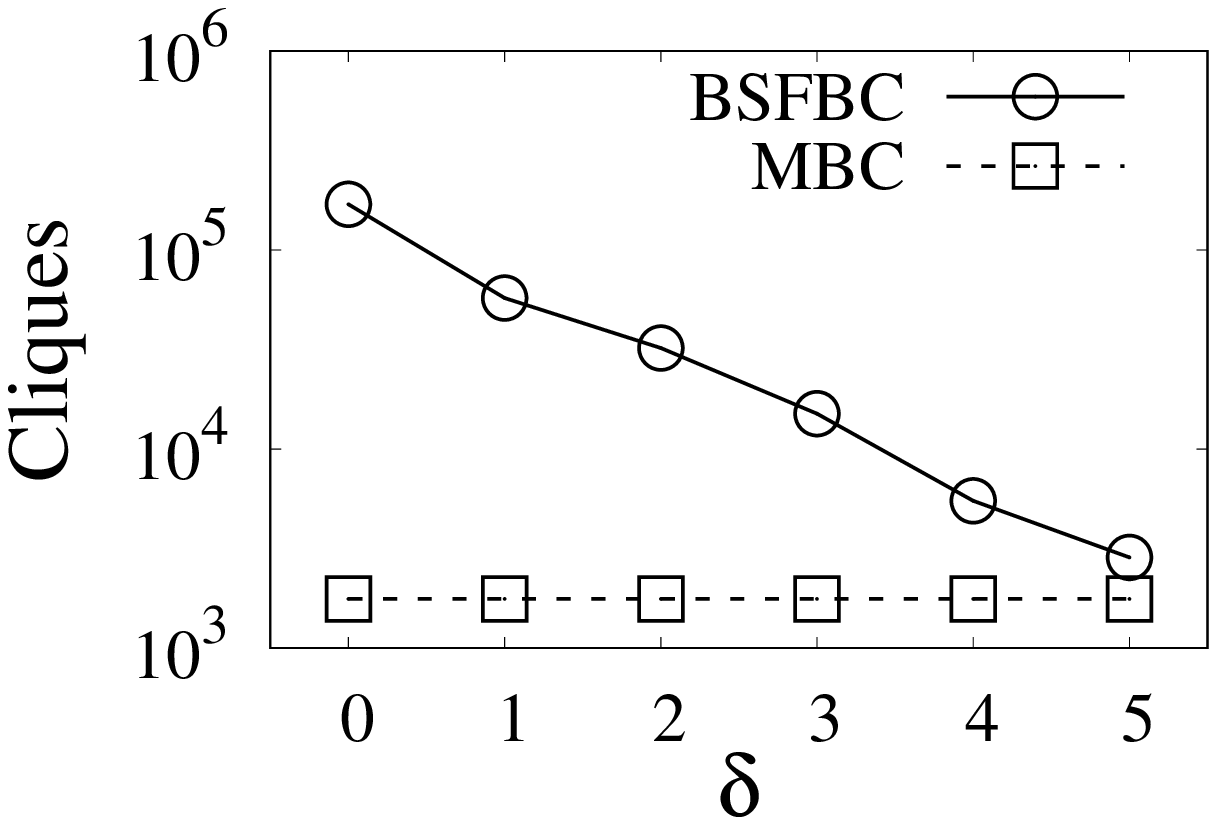}
      \end{minipage}
    }
	\vspace*{-0.3cm}
	\caption{The numbers of the maximal bicliques, {\osbc}s and {\tsbc}s.}
	\vspace*{-0.4cm}
	\label{fig:exp-onetwo-max-num}
\end{figure}

\stitle{Exp-3: Evaluation of \tsbc~enumeration algorithms.} We evaluate the runtime of \twosideFBCEM and \twosideFBCEMPP~with \degreeorder by varying $\alpha, \beta, \delta$. The results are depicted in \figref{fig:exp-twoside-alg-time}. As expected, the runtime of \twosideFBCEM and \twosideFBCEMPP~decreases as $\alpha,\beta,\delta$ increases, which is similar to that of \nonesidebc~enumeration algorithms. Moreover, we also observe that the \twosideFBCEMPP~algorithm~is almost 3-100 times faster than the \twosideFBCEM algorithm within all parameter settings on all datasets. For example, when $\beta=7$ with default $\alpha$ and $\delta$, the runtime of \twosideFBCEM and \twosideFBCEMPP~take 17 seconds and 1 second to output all {\ntwosidebc}s on \youtube, respectively. Obviously, the former is significantly faster than the latter. These results validate the efficiency of the proposed \twosideFBCEM and \twosideFBCEMPP~algorithms.

In addition, we compare the running time of \twosideFBCEM and \twosideFBCEMPP~algorithms armed with \idorder and \degreeorder under default $\alpha, \beta, \delta$. As seen in Table.\ref{table:order}, the \twosideFBCEM with \degreeorder significantly outperforms \idorder by a large margin. For example, in \imdb, the \twosideFBCEM algorithm with \idorder takes 253 seconds to find all {\ntwosidebc}s, while the algorithm with \degreeorder only needs 169 seconds. Similar results can also be found for \twosideFBCEMPP~algorithms with \idorder and \degreeorder. Again, the \twosideFBCEMPP~algorithm is faster than \twosideFBCEM over all datasets. These results also demonstrate the efficiency of \degreeorder ordering which is consistent with our previous findings.

Besides, we also evaluate the running time of \tsnaivesearchtree~with \twosideFBCEMPP~and \twosideFBCEMPP~on all datasets. We show the results on \dblp in \figref{fig:exp-twoside-alg-time} as \naivesearchtree cannot terminate with limited time on other datasets under parameter settings. We can see that \twosideFBCEM~is at least two orders of magnitude faster than \tsnaivesearchtree. These results confirm that our algorithms are significantly faster than the \tsnaivesearchtree algorithm.

\stitle{Exp-4: The number of {\osbc}s and {\tsbc}s.} \figref{fig:exp-onetwo-max-num} reports the number of {\nonesidebc}s and {\ntwosidebc}s with varying $\alpha, \beta, \delta$ on \wiki. Note that we find the maximal biclique $B(L,R)$ satisfying $|L| \ge \alpha$ and $|R| \ge 2 \times \beta$ for comparison with \nonesidebc. To compare with \ntwosidebc, we search the maximal biclique $B(L,R)$ with $|L| \ge 2\times \alpha$ and $|R| \ge 2 \times \beta$. Clearly, there are significant numbers of {\nonesidebc}s and {\ntwosidebc}s on \wiki. For example, in the case of $\alpha=6, \beta=6, \delta=2$ for \onesidebc enumeration problem, there are 9,548 maximal bicliques, 346,411 {\onesidebc}s. As the case of $\alpha=3,\beta=6,\delta=2$ for \twosidebc enumeration problem, there are 546,411 {\twosidebc}s, and 9,548 maximal biclique. In general, the number of {\nonesidebc}s and {\ntwosidebc}s is larger than that of maximal bicliques. This finding is consistent with our analysis in \secref{sec:preliminaries}, because any {\nonesidebc} or {\ntwosidebc} must be included in a maximal biclique. Additionally, we can see that the number of maximal bicliques, {\nonesidebc}s and {\ntwosidebc}s decreases as $\alpha, \beta, \delta$ increases. This is because with a larger $\alpha$/$\beta$/$\delta$, the fairness constraint and size constraint become stricter for \nonesidebc/\nonesidebc~models and maximal biclique model respectively.

\begin{figure}[t!]\vspace*{-0.5cm}
	\begin{center}
		\begin{tabular}[t]{c}
			\subfigure[{\scriptsize \dblp, \osbc~enumeration algorithms (vary $m$)}]{
                \label{fig:exp-scala-vary-m-oneside-dblp}
				\includegraphics[width=0.4\columnwidth, height=2.5cm]{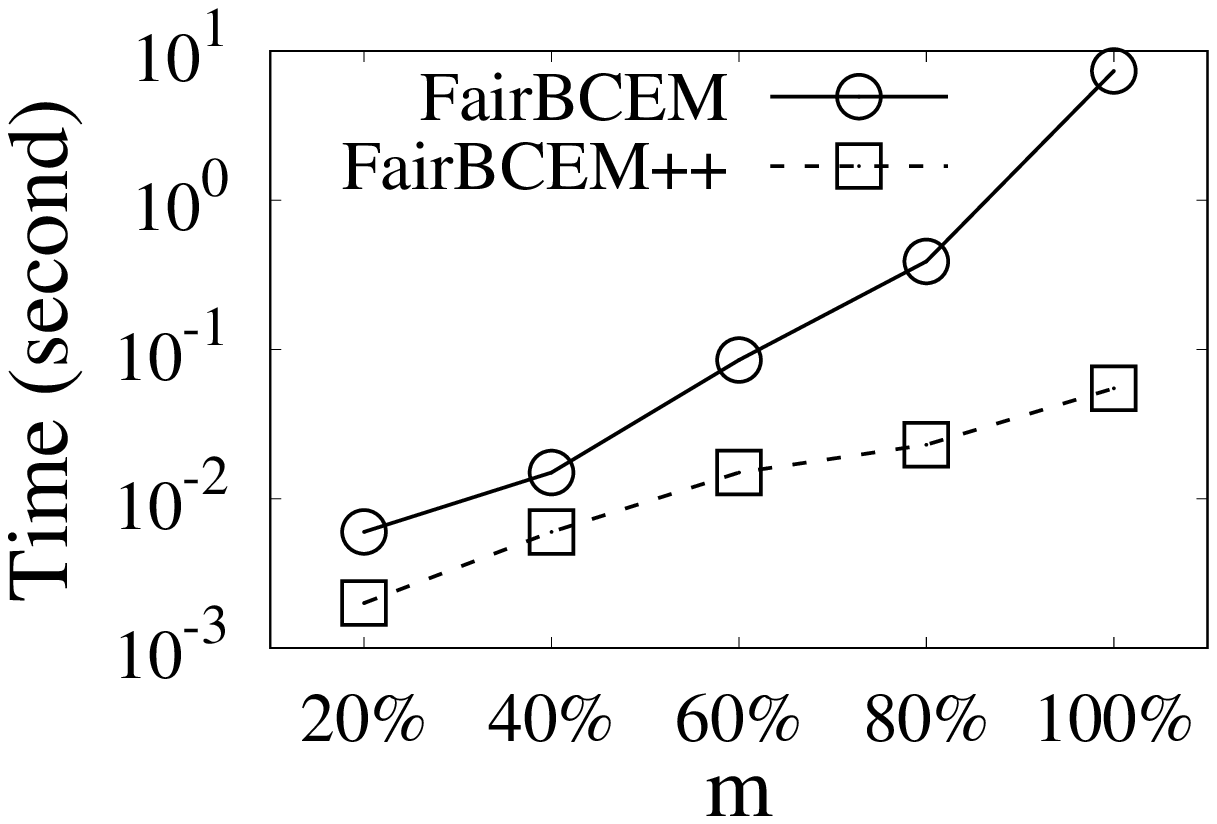}
			}
			\subfigure[{\scriptsize \dblp, \tsbc~enumeration algorithms (vary $m$)}]{
                \label{fig:exp-scala-vary-m-twoside-dblp}
				\includegraphics[width=0.4\columnwidth, height=2.5cm]{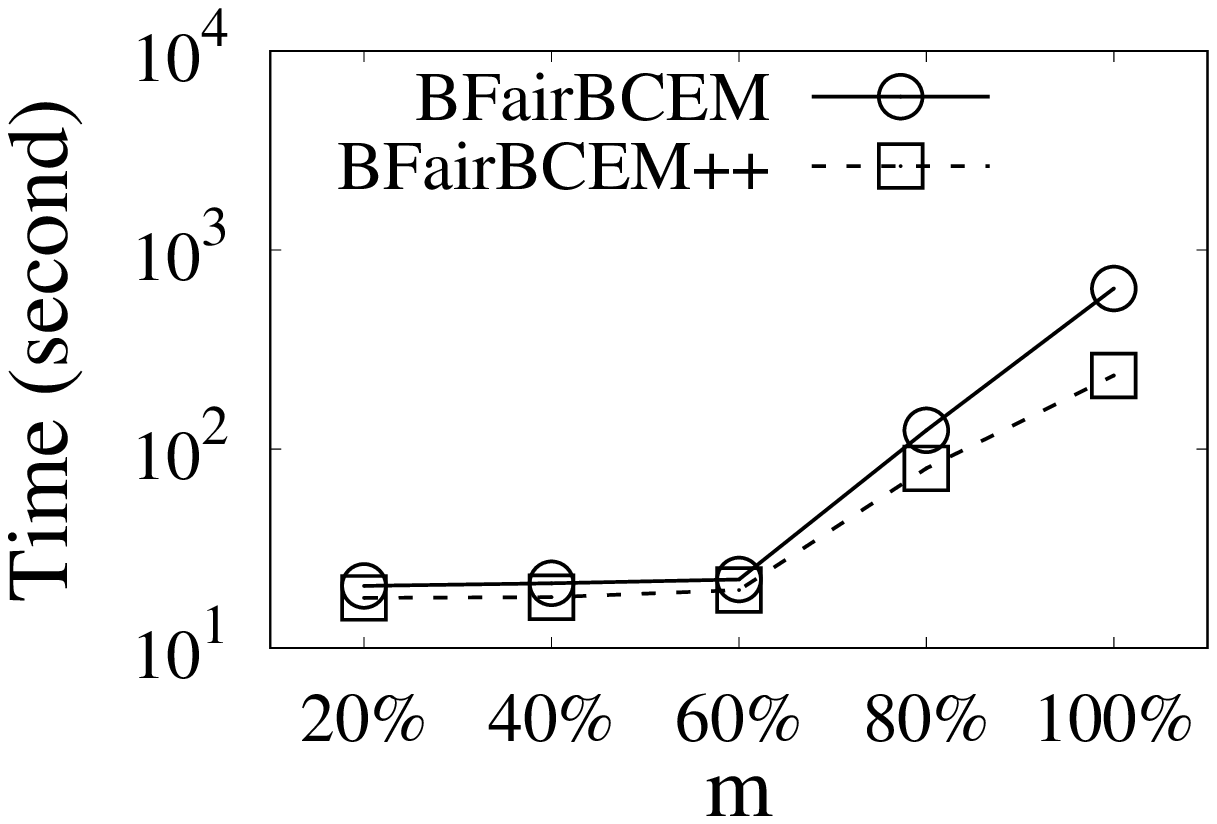}
			}
		\end{tabular}
	\end{center}
	\vspace*{-0.5cm}
	\caption{The scalability of the proposed algorithms.}
	\label{fig:exp-scalability-test}
\vspace*{-0.4cm}
\end{figure}

\begin{figure}[t!]\vspace*{-0.1cm}
	\begin{center}
		\begin{tabular}[t]{c}
			\subfigure[{\scriptsize \osbc~enumeration algorithms}]{
				\includegraphics[width=0.43\columnwidth, height=2.5cm]{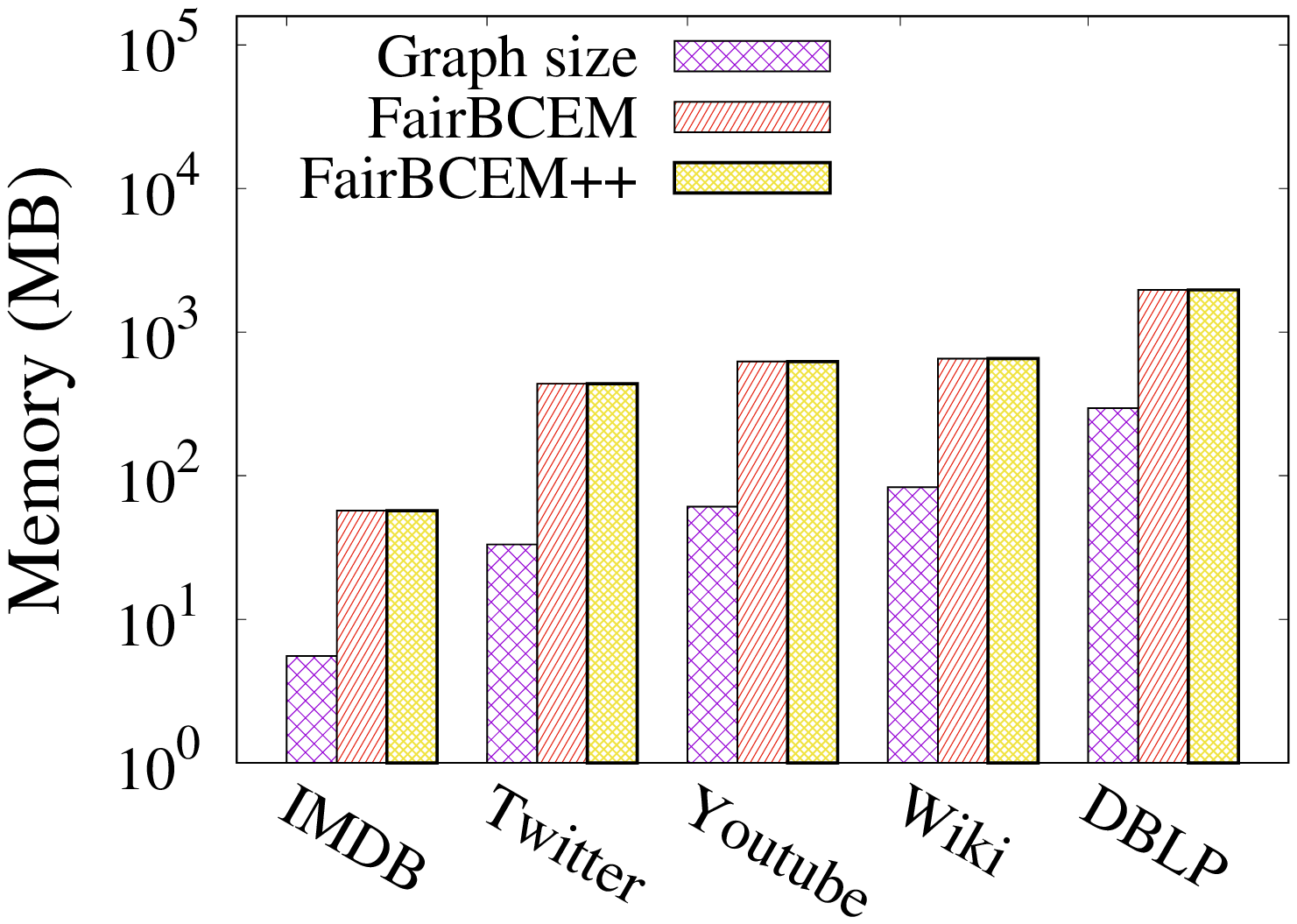}
			}
			\subfigure[{\scriptsize \tsbc~enumeration algorithms}]{
				\includegraphics[width=0.43\columnwidth, height=2.5cm]{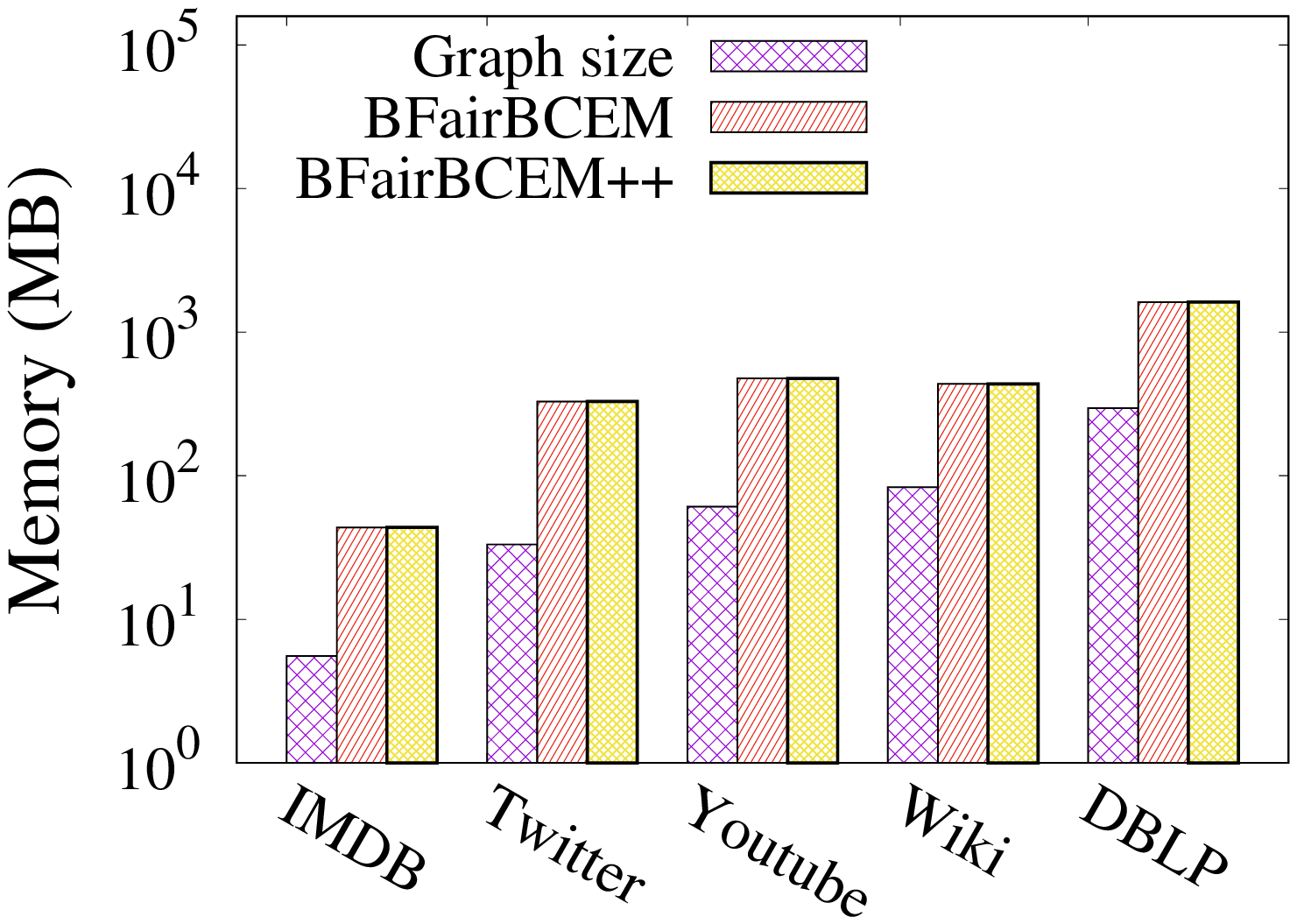}
			}
            \vspace*{-0.3cm}
		\end{tabular}
	\end{center}
	\vspace*{-0.2cm}
	\caption{The memory overhead.}
	\vspace*{-0.4cm}
	\label{fig:exp-Memory-overhead}
\end{figure}

\begin{figure*}[t!]\vspace*{-0.5cm}
    \subfigure[{\scriptsize \dbda, \osbc, {\protect\\} ($\alpha=3,\beta=3,\delta=2$)}]{
		\label{fig:dblpcase1}
		\raisebox{0.05\height}{
		    \centering
			\includegraphics[height=2cm]{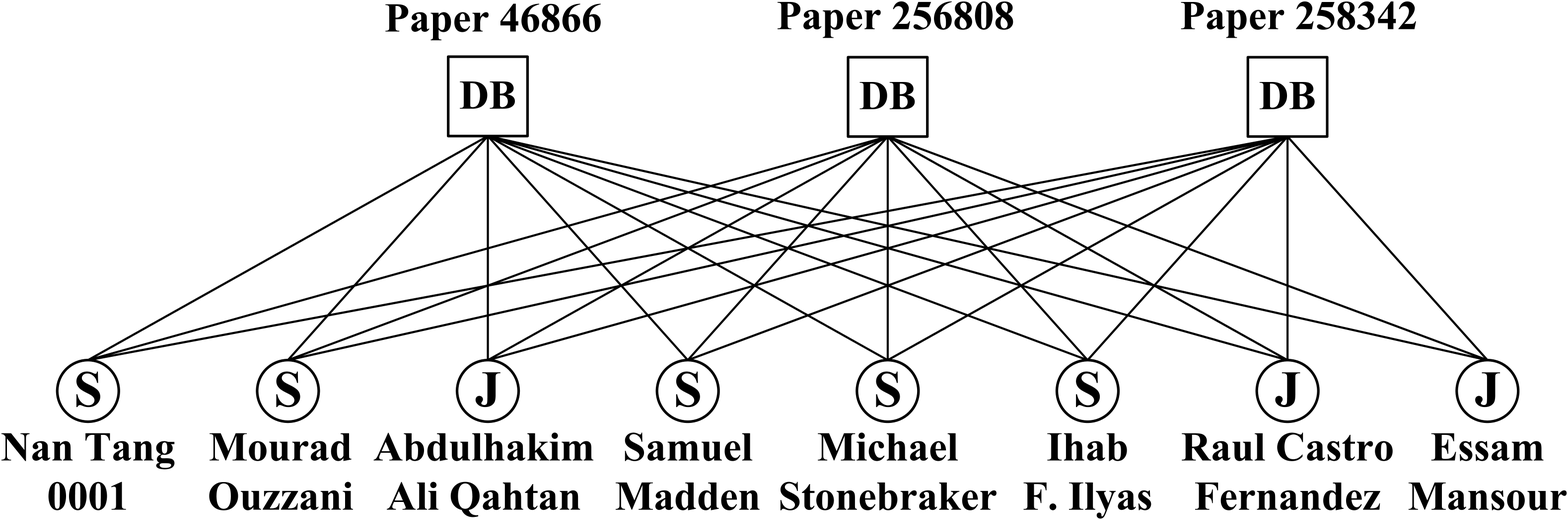}
		}
	}
	\subfigure[{\scriptsize{\dbda, \tsbc, {\protect\\} ($\alpha=1,\beta=2,\delta=2$)}}]{
		\label{fig:dblpcase2}
		\raisebox{0.05\height}{
    		\centering
    		\includegraphics[height=2cm]{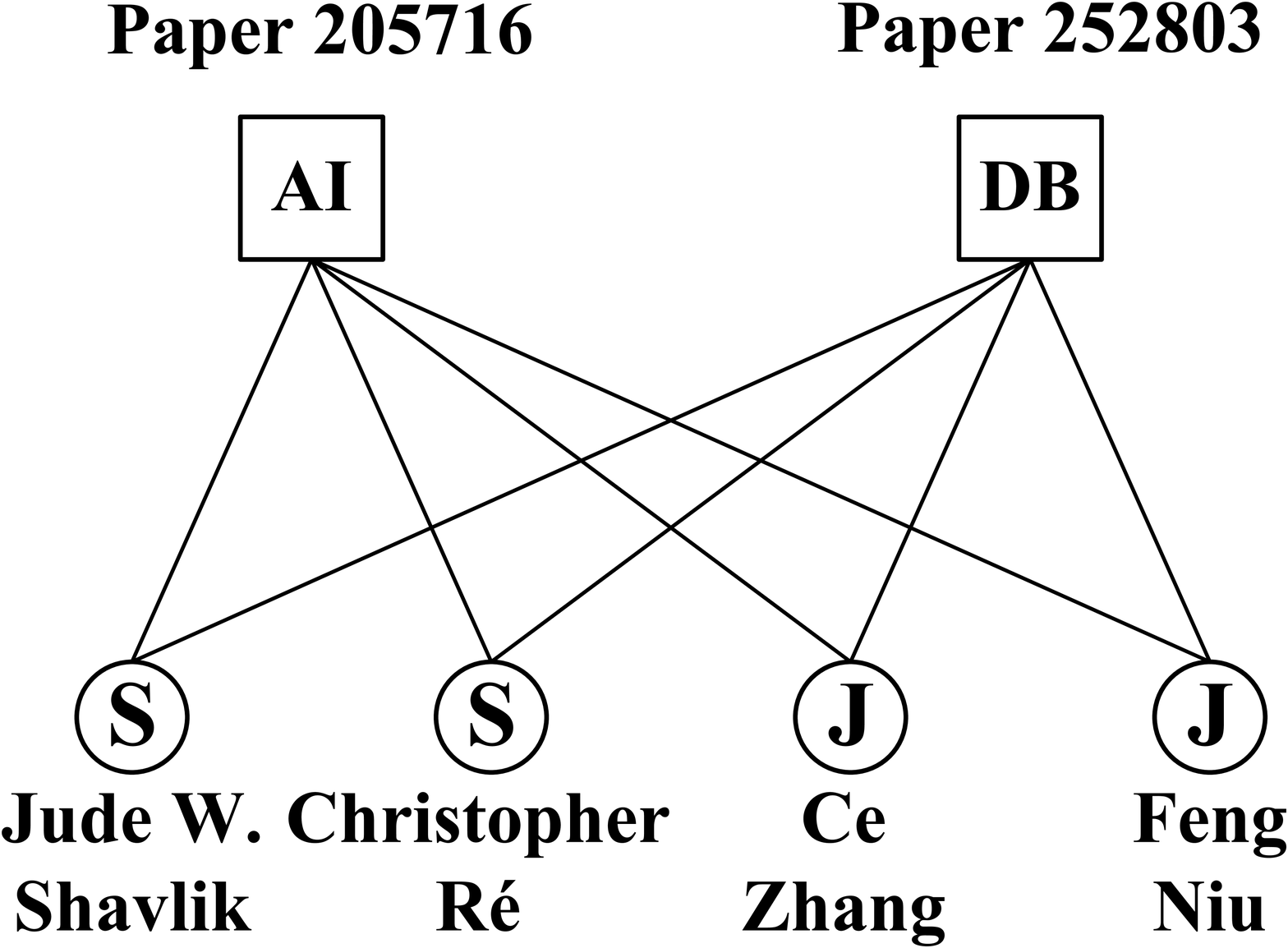}
        }
	}
	\subfigure[{\scriptsize{\dbds, \osbc, {\protect\\} ($\alpha=2,\beta=2,\delta=2$)}}]{
		\label{fig:dblpcase3}
		\raisebox{0.05\height}{
		    \centering
			\includegraphics[height=2cm]{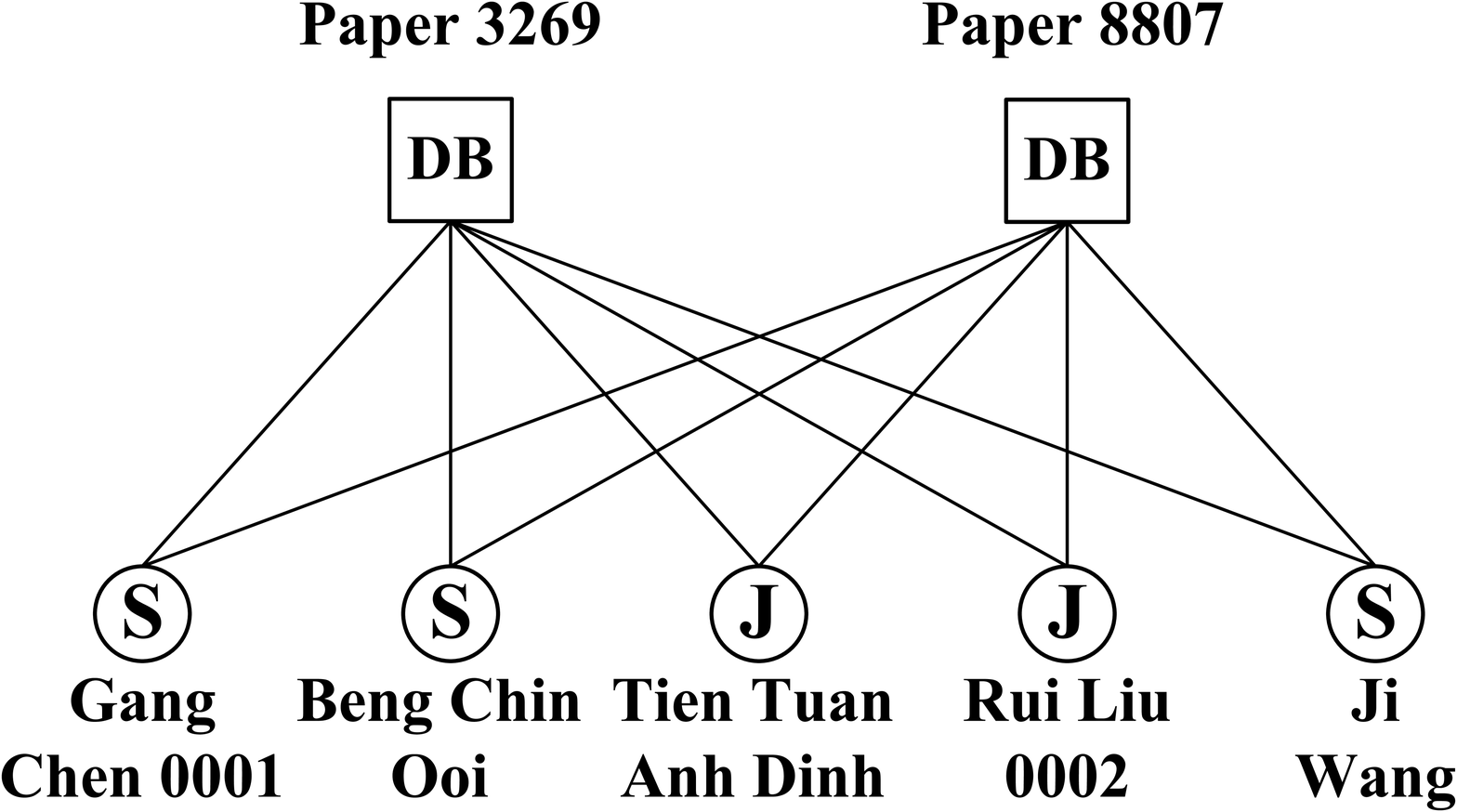}
		}
	}
	\subfigure[{\scriptsize{\dbds, \tsbc, {\protect\\} ($\alpha=1,\beta=2,\delta=2$)}}]{
		\label{fig:dblpcase4}
		\raisebox{0.05\height}{
		\centering
		\includegraphics[height=2cm]{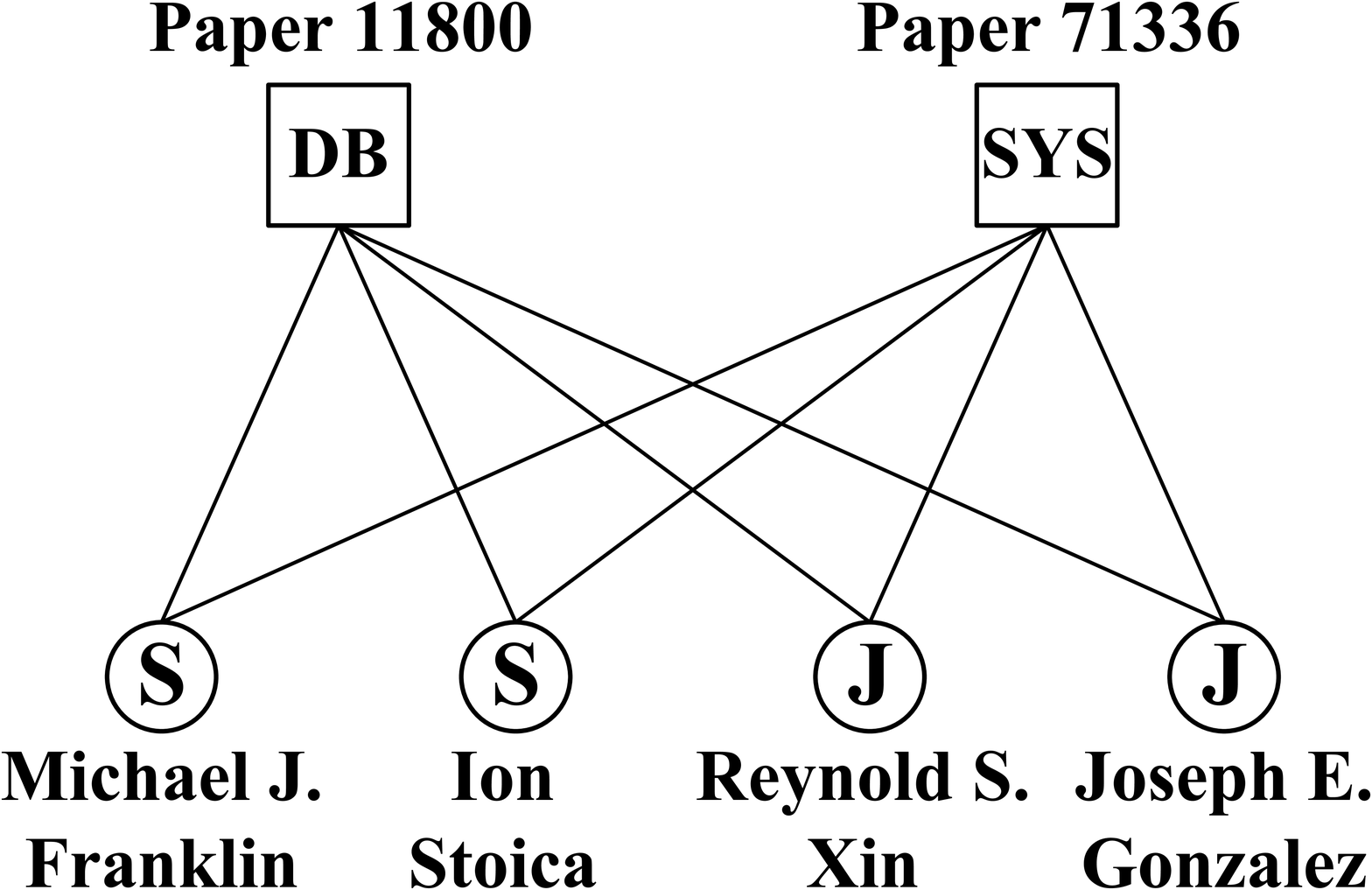}
            }
	}
	\vspace*{-0.3cm}
	\caption{Case studies on \dbda and \dbds.}
	\vspace*{-0.4cm}
	\label{fig:exp:csattrdblp}
\end{figure*}

\begin{figure*}[t!]
    \centering
    \subfigure[{\scriptsize{\job, the CF algorithm }}]{
		\includegraphics[height=1.6cm]{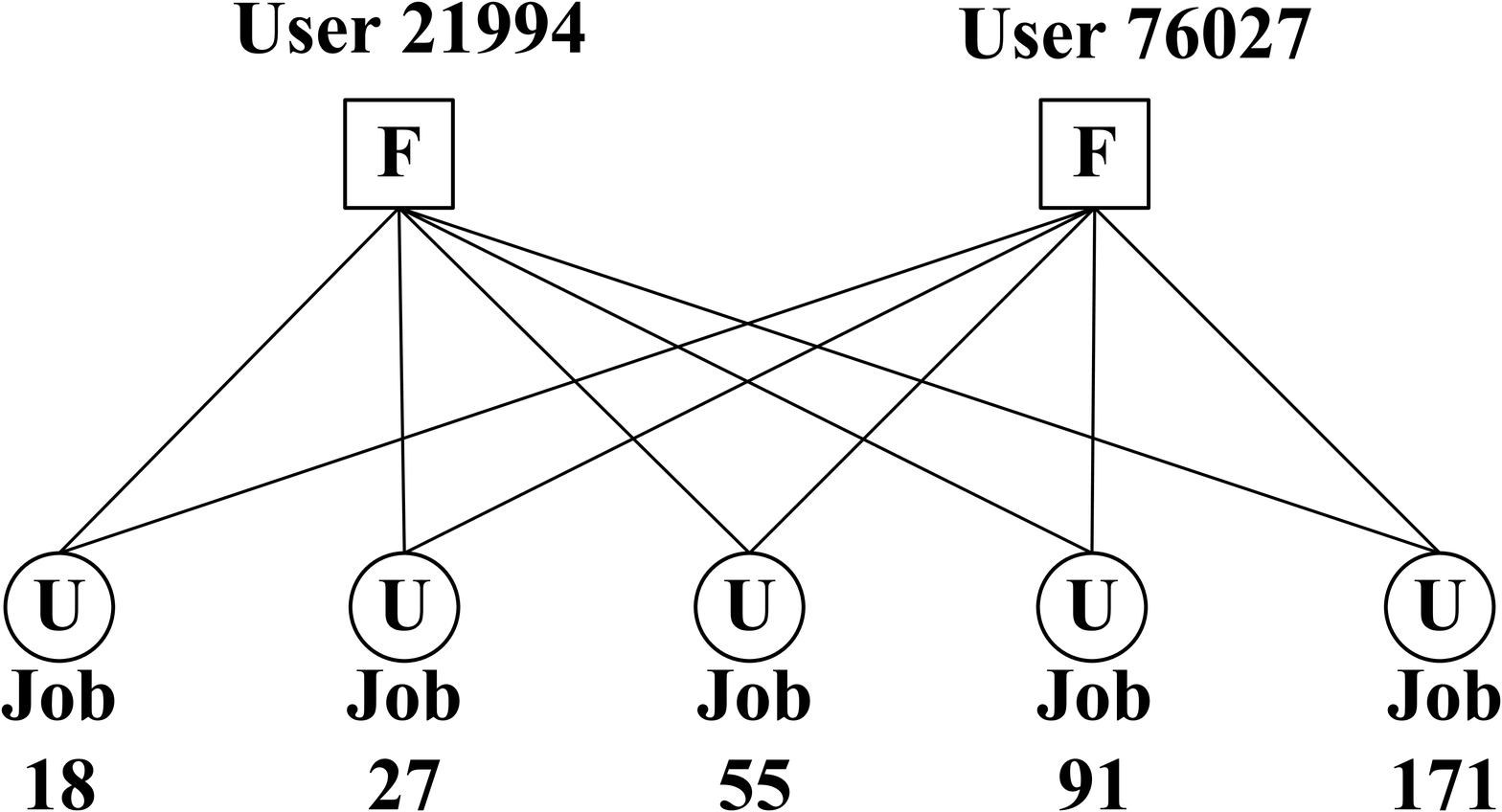}
		\label{fig:jobcase1}
  }    
    \subfigure[{\scriptsize{\job, \osbc, ($\alpha=2,\beta=2,\delta=1$)}}]{
	      \includegraphics[height=1.6cm]{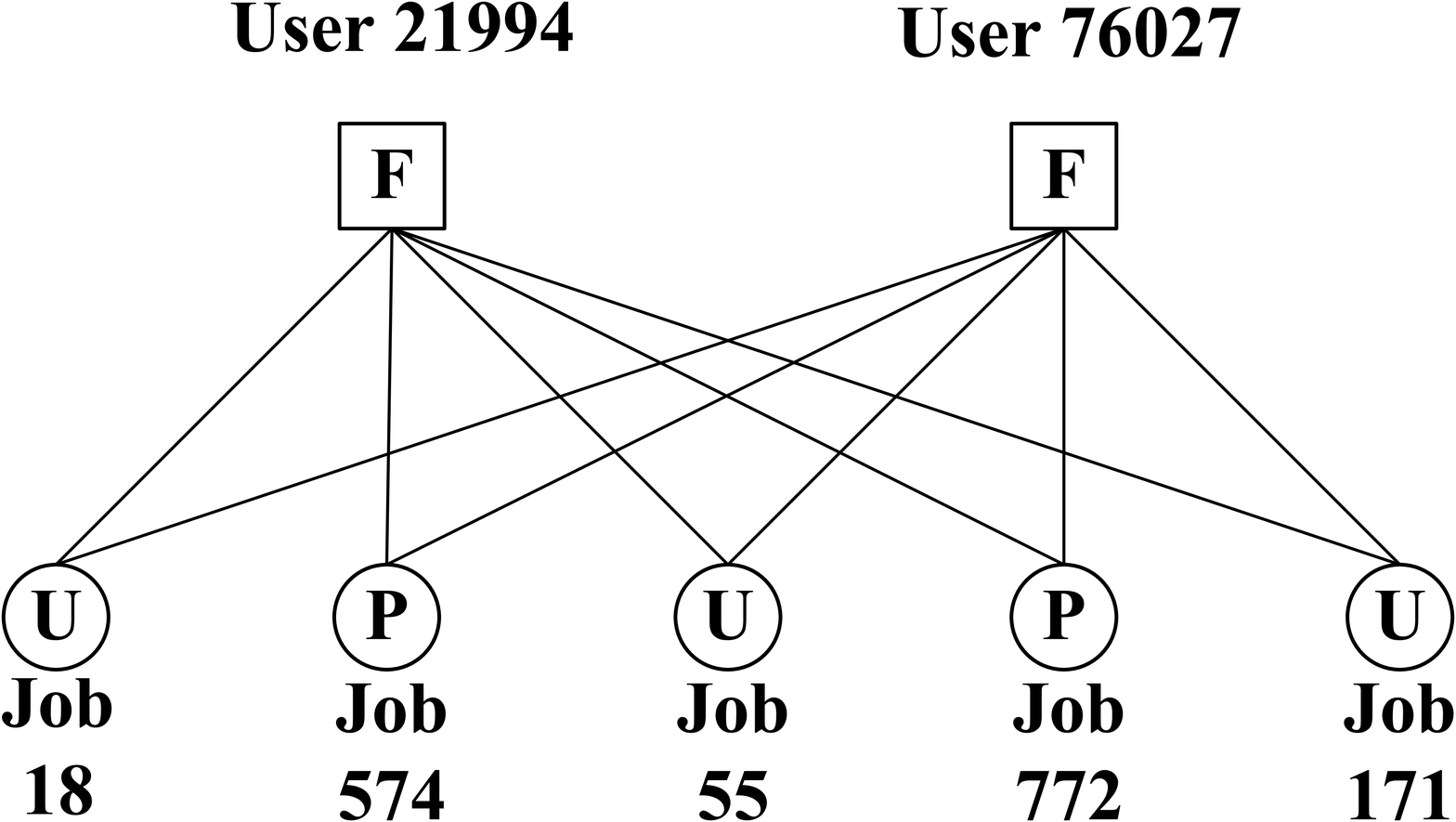}
	    \label{fig:jobcase2}
     }
    \subfigure[{\scriptsize{\movies, the CF algorithm}}]{
		\includegraphics[height=1.6cm]{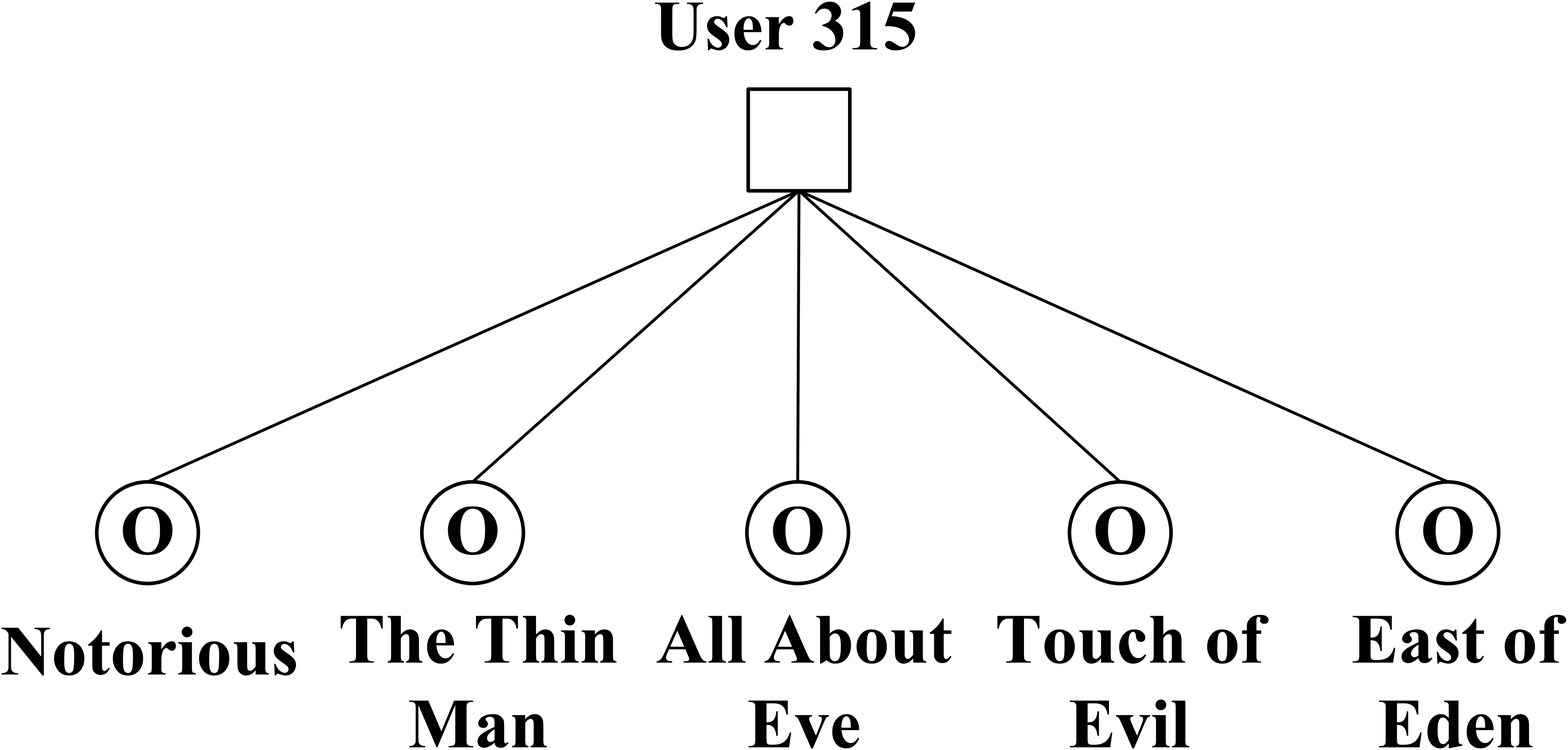}
		\label{fig:moviecase1}
  }
    \subfigure[{\scriptsize{\movies, the CF algorithm}}]{
		\includegraphics[height=1.6cm]{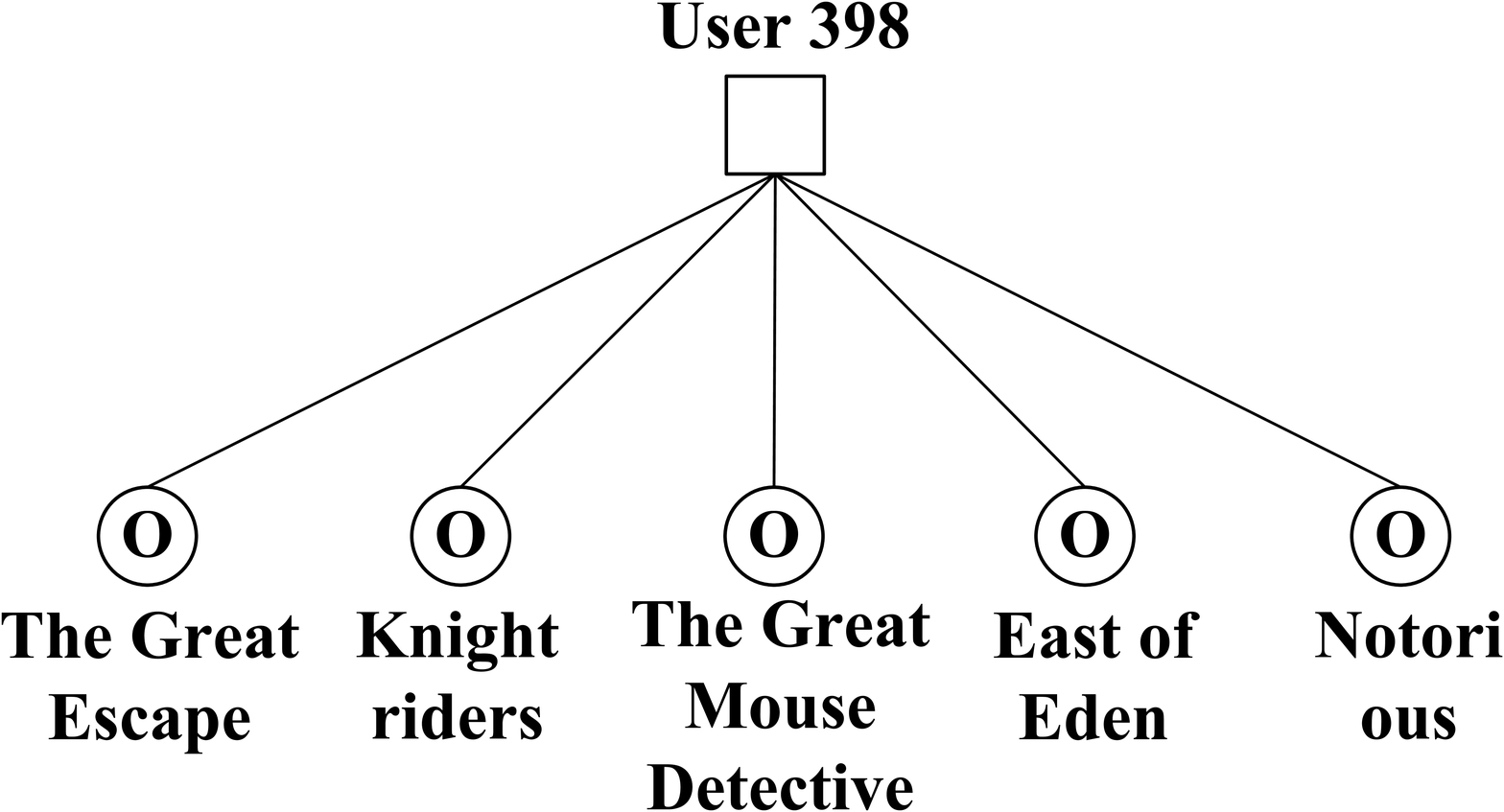}
		\label{fig:moviecase2}
  }    
    \subfigure[{\scriptsize{\movies, \osbc, ($\alpha=2,\beta=2,\delta=1$)}}]{
		\includegraphics[height=1.6cm]{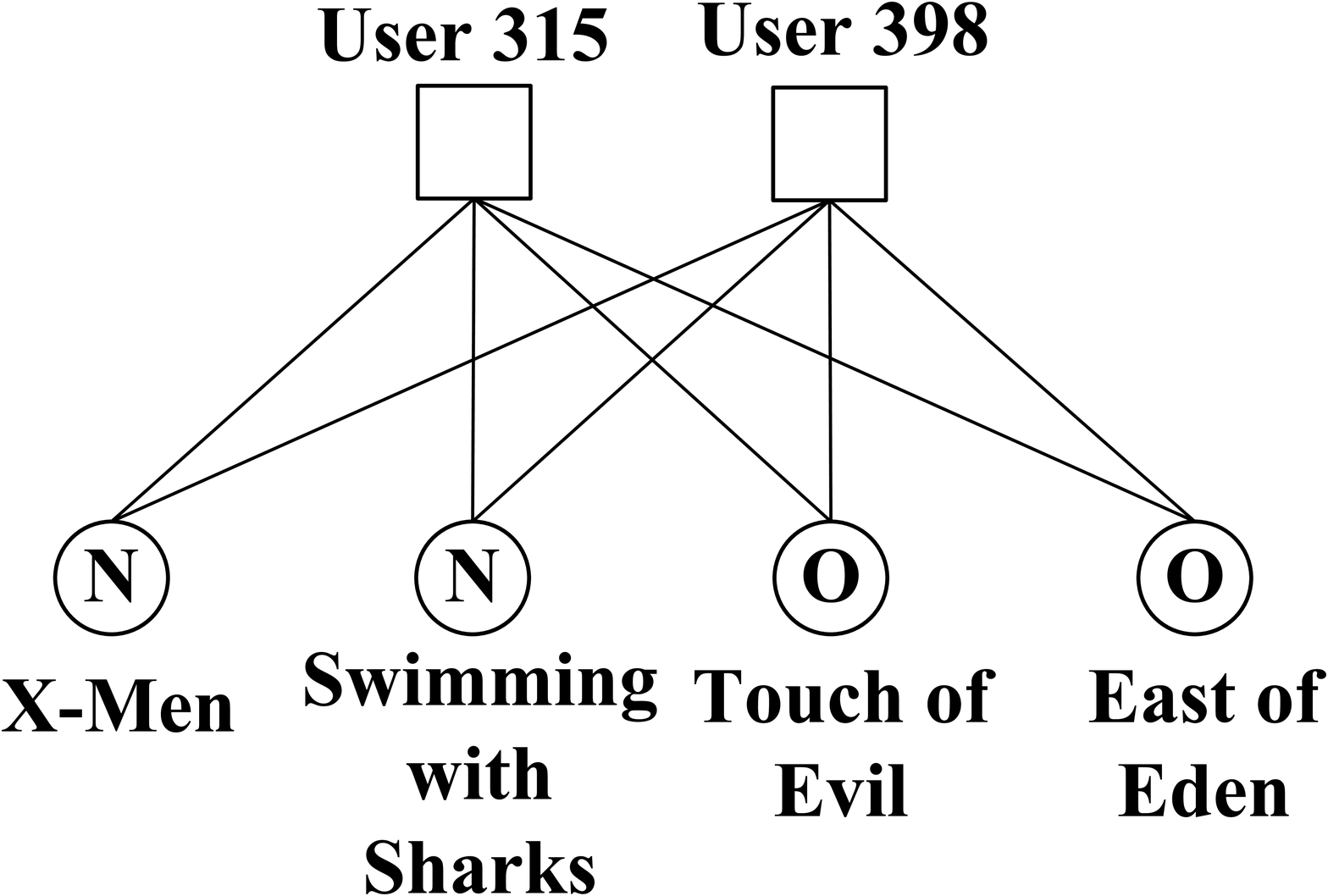}
		\label{fig:moviecase3}
  }
    \vspace*{-0.3cm}
    \caption{{Case studies on \job and \movies.}}
    \label{fig:exp:csattrmovie}
    \vspace*{-0.4cm}
\end{figure*}

\stitle{Exp-5: Scalability testing.} Here we evaluate the scalability of the proposed algorithms. To this end, we generate four subgraphs for each dataset by randomly picking 20\%-80\% of the edges, and evaluate the runtime of the algorithms for \nonesidebc~enumeration and \ntwosidebc~enumeration. \figref{fig:exp-scalability-test} illustrates the results on \dblp and the results on the other datasets are similar. For the \osbc~enumeration algorithms, as show in \figref{fig:exp-scala-vary-m-oneside-dblp}, the runtime of \onesideFBCEM increases smoothly as the graph size increases. while the runtime of \onesideFBCEMPP~keeps relatively stable with different values of $m$. Again, \onesideFBCEMPP~is at least 10 times faster than \onesideFBCEM~with all parameter settings, which is consistent with our previous findings. For the \osbc~enumeration algorithms, as can be seen from \figref{fig:exp-scala-vary-m-twoside-dblp}, the runtime of \twosideFBCEMPP~increases more smoothly w.r.t. the graph size than that of \twosideFBCEM. These results demonstrate the high scalability of the proposed algorithms.

\begin{figure}[t!]\vspace*{-0.2cm}
	\begin{center}
		\begin{tabular}[t]{c}
			\subfigure[{\scriptsize \youtube (vary $\theta$)}]{
				\includegraphics[width=0.4\columnwidth, height=2.5cm]{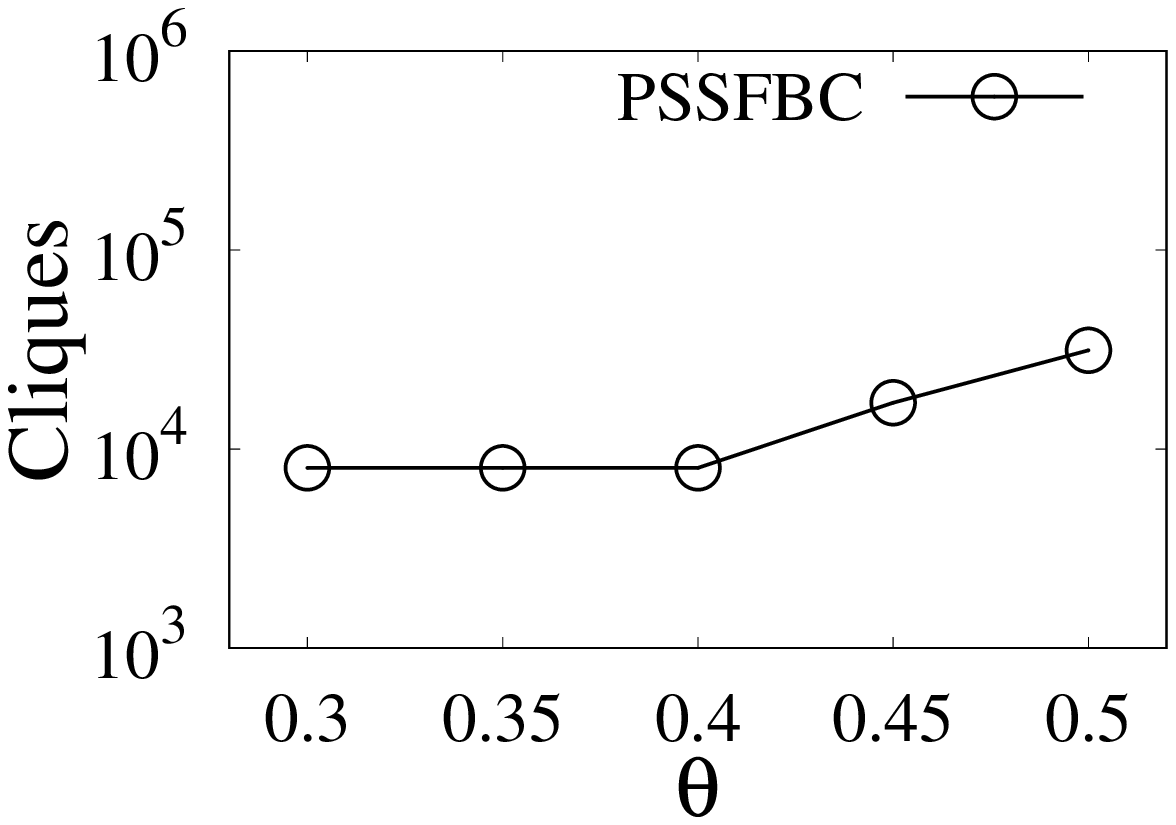}
			}
			\subfigure[{\scriptsize \youtube (vary $\theta$)}]{
				\includegraphics[width=0.4\columnwidth, height=2.5cm]{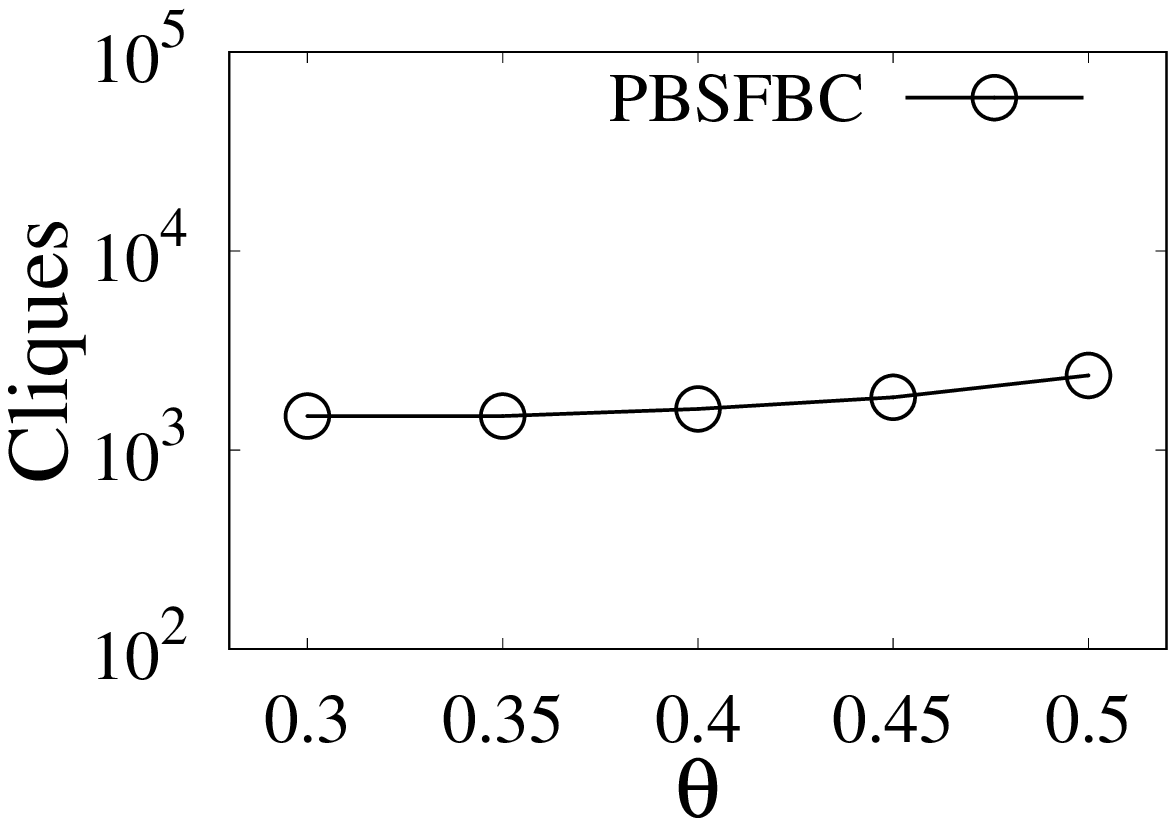}
			}
			\vspace*{-0.3cm} 
		\end{tabular}
	\end{center}
    \vspace*{-0.2cm}
    \caption{The number of {\osbcp}s~and {\tsbcp}s.}
    \vspace*{-0.4cm}
	\label{fig:exp-proportion-number}
\end{figure}
\begin{figure}[t!]\vspace*{-0.2cm}
	\begin{center}
		\begin{tabular}[t]{c}
			\subfigure[{\scriptsize \youtube, \onesideFBCEMPPPRO~algorithm (vary $\theta$)}]{
				\includegraphics[width=0.4\columnwidth, height=2.5cm]{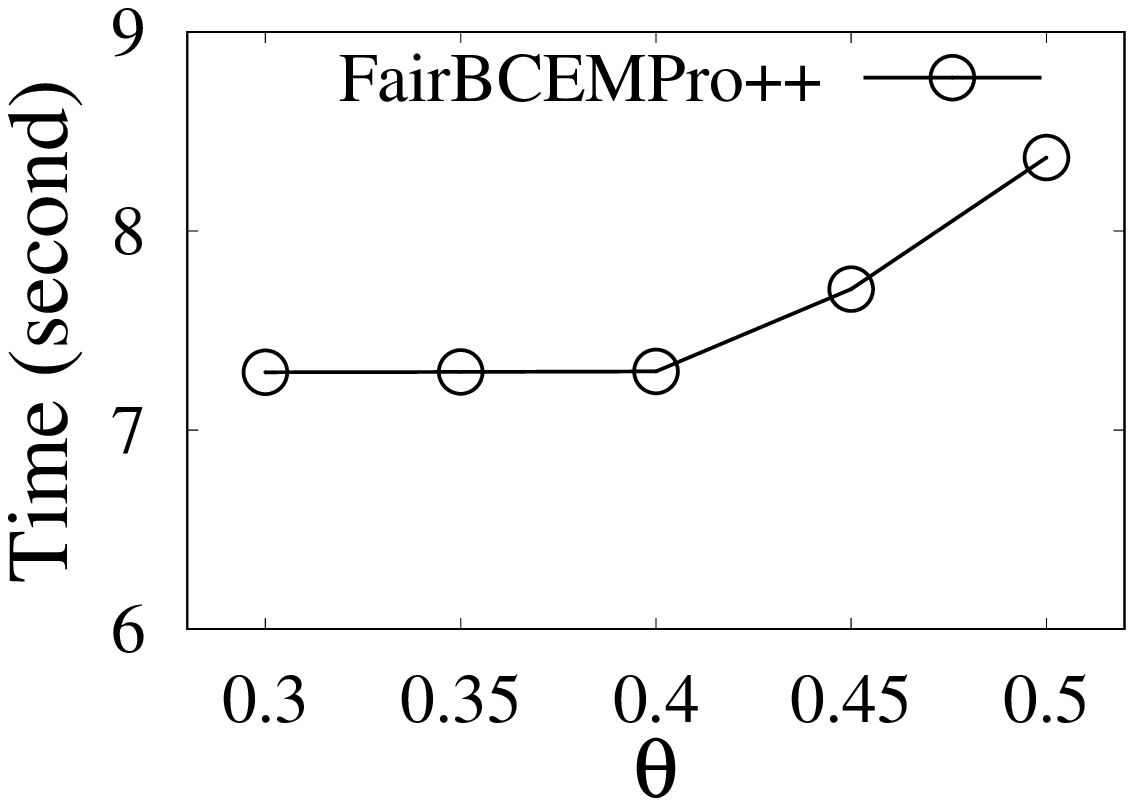}
			}
			\subfigure[{\scriptsize \youtube, \twosideFBCEMPPPRO~algorithm (vary $\theta$)}]{
				\includegraphics[width=0.4\columnwidth, height=2.5cm]{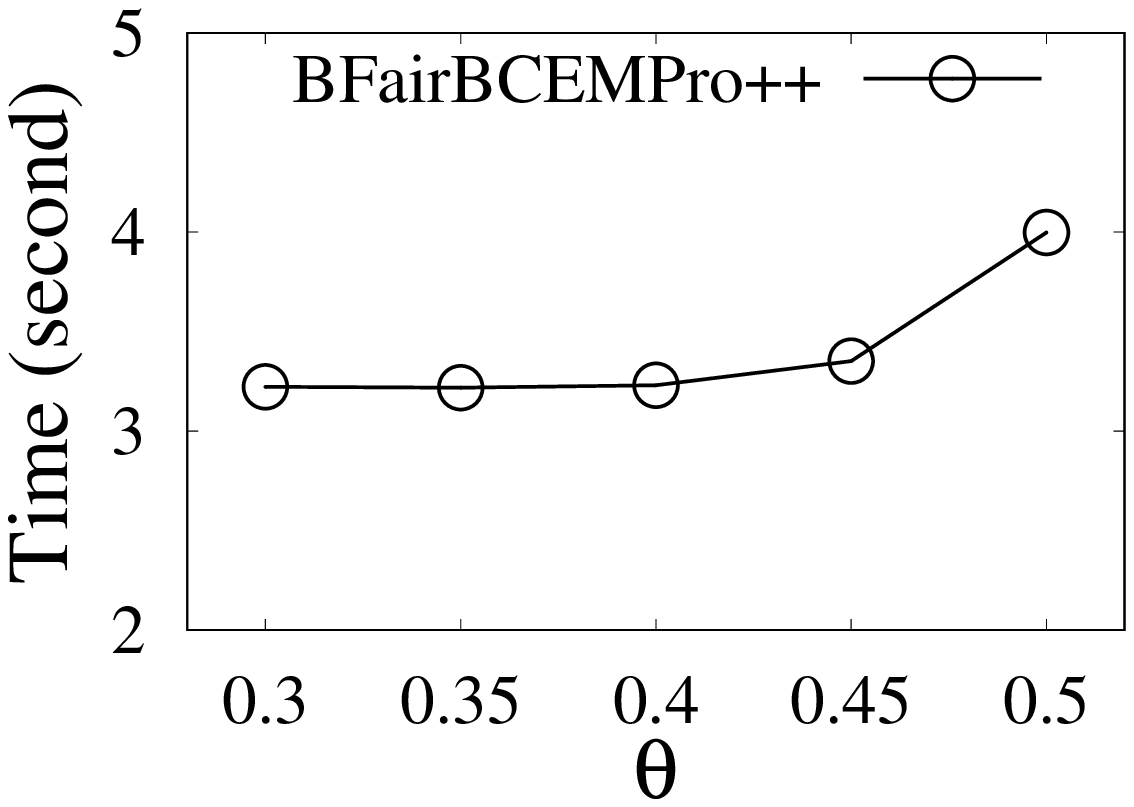}
			}
			\vspace*{-0.3cm}
		\end{tabular}
	\end{center}
	\vspace*{-0.2cm}
        \caption{The running time of \onesideFBCEMPPPRO~and \twosideFBCEMPPPRO.}
    \vspace*{-0.4cm}
	\label{fig:exp-proportion-time}
\end{figure}
\vspace{-0.1cm}
\stitle{Exp-6: Memory overhead.} \figref{fig:exp-Memory-overhead} shows the memory overheads of the enumeration algorithms on all datasets. Note that the memory costs of different algorithms do not include the size of the graph. From \figref{fig:exp-Memory-overhead}, we can see that the memory usages of \onesideFBCEM and \onesideFBCEMPP~are almost equal and are always larger than the original graph size. This is because they both perform the \cfcore pruning technique and enumerate {\onesidebc}s following a depth-first manner, thus the space overhead mainly depends on the data structures in \cfcore. These results are consistent with our analysis in \secref{sec:onesidealg}. Similar results can also be found for \twosideFBCEM~and \twosideFBCEMPP~algorithms.

\stitle{Exp-7: Evaluation of {\osbcp} and {\tsbc} enumeration algorithms.} Here we evaluate the \onesideFBCEMPPPRO~and \twosideFBCEMPPPRO~algorithms by varying the additional parameter $\theta$. Fig. 11 and Fig. 12 illustrate the number of {\osbcp}s and {\tsbcp}s and the running time of \onesideFBCEMPPPRO~and \twosideFBCEMPPPRO~on \youtube. The results on the other datasets are similar. As can be seen, the number of proportion fair bicliques and the runtime increase with the increasing $\theta$. When $\theta=0.5$, the {\osbcp}~enumeration problem degenerates to the {\osbc}~enumeration problem with $\delta=0$. Therefore, solving the \osbcp~enumeration problem takes a similar time as the \osbc~enumeration problem. The case is also similar to the \tsbcp enumeration problem. When $\theta$ approaches 0.5, more bicliques satisfy the definitions of proportion fair bicliques, thus the number of {\osbcp}s and {\tsbcp}s increases, and the running time of algorithms also increases.

{\comment{As expected, the running time of \onesideFBCEMPPPRO~and \twosideFBCEMPPPRO~increases with increasing $\theta$. This is because for a large $\alpha, \beta$, many vertices can be pruned by the \fcore and \cfcore pruning techniques and the search space can also be correspondingly reduced during the branch and bound procedure.}}

{\comment{
\stitle{Exp-8: Efficiency study.} 
We compare the naive search tree (\naivesearchtree) algorithm, which reserves all the pruning techniques such as Algorithm \ref{alg:fairalphabetacore} and Algorithm \ref{alg:colorfulprune} and drop off all pruning techniques in the search process such as Observation \ref{obs:obs1_maximal}, Observation \ref{obs:obs3_fulladdp} and Observation \ref{obs:obs4_alphabeta}, with our algorithm \onesideFBCEM~and \onesideFBCEMPP~to show our methods' efficiency. Due to the slow speed of the naive search tree algorithm, in most of the datasets, the \naivesearchtree algorithm costs more than 24 hours, so we only demonstrate it on the dataset \dblp which is the fastest among all datasets. The \naivesearchtree algorithm could be easily adapted for the \tsbc~enumeration problem, so we also the \naivesearchtree algorithm with \twosideFBCEM~and\twosideFBCEMPP~algotithms.
}}

{}

\subsection{Case study} \label{sec:casestudy} 
{\stitle{Case study on \dblp.}} We conduct a case study on a collaboration network \dblp to show the effectiveness of our algorithms. The \dblp dataset is downloaded from \url{dblp.uni-trier.de/xml/}. We construct a bipartite graph on \dblp by defining two type nodes, that is, the papers are on the upper side and the scholars are on the lower side. When a scholar is an author of a paper, there is an edge between them. Based on \dblp, We further construct two attributed bipartite subgraphs: \dbda and \dbds as follows. For \dbda, we keep the scholars that have published at least one paper on the database ($DB$), and artificial intelligence ($AI$) related conferences. Each scholar has an attribute $A_V$ with $A(V)=\{S, J\}$ where $S$ represents a senior scholar and $J$ indicates a junior scholar. We assign the attribute value for a scholar $v$ by identifying whether he/she has published papers for over 10 years. If yes, we set $v.val$ to $S$ otherwise the $v.val$ is $J$. Every paper is associated with an attribute $A_U$ with $A(U)=\{DB, AI\}$ to indicate that this paper is published in $DB$ and $AI$ related conferences. For \dbds, we only remain the scholars that have published at least one paper on the database ($DB$), and system ($SYS$) related conferences. Each scholar also has an attribute $A_V$ with $A(V)=\{S, J\}$ and We assign the attribute value for scholars by the method for \dbda. Each paper has an attribute $A_U$ with $A(U)=\{DB, SYS\}$ to indicate that this paper is published in $DB$ and $SYS$ related conferences. Finally, the \dbda has 260,605 papers and 240,420 scholars with 781,378 edges, i.e., $|U|=240,420$ and $|V|=260,605$. And the \dbds contains 163,545 papers and 139,703 scholars with 433,928 edges, i.e., $|U|=163,545$ and $|V|=139,703$. We perform \onesideFBCEMPP~and \twosideFBCEMPP~algorithms to find all {\nonesidebc}s and {\ntwosidebc}s. 

As examples, \figref{fig:exp:csattrdblp} (a)-(b) and \figref{fig:exp:csattrdblp} (c)-(d) show one {\nonesidebc}~and one {\ntwosidebc} on \dbda and \dbds respectively. We do not illustrate the title of papers since the title is too long. In \figref{fig:dblpcase1}, we can see that there are five senior scholars and three junior scholars, which is clearly a \nonesidebc~of \dbda with $\alpha=3, \beta=3, \delta=2$. From their homepages, all scholars in \figref{fig:dblpcase1} are interested in database-related areas, which is consistent with the attributes of papers they connected. The senior authors, such as Michael Stonebraker and Samuel Madden are indeed well-known scholars in the field of the database. This result indicates that our \onesideFBCEMPP~can find {\nonesidebc}s which guarantee the fairness of one side in real-world applications. The bipartite in \ref{fig:dblpcase2} is a \ntwosidebc which contains two senior scholars and two junior scholars in the lower side and one $AI$ paper\cite{ZhangNRS12} and one $DB$ paper \cite{NiuZRS12} in the upper side. Moreover, the professors Christopher Ré and Jude W. Shavlik are databases and artificial intelligence scientists, and Ce Zhang is relatively young compared with the former two scholars who are students of Christopher Ré. This result confirms that the proposed \twosideFBCEMPP~indeed can find {\ntwosidebc}s to ensure the fairness of two sides in real-world graphs. Similar results can also be found on \dbds. \figref{fig:dblpcase3} depicts a \nonesidebc~with five senior scholars and three junior scholars. In \figref{fig:dblpcase4}, there are two senior scholars and two junior scholars who have co-authored one $DB$ paper \cite{XinGFS13} published in SIGMOD and one $SYS$ paper \cite{GonzalezXDCFS14} published in OSDI. Among all scholars, the professors Michael Frankin and Ion Stoica are also well-known in data science and distributed systems areas. These results demonstrate the effectiveness of \nonesidebc~and \ntwosidebc~models and our proposed algorithms.

{\stitle{Case study on \job.}} We use a job recommendation dataset \job~to conduct a case study which can be downloaded from \url{https://www.kaggle.com/competitions/job-recommendation}. The dataset consists of 7 windows, and we consider window 1 for simplicity as each window is independent. We construct a bipartite graph $G$ by defining two type nodes, i.e., the user on the upper side and the job on the lower side. The attribute of jobs is popularity, which is set based on the number of applications for this position. In order to avoid cold start problem, we only reserve the top-1000 jobs with the highest number of applications and assign the top-500 jobs as more popular jobs (the attribute is $P$) and the others as less popular jobs (the attribute is $U$). We also assign each user an attribute value $A$ or $F$ to represent he/she is American or foreigner. Therefore, the bipartite graph $G$ contains 63,412 users and 1,000 jobs with $A_{V}=\{P, U\}$ and $A_{U}=\{A, F\}$. We use the Collaborative Filtering (CF) algorithm to calculate recommendation results which is shown in \figref{fig:jobcase1}. In \figref{fig:jobcase1}, there is an edge between a user and a job if the job lies in the top-5 recommendation jobs with the CF algorithm. From the information of \job, we can find that user 21,994 comes from India and has a master's degree with 9 years of work experience, and user 76,027 is a Canadian and has a master's degree with 23 years of work experience. Clearly, the two foreigners have similar education and work experience, but all the jobs recommended for them are less popular jobs. To eliminate the biases, we construct a bipartite graph $G'$ in which each edge represents that the job has the top-10 highest recommendation score computed by CF, i.e, $G$ contains 63,412 users, 1,000 jobs and 63,4120 edges. Then we perform \onesideFBCEMPP~to find {\osbc}s by setting the jobs as the fair side. A {\osbc} containing user 21,994 and user 76,027 is depicted in \figref{fig:jobcase2}. As expected, both more popular jobs and less popular jobs are recommended to the two foreigners. These results demonstrate the effectiveness of our fair biclique models and proposed algorithms.

{}

{\stitle{Case study on \movies.}} We also conduct a case study on a movie recommendation dataset \movies which can be downloaded from \url{https://www.kaggle.com/code/rounakbanik/movie-recommender-systems}. We construct a bipartite graph including the user on the upper side and the movies on the lower side. For each movie, we assign its attribute to $O$ to represent an old movie which is published before 1990, and otherwise, its attribute is set to $N$ to indicate a new movie. The bipartite graph consists of 9,000 movies and 700 users, i.e., $|U|=700$ and $|V|=9,000$. The recommendation result by the traditional CF algorithm is shown in \figref{fig:moviecase1} and \figref{fig:moviecase2}, an edge means that a movie lies in the top-5 recommendation answers for a user. As can be seen, for two users of similar interests, all five movies in \figref{fig:moviecase1} and \figref{fig:moviecase2} are old movies. The CF algorithm suffers from explosion bias, that is, already popular movies will get more chance to be recommended and relatively new movies get less recommendation chance even if they are of comparable quality, which is generally called cold start problem. To solve this problem, We connect each user with top-10 movies according to the personalized recommendation scores computed by CF and invoke \onesideFBCEMPP~to find {\osbc}s. A {\osbc} containing user 310 and user 512 is shown in \figref{fig:moviecase3}. By introducing fairness into the movie recommendation task, the new recommended movie ``X-men" is more desirable and famous compared with old movies. This result indicates that fair biclique models can relieve the problem of explosion bias.


\section{Related work} \label{sec:relatedwork}
\vspace{-0.1cm}
\stitle{Cohesive bipartite subgraph mining.} Our work is related to cohesive subgraph mining in bipartite graphs which has attracted much attention in recent years. For example, Zhang \etal \cite{ZhangPRBCL14} proposed a branch and bound algorithm, i.e., MBEA, to search all maximal bicliques. To accelerate the search efficiency, Abidi \etal \cite{AbidiZCL20} further presented a pivoting enumeration algorithm called PMBE which is based on the Containment Directed Acyclic Graph (CDAG). Yang \etal \cite{YangPZ21} investigated the problem of $(p,q)$-clique counting and proposed BCList and BCList++ algorithm which applies a layer-based exploring strategy and cost model to accelerate the searching process. Lyu \etal \cite{LyuQLZQZ20} presented a new algorithm to search maximum bi-clique which can be used to process bipartite graphs of billion scale. Wang \etal \cite{Wang00ZQZ21} developed a novel index structure to help finding the $(\alpha,\beta)$-community which is a minimum edge weight $(\alpha, \beta)$-core. Wang \etal \cite{abs181200283} proposed a vertex-priority-based paradigm BFC-VP to accelerate butterfly counting by a large margin. All the algorithms mentioned above do not consider the fairness of cohesive subgraphs and they are mainly tailored to non-attributed bipartite graphs. To the best of our knowledge, the definition of fairness-aware biclique is proposed for the first time, and also our work is the first to study the problem of finding fairness-aware biclique in bipartite graphs.

\comment{
There are also some important maximal clique researches, \cite{MakinoU04} propose three new algorithms to search maximal cliques which are both excellent in the sparse and dense graph. 
\cite{DanischBS18} proposes the most efficient parallel algorithm to list all k-cliques, \cite{abs210710025} first propose the fairness concept in maximal clique search area. In this paper, we propose a fairness-aware bi-clique model, it's notable that we are the first to model a fairness-aware bi-clique. There are also some other applications for bipartite graphs, such as \cite{ChenWLZQZ21} study the query-answer problem in temporal bipartite graphs. develops novel fa algorithms cohesive subgraph model of a to compute maximal fair cliques in attributed bipartite graphs with several
non-trivial pruning techniques.
}
\stitle{Fairness-aware data mining.} Our work is inspired by a concept called fairness which has been widely studied in machine learning communities. Verma \etal \cite{verma2018fairness} proposed many concepts to better measure fairness. Zehlike \etal \cite{zehlike2017fa} presented a method to generate a ranking with guaranteed group fairness, which can ensure the proportion of protected elements in the rank is no less than a given threshold. Serbos \etal \cite{serbos2017fairness} investigated a problem of fairness in the package-to-group recommendation, and propose a greedy algorithm to find approximate solutions. Beutel \etal \cite{beutel2019fairness} also studied the fairness in recommendation systems and presented a set of metrics to evaluate algorithmic fairness. Another line of research on fairness is studied in classification tasks. Some notable works include demographic parity \cite{dwork2012fairness} and equality of opportunity \cite{hardt2016equality}. For instance, Hardt \etal \cite{hardt2016equality} proposed a framework that can optimally adjust any learned predictor to reduce bias. Our definition of fairness which requires the equality of different attribute values in a group is different from those in the above studies in the machine learning literature. In the field of data mining, Pan \etal \cite{abs210710025} introduced the fairness into clique model and proposed several algorithms to find fair cliques. Unlike their work, we focus on studying the fairness-aware biclique enumeration problem on bipartite graphs, and our techniques are significantly different from their techniques. 

\vspace{-0.1cm}
\section{Conclusion} \label{sec:conclusion}
In this paper, we study the problem of enumerating fairness-aware bi-cliques in bipartite graphs. We propose a \nonesidebc~model and a \ntwosidebc~model to introduce fairness to bipartite graphs. To enumerate all {\nonesidebc}s, we first present the \fcore and \cfcore pruning techniques to prune unpromising vertices, and then develop a branch and bound algorithm \onesideFBCEM to enumerate all {\nonesidebc}s in the pruned graph. To improve the efficiency, we present the \onesideFBCEMPP~algorithm to search all {\nonesidebc}s by using maximal cliques as candidates to reduce search space. For the \ntwosidebc~enumeration problem, we also propose \bfcore and \bcfcore pruning techniques and develop the \twosideFBCEM algorithm with a branch and bound technique. The improved algorithm, i.e., \twosideFBCEMPP, is also presented to find all {\ntwosidebc}s. We also consider the ratio of the number of vertices of each attribute to the total number of vertices and propose the \nonesidebcpro~and \ntwosidebcpro~models and enumeration algorithms. We conduct extensive experiments using five large real-life graphs, and the results demonstrate the efficiency, effectiveness, and scalability of the proposed solutions.
\vspace{-0.15cm}
\section*{Acknowledgement}
This work was partially supported by (i) National Key R$\&$D Program of China 2021YFB3301300, (ii) NSFC Grants U2241211, 62072034, U1809206, and (iii) CCF-Huawei Populus Grove Fund. Rong-Hua Li is the corresponding author of this paper.

\balance
\bibliography{fairbiclique}
\bibliographystyle{IEEEtran}



\end{document}